\documentclass[a4paper,12pt]{article}
\usepackage{jheppub} 
\usepackage{lineno}
\usepackage{braket}
\usepackage{caption}
\usepackage{subcaption}
\usepackage{eucal}
\usepackage{physics}
\usepackage{float}
\usepackage{mathpazo}

\def\Tr{\mathrm{Tr}}
\def\k{k_{\textsc{b}}}
\def\te{t_\textsc{e}}
\def\lg{\lambda_{\mathrm{g}}}
\def\lc{\lambda_{\mathrm{cl}}}
\def\lq{\lambda_{\mathrm{qu}}}
\def\It{I_{\mathrm{trunc}}}

\keywords{OTOC, Chaos, Lyapunov Exponent, Ehrenfest Time}

\title{\boldmath Out-of-time-order Correlators and Chaos in Quantum Billiards}

\author{Tasnim Anzum Ador, Nayeem Farid, and Tibra Ali}

\affiliation{Department of Mathematics and Natural Sciences, School of Data and Sciences,\\ Brac University, Dhaka 1212, Bangladesh.}

\emailAdd{tasnim.anzum@bracu.ac.bd, nayeemfarid919@gmail.com, tibra.ali@bracu.ac.bd}

\abstract{We examine three billiard systems -- the cardioid, diamond (Superman), and Sinai billiards -- all of which are known to be classically chaotic. We compute their classical Lyapunov exponents, and using out-of-time-order correlators (OTOCs) in the semi-classical regime, we also derive their quantum Lyapunov exponents. We observe that the classical and quantum Lyapunov exponents are in agreement, strengthening the role of OTOCs as a diagnostic for quantum chaos in billiard systems. At very low temperatures, the OTOC of the Sinai billiard shows sharp growth, a phenomenon absent in the other two billiards. We identify the source of this anomalous behaviour in the geometry of the ground state wave function of the Sinai billiard, which is more sensitive to the curvature of the billiard compared to the other billiards. We also remark on the late-time behaviour of the OTOCs and how the scrambling/Ehrenfest time is related to the temperature of quantum billiards.}

\dedicated{Dedicated to the memory of all the Bangladeshi students who were either martyred, maimed, or injured in the struggle to free their country from dictatorship.}

\begin{document}
\maketitle
\flushbottom

\section{Introduction and Overview}\label{sec:intro}
Originally introduced as a way of examining the semi-classical behaviour of superconductors by Larkin and Ovchinnikov more than five decades ago \cite{larkin1969quasiclassical}, the out-of-time-order correlator (OTOC) has featured prominently in the exploration of quantum chaos in recent years. In its recent renaissance, it has received significant attention initially motivated by the holographic description of black hole physics \cite{Shenker_2014,Shenker2_2014,kitaev,shenker2015stringy,Maldacena_2016}. OTOCs have been useful in capturing chaotic behaviour in a wide array of quantum systems such as the SYK model \cite{kitaev}, the kicked rotor \cite{Rozenbaum_2017}, spin chains \cite{Gharibyan_2019}, and quantum billiards \cite{Hashimoto_2017,Rozenbaum_2017,Jalabert_2018}, to name a few.

In general, OTOCs represent the growth of Heisenberg operators and measure the way quantum information spreads and gets scrambled over time \cite{swingle2018unscrambling}. The Lyapunov exponent, on the other hand, is an important quantity in classical chaos that measures the time scale (via the Lyapunov time) over which predictability breaks down. In many quantum systems there exist semiclassical regimes in which the growth of the OTOC has a similar exponential growth region and a ``quantum" Lyapunov exponent \cite{kitaev} can be identified. In \cite{Maldacena_2016}, Maldacena, Shenker and Stanford conjectured, from general grounds, that the quantum Lyapunov exponent is bounded from above by the temperature of the system. Black holes, believed to be nature's fastest scramblers, saturate this bound. Other systems that saturate the bound include the SYK model which has a holographic description in terms of black holes.

Chaotic billiards are some of the simplest examples of classically chaotic systems. On general grounds one expects their quantum counterparts to exhibit chaos in the semiclassical limit: Ehrenfest's theorem guarantees chaotic dynamics in the quantum versions of chaotic billiards well before Ehrenfest time. But some of the first attempts to identify an exponential region and derive a Lyapunov exponent from the quantum stadium billiard proved to be difficult \cite{Hashimoto_2017}. However, Jalabert et.\ al.\ \cite{Jalabert_2018} were able to develop a method for dealing with the semiclassical limit of low-dimensional systems and they identified a region in which the thermal OTOC exhibits exponential growth.  Therefore it was possible to derive a quantum Lyapunov exponent for the quarter stadium billiard. (For a review of their method and the role of OTOC in quantum chaos in general see \cite{garciamata2022outoftimeorder}.) Rozenbaum et.\;al.\;\cite{Rozenbaum_2019} introduced a state-independent way of characterizing quantum chaos by studying the level statistics of a new class of ``Lyapunovian" operators. Based on this, they were able to derive the Lyapunov exponent using a non-thermal OTOC. 

In this paper, building on the methods of \cite{Hashimoto_2017,Jalabert_2018,garciamata2022outoftimeorder,Rozenbaum_2019}, we examine three billiard systems, other than the stadium billiard, which are known to be classically chaotic: the cardioid, diamond, and Sinai billiards. While there are many different methods for approaching quantum versions of classically chaotic systems, our motivation in this paper is to utilize the new and emerging tool of OTOC to derive the Lyapunov exponents of the simplest chaotic systems. We compute the quantum Lyapunov exponents using the semi-classical expansion of the OTOCs\footnote{We are grateful to the authors of \cite{Jalabert_2018,garciamata2022outoftimeorder} for explaining details of their calculation.}. We also compute the classical Lyapunov exponents to compare them to their quantum counterparts. For the Sinai and cardioid billiards, the quantum Lyapunov exponents agree well with their classical counterparts, but the agreement for the case of the diamond billiard is not as good. We rechristen the diamond billiard as the Superman billiard as its shape reminds us of the comic book hero's logo.

We also find that while at low temperatures the OTOCs do not have sharp growths in general, the Sinai billiard is an exception. By examining the \emph{microcanonical} OTOCs, we show that the origin of this anomalous behaviour lies in the geometry of the ground state wave function of the Sinai billiard which is sensitive to the curvature of the billiard in contrast to the ground states of the other two billiards under study.

Lastly, we give a couple of heuristic arguments as to why the Ehrenfest time $\te$ (\emph{i.e.,} the scrambling time $t^*$) of chaotic billiards should have $1/\sqrt{T}$ dependence on the temperature.

\medskip
\noindent\textbf{Outline:}
\medskip

\noindent The outline of the paper is as follows. In Section \ref{sec:OTOC}, we introduce the OTOC and explain its connection to the quantum Lyapunov exponent in the semi-classical limit. This is followed in Section \ref{sec:review} by fundamental concepts from quantum chaos that we shall be using for computation. We then present, in Section \ref{sec:three-billiards}, a review of the three classical billiards which we shall be studying. We use the cardioid billiard as an example for demonstrating some of the fundamental concepts. In Section \ref{sec:classical}, we compute the classical Lyapunov exponents of these billiards. In Section \ref{sec:numerics}, we present the numerical calculations of the OTOCs and we examine the ground states of the three billiards. In Section \ref{sec:lyapunov} we calculate the quantum Lyapunov exponents and compare them with our classical computation, and in Section \ref{ehrenfest} comment on the temperature dependence of the Ehrenfest time. We conclude our paper with some observations in Section \ref{sec:discussion}.

\section{Review of Quantum Chaos and OTOC} \label{sec:OTOC}

The \emph{Lyapunov exponent} is an important measure of chaos in classical systems. Let us consider a system with a one-dimensional phase space \(x\). Two states of this system, starting from initial points \(x_0\) and \(x_0 + \delta x(0)\), evolve with time. Here \(\delta x(0)\) is very small, and it is the separation of the two states at time \(t = 0\). For a chaotic system, this separation increases exponentially with time,
\begin{equation}\label{deltax}
\delta x(t) \approx e^{\lambda t} \delta x(0)
\end{equation}
where \(\delta x(t)\) is the separation of the two states at time \(t\), and \(\lambda\), a positive, real constant, is the \emph{Lyapunov characteristic exponent}, or the Lyapunov exponent for short. When considering a chaotic system with a multidimensional phase space, one can associate a Lyapunov exponent with each direction of phase space.

In the study of quantum chaos, however, the Lyapunov exponent may not seem like a useful or even a measurable quantity at first. This is because for quantum systems the notions of trajectories and phase spaces are not generally well-defined concepts. But using out-of-time-order correlators (OTOCs) \cite{larkin1969quasiclassical}, one \emph{can} extract in many cases the Lyapunov exponent of quantum chaotic systems and compare them to their classical exponents. As mentioned in the introduction, OTOCs have captured significant attention recently due to their relevance in studying the relationship between black-hole horizon geometry and chaos \cite{Shenker_2014,Shenker2_2014,shenker2015stringy}. Black holes were found to exhibit the same early exponential growth of OTOCs observed in classically chaotic systems, which led to terms such as ``scrambling", ``quantum butterfly effect" and ``quantum Lyapunov exponent". This behaviour classifies black holes as fast scramblers, where information is quickly spread. Furthermore, Maldacena, Shenker, and Stanford \cite{Maldacena_2016} conjectured in 2016 that there is an upper limit to the OTOC growth rate, which is essentially determined by the temperature of the system. Explicitly, this bound is given by $\lambda \le 4 \pi \k T/\hbar$.

The recognition of OTOCs and the quantum Lyapunov exponent in the arena of black holes and holography has led to their usefulness in quantifying quantum chaos in billiard systems which are known to be classically chaotic. Recently examinations of the stadium billiard \cite{Hashimoto_2017,Jalabert_2018,Rozenbaum_2019} have paved the way for the application of these methods to other billiard systems which are known to be classically chaotic.

\subsection{Definition of OTOC} 

The out-of-time-order correlator is defined as
\begin{equation}\label{otoc1}
    C^{(AB)}_T(t) = \langle [{B_t},{A}]^\dag [{B_t},{A}] \rangle.
\end{equation}
In the above equation, \({A}\) and \({B}\) are Heisenberg operators (time-dependent operators), where the subscript \(t\) means that the respective operator has been time-evolved to time \(t\) with respect to the other operator. The angular brackets denote the thermal average of the quantity inside the brackets, which is calculated as \(\langle{\mathcal{O}}\rangle \equiv \Tr[\rho\,{\mathcal{O}}]\) for an operator \({\mathcal{O}}\) in the canonical ensemble. Here, \(\rho=Z^{-1}e^{-\beta {H}}\), where \(Z\) is the partition function, and \(\beta = {1}/{\k T}\), where \(\k\) is the Boltzmann constant and \(T\) is the temperature of the system. It is also possible to consider temperature-independent OTOCs as was done in \cite{Rozenbaum_2019}, but in this paper we restrict ourselves to thermal OTOCs (see Section \ref{sec:lyapunov}).

Our goal is to use OTOCs to analyse the behaviour of measurable physical quantities in quantum systems. As these quantities are represented by Hermitian operators, we can limit our definition of the OTOC in Eq.\ (\ref{otoc1}) (which is valid for any two operators on the Hilbert space of the system) to the scenario where the operators \({A}\) and \({B}\) are Hermitian. The definition (\ref{otoc1}) then becomes
\begin{equation}\label{otoc2}
    C_T(t)=-\langle[{B}_t,{A}]^2 \rangle
\end{equation}
where $T$ refers to the temperature of the system. For the sake of notational clarity we shall drop the superscripts as it will be obvious from the context which operators are involved in a particular OTOC.  In quantum systems the growth of the OTOC is linked to the diffusion of quantum information known as information scrambling \cite{swingle2018unscrambling}.

\subsection{Connection between OTOC and classical chaos}
When studying the behaviour of an OTOC in the context of the classical limit, it is possible to utilize the quasi-classical limit of $\hbar \rightarrow 0$, as was originally noted when introducing the OTOC \cite{larkin1969quasiclassical}. In this limit, we can replace the commutator of two operators with the Poisson bracket of the corresponding classical quantities,
\begin{equation}
    \lim_{ \hbar \to 0} \dfrac{1}{i\hbar}[{B},{A}]\rightarrow\{B,A\}. 
\end{equation}
By selecting the position and momentum operators, ${B}={x}$ and ${A}={p}_x$, and utilizing the quasi-classical limit, we get 
\begin{equation}
   \lim_{ \hbar \to 0} [{x}_t, {p}_x] \rightarrow i\hbar \{x(t),p_x(0)\} = i\hbar \dfrac{\partial x(t)}{\partial x(0)}.
\end{equation}
In the case of a fully classically chaotic system, we have $\partial x(t)/\partial x(0) \thicksim \exp(\lambda t)$, denoting the exponential sensitivity to initial conditions.
Thus, the OTOC in the quasi-classical limit is given by
\begin{equation}\label{quasi-classical}
    C_T^{\text{qc}}(t) \thicksim \hbar^2 \exp(2 \lambda t).
\end{equation}
The above equation is noteworthy, as it relates the OTOC (a quantum entity) on the left-hand side, to the Lyapunov exponent (a classical characteristic) on the right-hand side. Thus, it links the classical and quantum counterparts of a system.

\section{Out-of-Time-Order Correlators for Billiard Systems}\label{sec:review}

\subsection*{Units for the rest of this paper}
Going forward, we adopt the units: $\hbar=\k= 2m = 1$, where $m$ denotes the mass of the particle. We consistently set the area of the billiard to \(A=1\).

\subsection{Formalism for OTOC computations}\label{sec:otoc}

Let \(H\) be a time-independent Hamiltonian where $H=H(x_1,\dots,x_n,p_1,\dots,p_n)$. As in \cite{Hashimoto_2017} with the choice of ${B}_t=x(t)$ and ${A}=p(0)$ as operators, the OTOC in eq.\ (\ref{otoc2}) becomes:
\begin{equation}
    C_T(t)=-\langle[x(t),p(0)]^2 \rangle.
\end{equation}
In the following, we will leave out the argument of Heisenberg operators when $t = 0$ and denote them as $\mathcal{O}\equiv \mathcal{O}(0)$. We reformulate the OTOC in the basis of energy eigenstates of $H$:
\begin{equation}\label{cT}
    C_T(t)=\dfrac{1}{Z}\sum_n e^{-\beta E_n}c_n(t)\: \text{,}\qquad c_n(t)\equiv \bra{n}[x(t),p]^2\ket{n}\: ,
\end{equation}
where $H\ket{n}=E_n\ket{n}$. The OTOC associated with a specific energy eigenstate, $c_n(t)$, will be called a \emph{microcanonical OTOC}. $C_T(t)$ will be referred to as a thermal OTOC \cite{Hashimoto_2017}.
\par
With the aid of the completeness relation, a microcanonical OTOC can be expressed as
\begin{equation}\label{cn}
    c_n(t)=\sum_m b_{nm}(t) b^*_{nm}(t)\text{ ,\hspace{0.4cm} }b_{nm} \equiv -i \bra{n}[x(t),p(0)]\ket{m}
\end{equation}
$b_{nm}(t)$ is Hermitian: $b_{nm}(t) = b^*_{mn}(t)$. We substitute $x(t)=e^{iHt}xe^{-iHt}$ and use the completeness relation again to write $b_{nm}(t)$ as
\begin{equation}\label{bmn}
    b_{nm}(t)=-i\sum_k\left(e^{iE_{nk}\,t}x_{nk}p_{km}-e^{iE_{km}\,t}p_{nk}x_{km}\right),
\end{equation}
where $E_{nm}\equiv E_n-E_m$, $x_{nm}\equiv \bra{n}x\ket{m}$, and $p_{nm}\equiv \bra{n}p\ket{m}$.

The expression in (\ref{bmn}) contains matrix components of $p$, which are not ideal because numerical derivatives of wave functions can lose accuracy. For a Hamiltonian of the form
\begin{equation}
    H=\sum_{i=1}^N p^2_i+U(x_1,\dots , x_N)\text{,}
\end{equation}
$p_{nm}$ can be calculated using $x_{nm}$. We apply $\bra{m}\dots\ket{n}$ to both sides of $[H,x]=-2\mathrm{i}p$, and obtain
\begin{equation}\label{pnm}
    p_{mn}=\dfrac{i}{2}E_{mn}x_{mn}.
\end{equation}
Substituting Eq. (\ref{pnm}) into Eq. (\ref{bmn}), we get
\begin{equation}\label{bnm}
    b_{nm}(t)=\dfrac{1}{2}\sum_k x_{nk}x_{km}\left(E_{km}e^{iE_{nk}\,t}-E_{nk}e^{iE_{km}\,t}\right).
\end{equation}
Once we know the matrix elements of $x$ and the energy spectrum $E_n$, OTOCs can be computed using Eqs.(\ref{cT}), (\ref{cn}), and (\ref{bnm}). However, in actual numerical calculations a sufficiently large cut-off for the number of energy eigenvalues and eigenstates considered  must be chosen. For the circle and stadium billiards the OTOCs were calculated in ref.\ \cite{Hashimoto_2017}.

\subsection{Time windows}
\begin{figure}[htbp]
    \centering
    \includegraphics[scale=0.4]{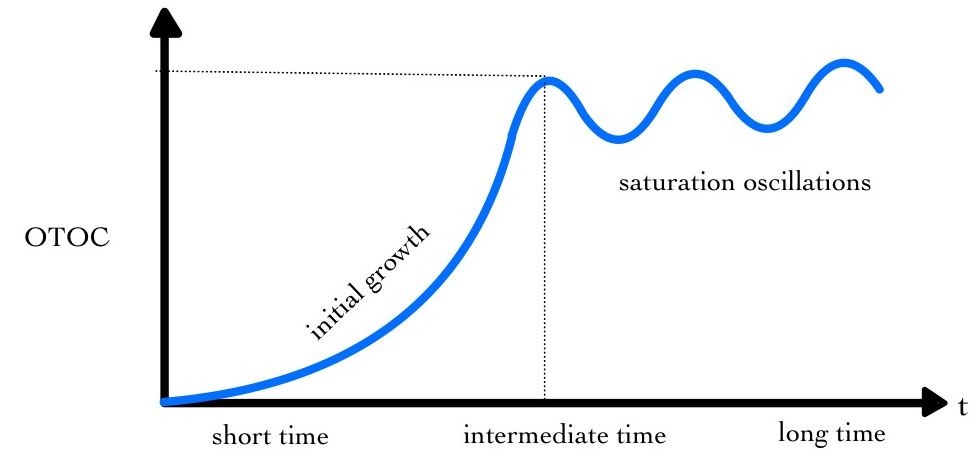}
    \caption{A blueprint for the time-windows of the OTOC. If the system is chaotic, it is anticipated that the initial growth will be exponential. The growth may take on other shapes if the system is not chaotic. The OTOC will stabilize and exhibit oscillations around a constant value after the scrambling time $t^*$. In the chaotic cases, the oscillations are highly suppressed, and the OTOC progresses toward an almost constant value. The figure shown here is a replica of a figure in \cite{garciamata2022outoftimeorder}.}
    \label{time-window}
\end{figure}

The quasi-classical form (\ref{quasi-classical}) of the OTOC suggests that classical properties can emerge in quantum systems. However, this behaviour is only present in a specific type of system and within a particular time-frame as discussed in \cite{garciamata2022outoftimeorder,time-window}. (Also see \cite{Maldacena_2016} which discusses the hierarchies of different time-frames.) Therefore, it is essential to understand the various time-frames that can be observed in the evolution of the OTOC. Many previous studies on the time evolution of various systems have shown that the OTOC can generally be classified into three different time-windows, as illustrated in Fig.\;\ref{time-window}.
\subsubsection*{Short times}
The short-time window signifies the start of the process where the influence of the time-dependent Heisenberg operator expands, driven by the dynamics of the Hamiltonian. Using the BCH formula, at the beginning, the OTOC is predicted to increase following a power law. Following this initial stage, the OTOC will continue to increase steadily for a brief period before reaching scrambling time $t^*$. This behaviour has been extensively analyzed for short times, and predictions such as Eq.\;(\ref{quasi-classical}) have been made.\par
In \cite{Jalabert_2018}, the authors developed a semi-classical approach to computing the OTOC of systems with low number of degrees-of-freedom based on earlier ideas of semi-classical theory of chaos \cite{gutzwiller}. As a result they found a region of exponential increase of the OTOC characterized by a positive Lyapunov exponent.  This behaviour emerges after a much faster growth during an extremely short time window. The time window for detecting the Lyapunov regime is temperature-dependent, with longer intervals observed at lower temperatures. See Section \ref{ehrenfest} for a discussion of the temperature dependence of the scrambling time. 
\subsubsection*{Intermediate times}
At around the scrambling time $t^*$, the initial growth stops, and the OTOC becomes relatively stable, maintaining an approximately constant value, with some minor fluctuations in certain cases. This time interval is commonly referred to as ``intermediate times" and is particularly significant for highly chaotic systems, as it represents the duration between $t^*$ and complete relaxation.\par
The scrambling time $t^*$ marks the point at which information about the initial state becomes widely spread throughout the available space. For a single particle system with a bounded space and classical chaotic dynamics, this time corresponds to the Ehrenfest time $\te$. The Ehrenfest time is defined as the time it takes for a narrow, coherent wave-packet to spread over whole system. It is given by $\te = \lambda^{-1} \ln{(a/\hbar)}$ \cite{Maldacena_2016}, where the constant $a$ is determined by the system size and the initial wave-packet size.
\subsubsection*{Long times}
One aspect of the OTOC in strongly chaotic systems is that after the scrambling time, the value of the OTOC becomes constant. It has been observed that the system's dynamics greatly impact the long-time behaviour of the OTOC, whether it is chaotic or regular. Therefore, studying fluctuations in the long-time regime can provide a deeper understanding of the system's behaviour.\par
For one-body systems with fully chaotic classical dynamics, the saturation value of the OTOC scales with the system size and temperature in a linear fashion when $B = X$ and $A = P_X$. However, the temperature scaling can differ for other operator choices. A study described in ref.\ \cite{Markovi__2022} suggests that the integrability of the one-body dynamics and the complexity of the selected operators might alter the temperature scaling of the OTOC saturation value.\par
The long-time behaviour of the OTOC can provide accurate measures of quantum chaos, and analyzing its oscillations can quantify the shift from regular to chaotic dynamics. For non-chaotic systems, it is expected that OTOCs at  long times will exhibit strong oscillations, whereas for strongly chaotic systems these oscillations are expected to be suppressed  \cite{garciamata2022outoftimeorder}. This method is consistent with other indicators of chaos commonly used \cite{PhysRevResearch.3.023214}.

\section{The Three Billiards} \label{sec:three-billiards}
We now introduce the three classically chaotic systems which are the main characters of our paper: the \textbf{Sinai, cardioid}, and \textbf{diamond} billiards. In the subsequent sections, we calculate the classical Lyapunov exponents of these systems and then move on to compute their OTOCs. This facilitates a comparison between classical and quantum dynamics. As mentioned in the introduction, our examination of the OTOCs of these billiards is done along the lines of the analyses of the stadium billiard carried out in \cite{Hashimoto_2017,Jalabert_2018}. We will examine the growth rate of OTOC at low temperatures and its saturation behaviour at late time. The upper limit that we take for the temperature will be dictated by the existence of a semi-classical region which, as we shall see, starts to disappear at high temperatures. We will compare semiclassical results with numerical quantum calculations done on our selected billiards and connect these results with predicted limits on OTOC growth rate.

\subsection{Sinai billiard}
The Sinai billiard is based on the Lorentz gas system, proposed originally by Lorentz in 1905 as a model for the behaviour of a dilute electron gas in a metal. In this model, the electrons are assumed to interact with the fixed heavier atoms in a lattice but not with each other. 
\par
\begin{figure}[htbp]
    \centering
    \includegraphics[scale=0.2]{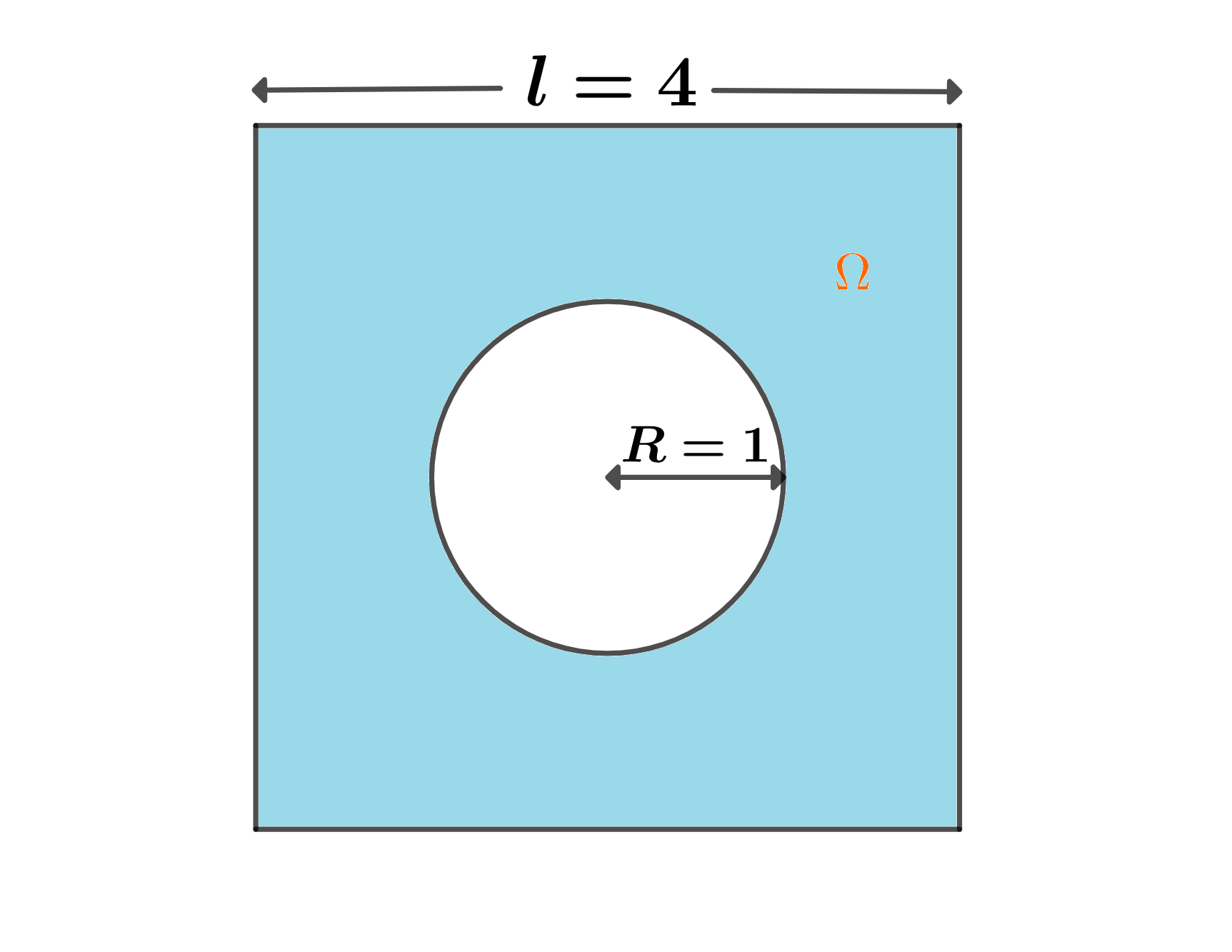}
    \caption{The Sinai billiard enclosure. The particle is confined within the shaded region $\Omega$, and the deformation parameter is $\ell/R=4$.}
    \label{sinai shape}
\end{figure}
The Sinai billiard is a simplified version of this system consisting of a single hard disk placed at the centre of a square box \cite{Sinai_1970}. The point-like particle moves freely within the region bounded by the box and the disk, and undergoes elastic reflections on the walls of this region. The deformation parameter is determined by $\ell/R$, where $\ell^2$ is the area of the box and $R$ is the radius of the disk. We fix the deformation parameter at $\ell/R=4$ for our numerical calculation, as depicted in Fig.\ \ref{sinai shape}. The units of length used here are arbitrary. They are fixed by the requirement that the area $A = 1$.
\par

The presence of a circular dispersing wall in Sinai billiards makes them chaotic for a finite deformation parameter $\ell/R$. They are also ergodic but non-hyperbolic. However, if we consider only particle collisions with disks and no flat walls, these billiards are known to be hyperbolic and are called infinite horizon billiards \cite{infinite_horizon_billiards}. In fact, any billiards with everywhere dispersing walls are chaotic and hyperbolic \cite{Sinai1970}. Our one-disk Sinai billiard 
exhibits the stickiness property due to the presence of flat walls. Roughly speaking, stickiness of a system is the propensity of some of its chaotic paths to spend considerable chunks of time very close to the non-chaotic regions of the phase space \cite{altmann2007intermittent}. This property has an impact on the system's dynamics, and as a result, accurately calculating the Lyapunov exponent is a challenging task.

\subsection{Cardioid billiard}
\begin{figure}[htbp]
    \centering
    \includegraphics[scale=0.3]{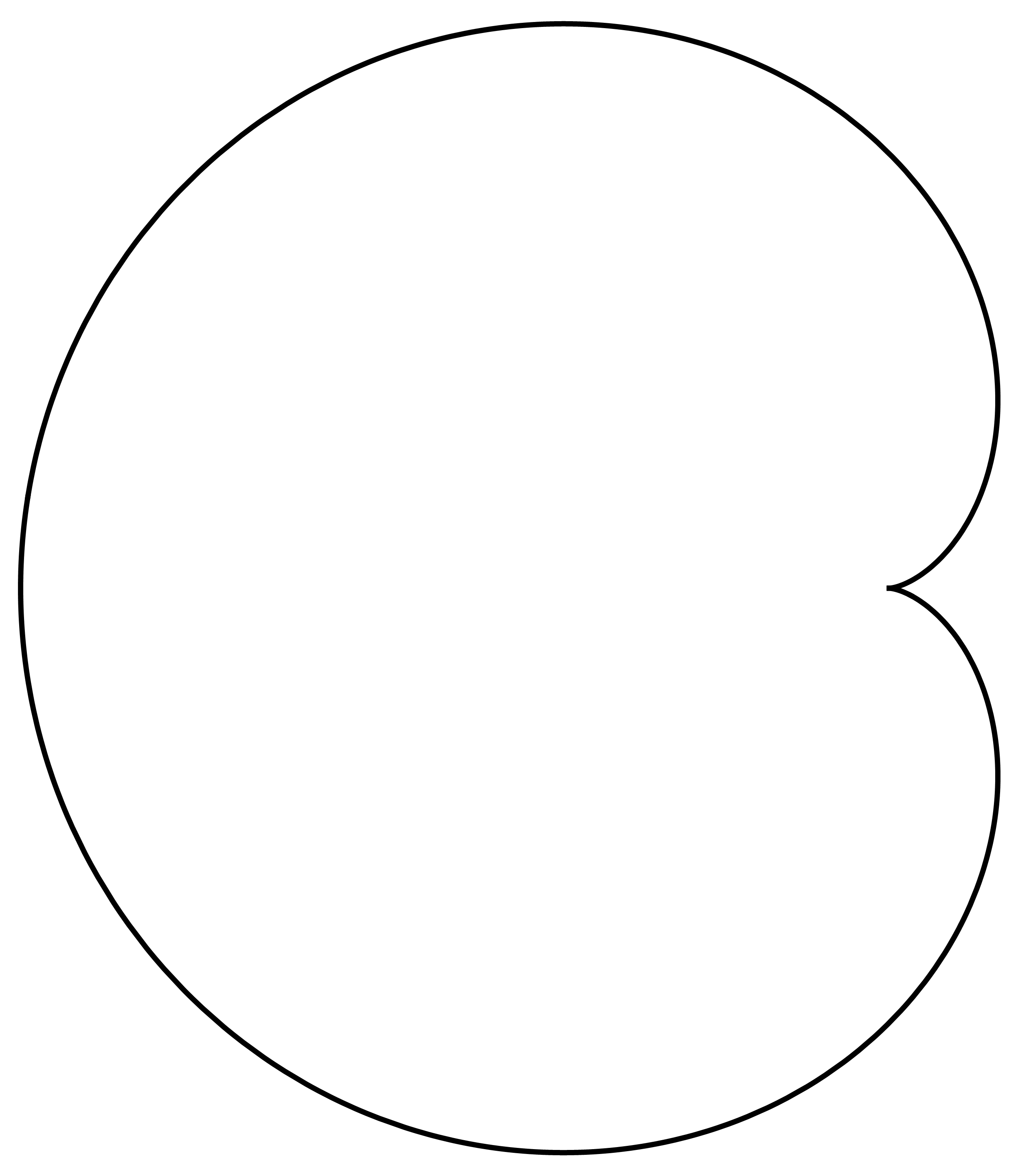}
    \caption{The cardioid shape}
    \label{cardioid_shape}
\end{figure}

Our next classically chaotic billiard system of interest is the cardioid billiard. The cardioid billiard consists of a particle confined within a cardioid, which is a 2-dimensional ``heart-like" shape, depicted in Fig.\ \ref{cardioid_shape}.\par

The cardioid billiard is part of the family of cos-billiards described by the parametric equation
\begin{equation}\label{cos-polar}
    r(\gamma)=1+\varepsilon \cos{\gamma},
\end{equation}
 specifically for the deformation parameter $\varepsilon=1$. Its boundary is made up entirely of focusing walls that intersect at a single point, forming a cusp. It has been proven that the cardioid billiard is both ergodic and mixing\footnote{In ergodic motion, the trajectory successively fills the phase space. The nature of the movement during mixing is different. Initially, the system covers the whole space with a grid of trajectories for a particular duration of \(t=T/2\). Then after a time \(t=T\), the phenomenon is repeated, and the cell sizes of the grid are roughly halved. Mixing is more potent than ergodicity, meaning that the presence of mixing guarantees ergodicity, but the inverse is not necessarily true.} \cite{cardioid1,cardioid2,cardioid3}.\par

\subsection{Diamond/Superman billiard}
\begin{figure}[htbp]
     \centering
     \includegraphics[scale=0.6]{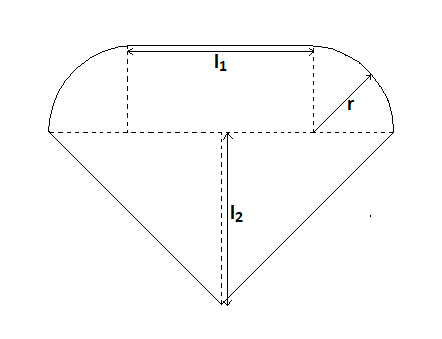}
     \caption{The diamond/Superman shape}
     \label{diamond shape}
 \end{figure}

The final billiard system we will be exploring in this paper is the diamond billiard. The diamond billiard is a classically chaotic system that is non-integrable, where a particle is confined within a two-dimensional enclosure that has a shape resembling a diamond or the outline of the logo of Superman. The enclosure is made up of a half-stadium combined with a triangular region at the bottom, as illustrated in Fig.\ \ref{diamond shape}.\par

We denote the radii of the two quarter-circles on each side of the shape by \(r\), the length of the horizontal line making up the very top of the boundary of the enclosure by \(\ell_1\), and the length of the vertical line bisecting the triangular region as \(\ell_2\), as shown in Fig.\ \ref{diamond shape}. We shall define a deformation parameter \(\sigma\) in the same way as was done in ref.\ \cite{salazar2012classical}. The relationships between \(\sigma\) and the quantities \(r\), \(\ell_1\) and \(\ell_2\) are given below.
\begin{align}
\begin{split}
r(\sigma) & = 1-\sigma\\ 
\ell_1(\sigma) & = \frac{5}{2}+\sigma\\ 
\ell_2(\sigma) & = \sqrt{\frac{3}{4}}\, \ell_1(\sigma)\\ 
\end{split}
\end{align}
Thus, the value of \(\sigma\) determines the shape of the enclosure. It takes values in the range \(0\le\sigma\le 1\), and the shape changes from a diamond to an equilateral triangle as \(\sigma\) goes from \(0\) to \(1\). In our calculations, we took \(\sigma\) to be \(0\), thus giving us the diamond/Superman shape.
 
\section{Numerical Calculation of Classical Lyapunov Exponents}\label{sec:classical}
\subsection{Lyapunov exponents and hyperbolicity}\label{hyperbolic}
Lyapunov exponents describe a system's sensitivity to initial conditions. The Lyapunov exponents associated with a specific point in the phase space determine how rapidly the nearby trajectories diverge with time. In billiard systems, particles move in a straight line between collisions. As a result, the trajectories diverge linearly between collisions. Thus, the chaotic dynamics of a billiard system are contained within its collisions. For this reason, we use the collision index $n$, which denotes the number of collisions, as the parameter to measure sensitivity to initial conditions instead of time  \cite{salazar2012classical,salazar2}.\par 

\begin{figure}[htbp]
    \centering
    \includegraphics[scale=0.8]{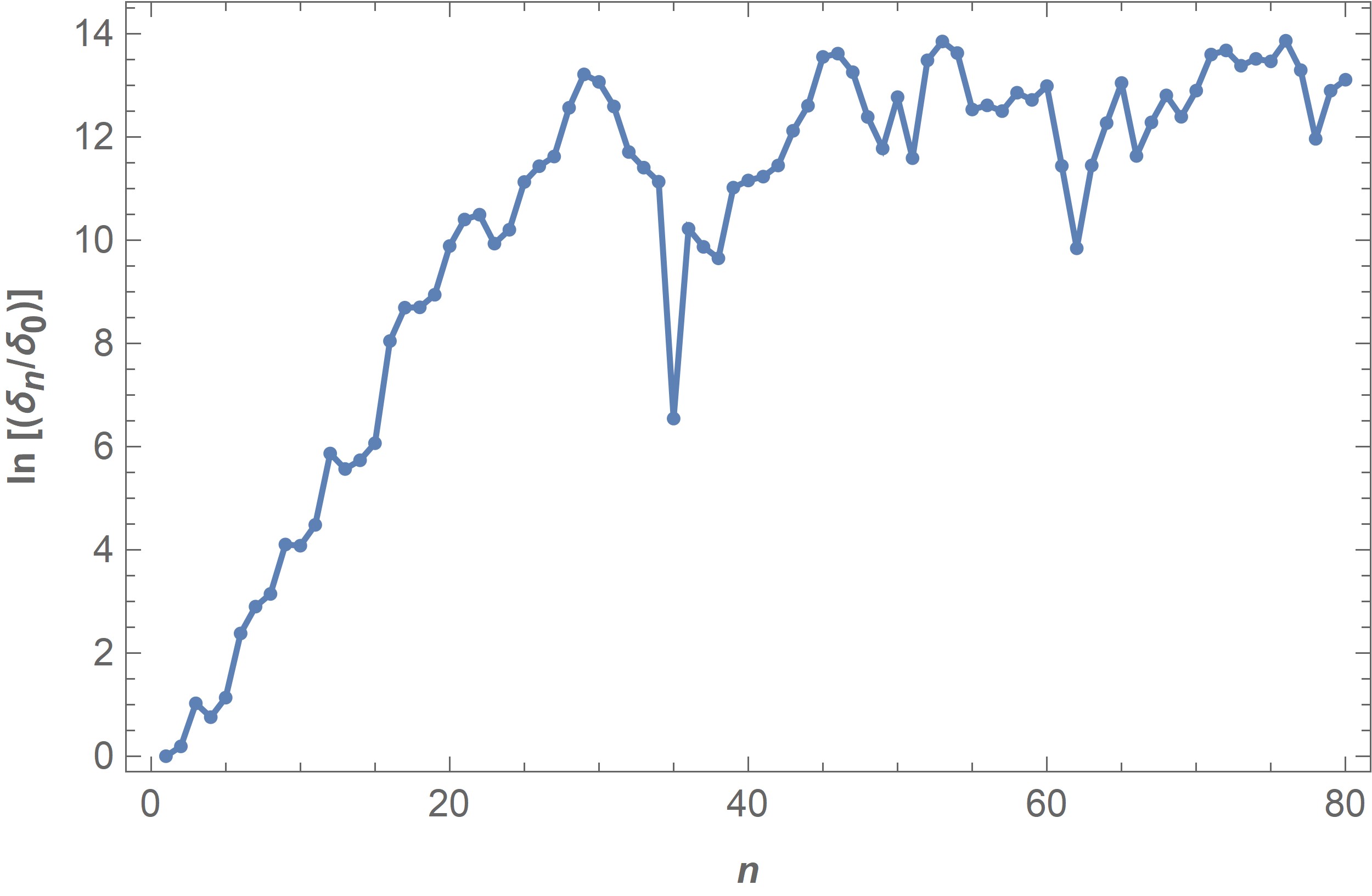}
    \caption{Saturation of Lyapunov exponent for a cos-billiard for $\varepsilon=0.5$. For this particular pair of typical trajectories, 80 collisions are taken and saturation starts at around 30 collisions. The Lyapunov exponent corresponds to the gradient of the unsaturated region.}
    \label{lyapunov}
\end{figure}

We consider a pair of particles that start out very close to one another. Furthermore, we choose the difference between the angles of incidence of the collisions of these particles with the boundary walls as our measure for the separation of the trajectories. Eq.\ (\ref{deltax}), which was used to express the sensitivity condition, is now expressed by
\begin{equation}\label{eq.lyapunov}
    \delta_n = \delta_0 \mathrm{e}^{\lambda n},
\end{equation}
where $\delta_n$ refers to the modulus of the difference between the angles of incidence of trajectories after $n$ collisions. As \(\delta_n\) is calculated \emph{after} \(n\) collisions, it is calculated precisely when the particles are on the verge of the \((n+1)\)th collision. Thus, it follows that \(\delta_0\) is the modulus of the difference between the angles of incidence of trajectories at the first collision. Our aim is to find the Lyapunov exponent of the billiard system. We write the necessary code to determine the incident angles at each collision and obtain the set of differences in incident angles between two typical trajectories that started arbitrarily close to each other. In Fig.\ \ref{lyapunov}, we plot $\ln{(\delta_n/\delta_0)}$ against collision points $n$ for two of these typical trajectories. The Lyapunov exponent is the slope of the unsaturated part of the graph. To compute the average Lyapunov exponent, we calculate $\lambda$s with many different initial conditions and then take the average. This is done to minimize dependence on initial conditions.
 \par

A point in the phase space of a billiard system is considered hyperbolic if its Lyapunov exponent is non-zero. The entire billiard system is referred to as a hyperbolic billiard when the probability of a non-hyperbolic point in the phase space approaches zero. In other words, almost all the points in the phase space of a hyperbolic system are hyperbolic. Because of the presence of multiple flat walls on which the particle can repeatedly bounce off of without encountering a dispersing wall, the Sinai and Superman billiards are non-hyperbolic. The cardioid billiard, on the other hand, is made entirely of dispersing walls and thus, it is a hyperbolic billiard.

\subsection{Classical Lyapunov exponents of the three billiards}\label{ch. three-billiards}
\begin{figure}[htbp]
    \centering
    \includegraphics[scale=0.4]{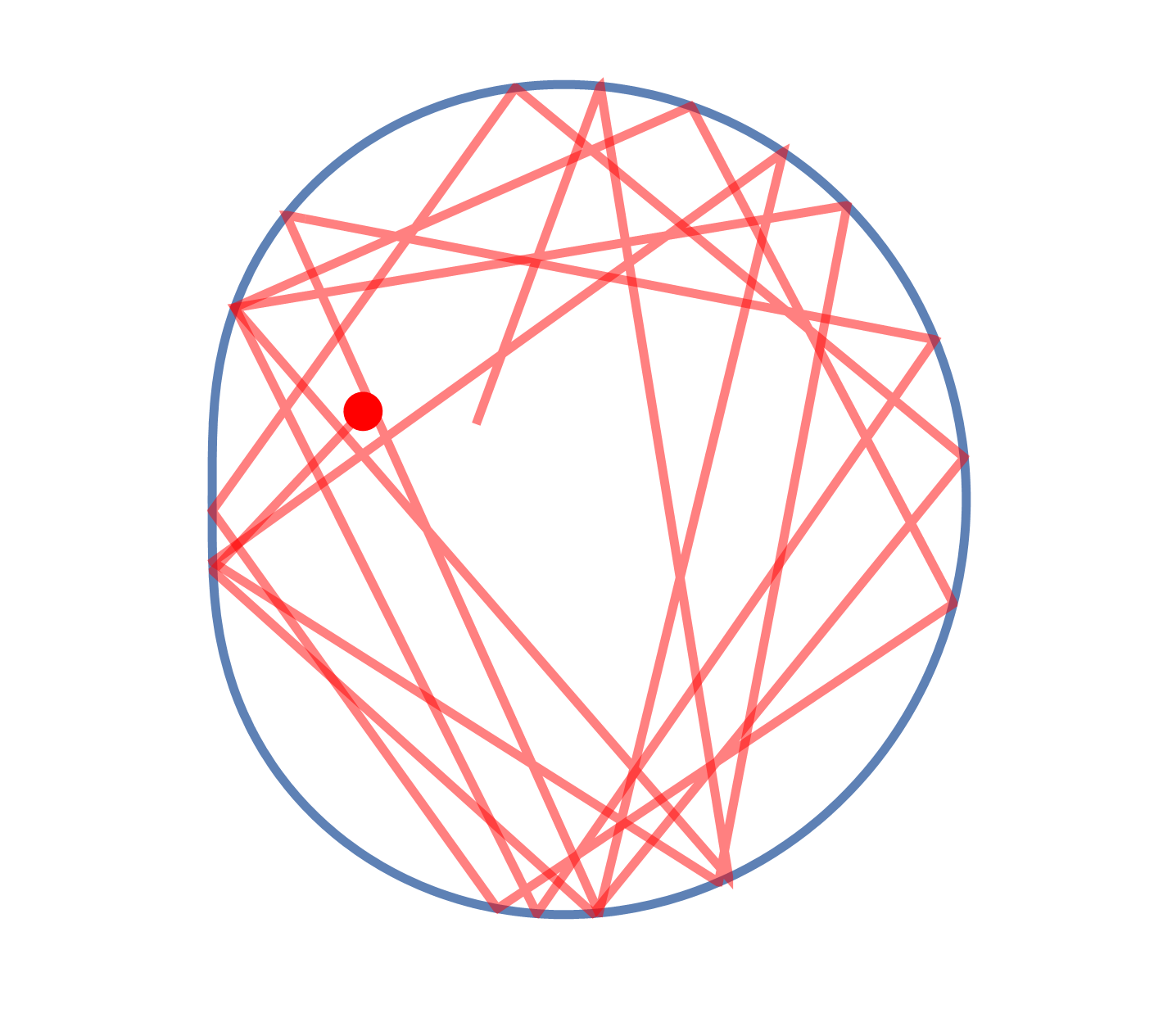}
    \includegraphics[scale=0.4]{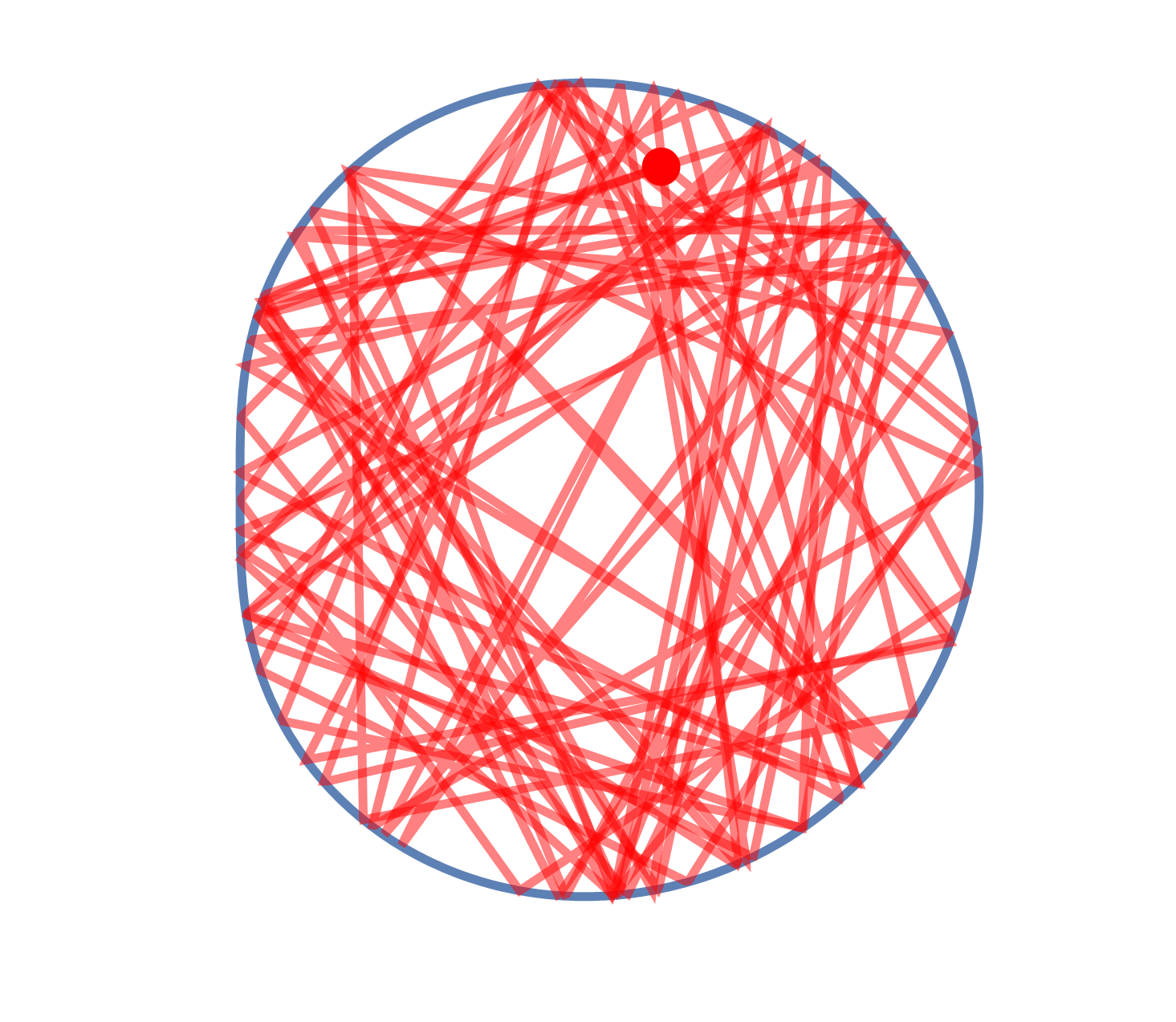}
    \includegraphics[scale=0.4]{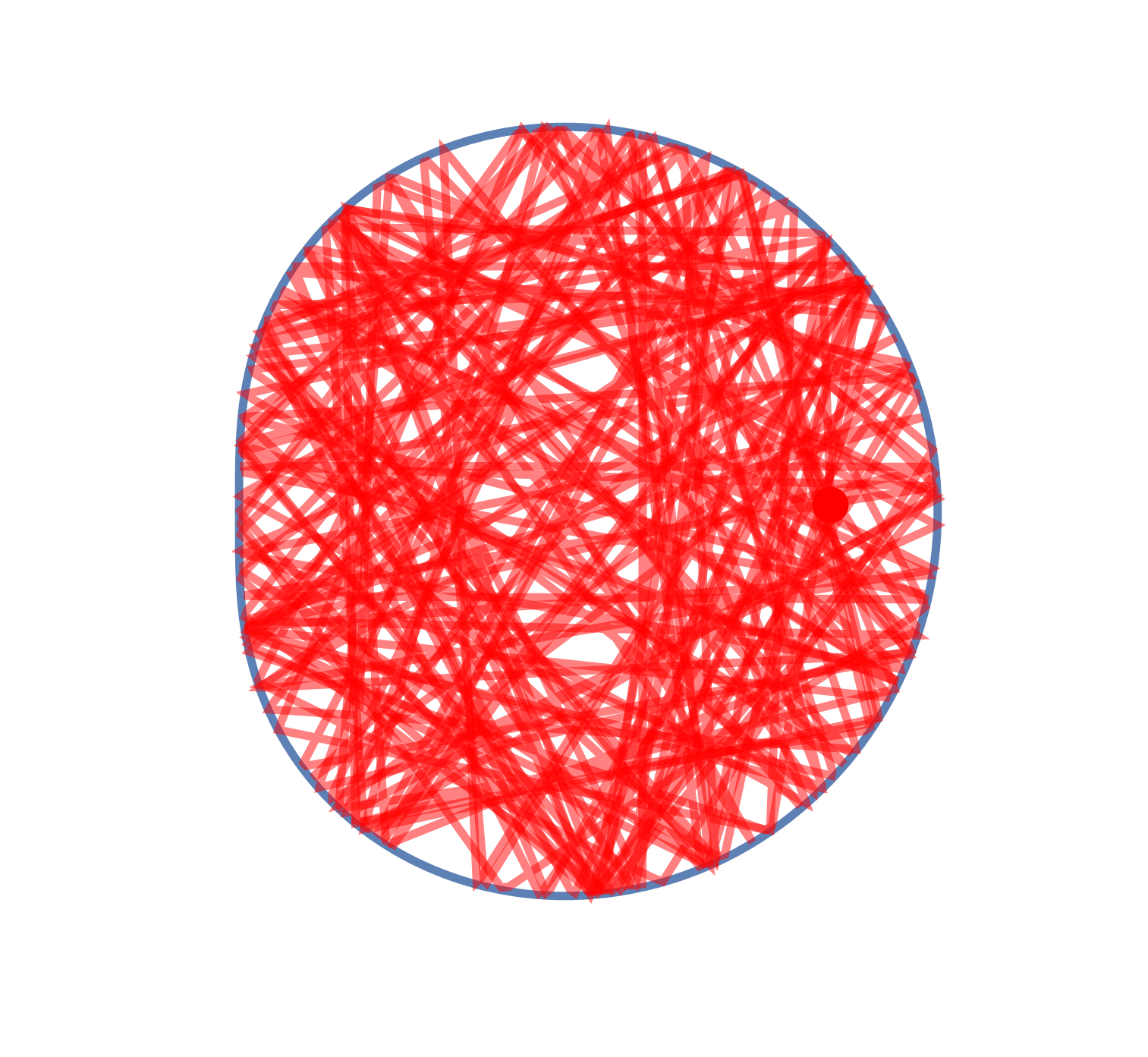}
    \includegraphics[scale=0.4]{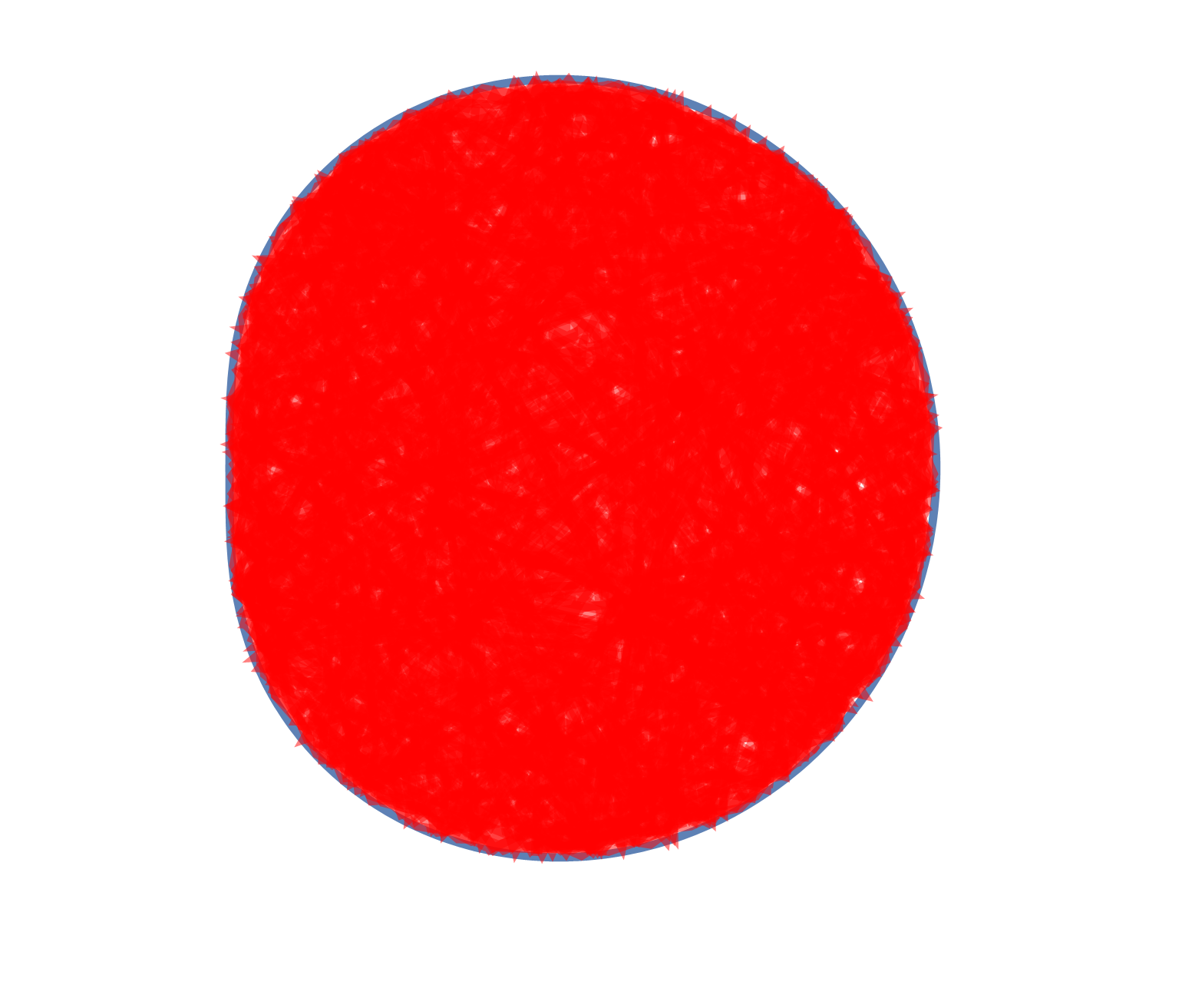}
    \caption{Growth of a typical trajectory inside a cos-billiard for $\varepsilon=0.5$.}
    \label{a-typical}
\end{figure}

We will employ the numerical method described above to calculate the classical Lyapunov exponents of the Sinai, cardioid, and diamond billiards. However, to compute the exponents, we have to obtain typical trajectories. In numerical simulations,  one generates a trajectory of the system, usually by selecting an initial point $X_0$ randomly. We call these types of trajectories \textit{typical} trajectories. To obtain the particle's trajectory, we numerically solve Newton's equations with elastic reflection condition as boundary conditions. In Fig.\ \ref{a-typical}, we show the growth of a typical trajectory of a particle inside a cos-billiard over a period.

\par

\begin{figure}[htbp]
    \centering
    \begin{subfigure}[b]{0.45 \textwidth}
        \includegraphics[width=\textwidth]{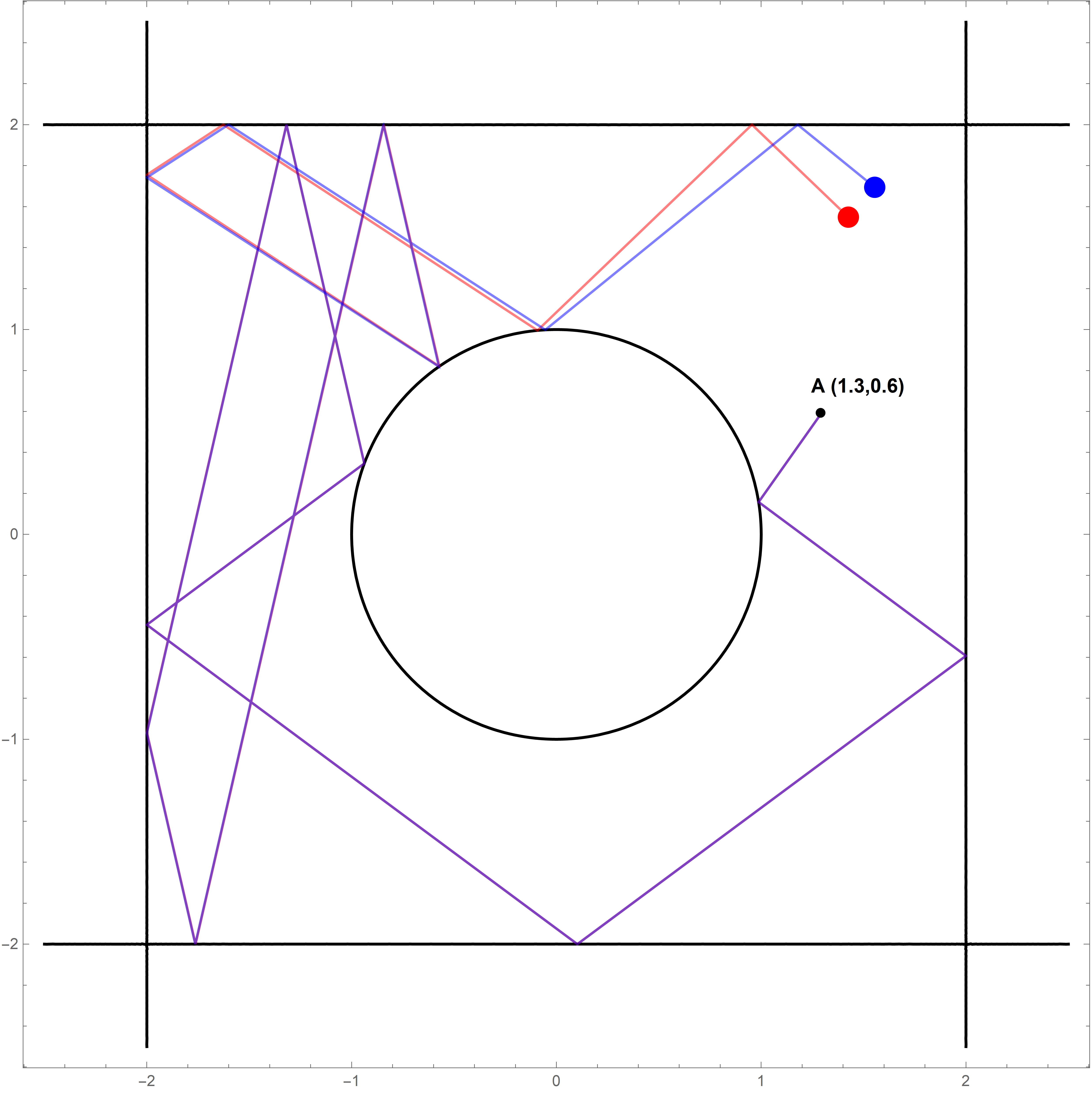}
        \caption{}
        \label{sinai trajectory1}
    \end{subfigure}
    \hfil
    \begin{subfigure}[b]{0.45 \textwidth}
        \includegraphics[width=\textwidth]{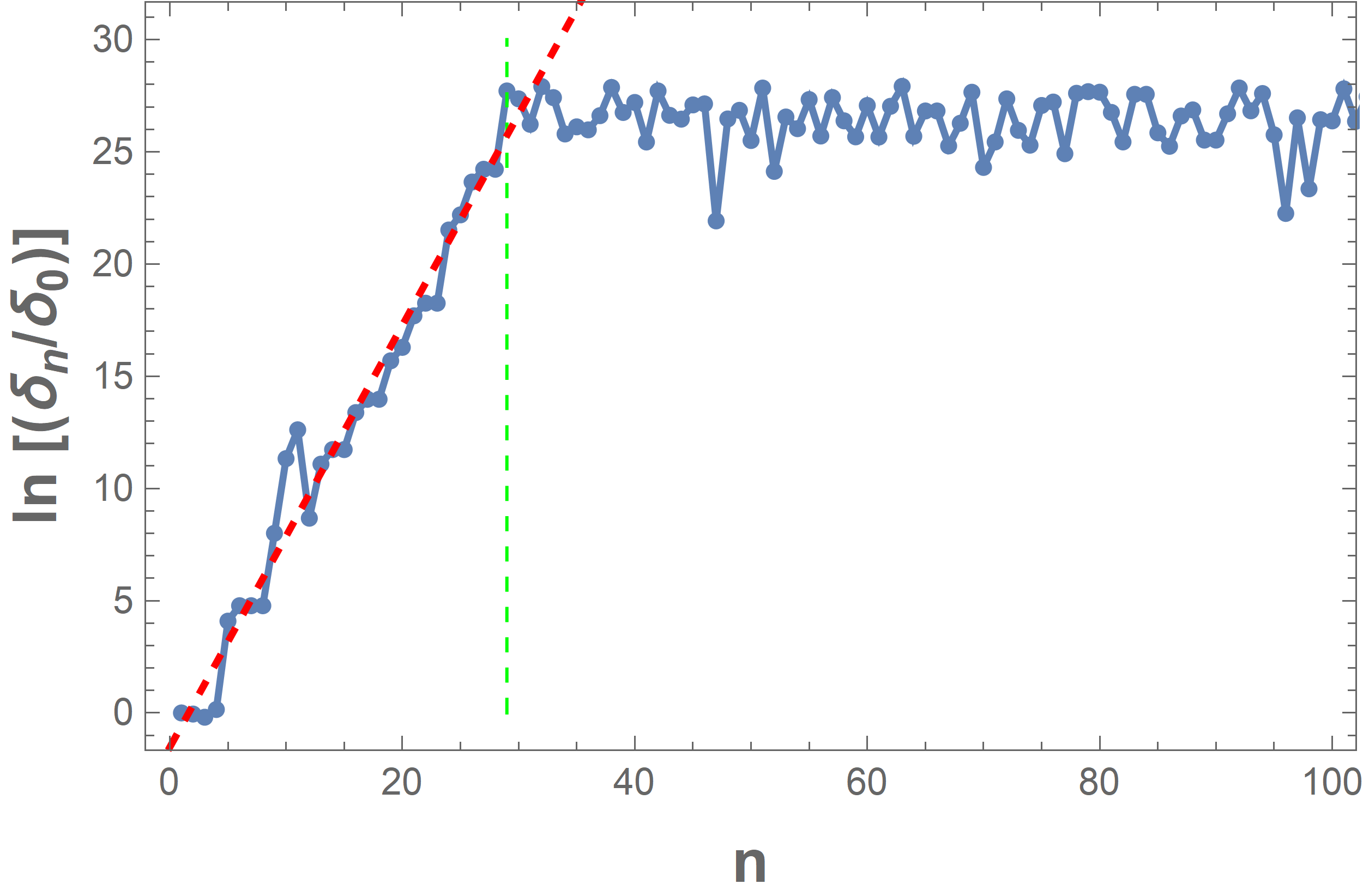}
        \caption{ }
        \label{sinai trajectory2}
    \end{subfigure}
    \caption{(a) The trajectories of two particles in the Sinai enclosure starting from initial points very close to one another. (b) The growth rate of separation between the trajectories in (a), with respect to number of collisions, \(n\). The dashed green line indicates the saturation point at \(n=29\), while the dashed red line represents the best fit for the unsaturated region. The slope of this red line corresponds to the Lyapunov exponent for this specific pair of trajectories.}
    \label{sinai trajectory}
\end{figure}

In Fig.\ \ref{sinai trajectory1}, we display, for the Sinai billiard, the typical trajectories (coloured red and blue) of two point particles that start from two extremely close points, $A (1.3, 0.6)$ and $A' (1.3 + 10^{-10}, 0.6 + 10^{-10})$, with very similar velocity components. After just a few collisions, the particles become separated.

Fig.\ \ref{sinai trajectory2} illustrates the growth rate of trajectory separation for the trajectories shown in Fig. \ref{sinai trajectory1} as a function of the number of collisions, denoted by $n$. In this case, we have considered approximately $100$ collisions, and the saturation point occurs at $n = 29$. The Lyapunov exponent is the slope of the unsaturated part, as given by eq.\ (\ref{eq.lyapunov}).\par

\begin{figure}[htbp]
    \centering
    \begin{subfigure}[b]{0.45 \textwidth}
        \includegraphics[width=\textwidth]{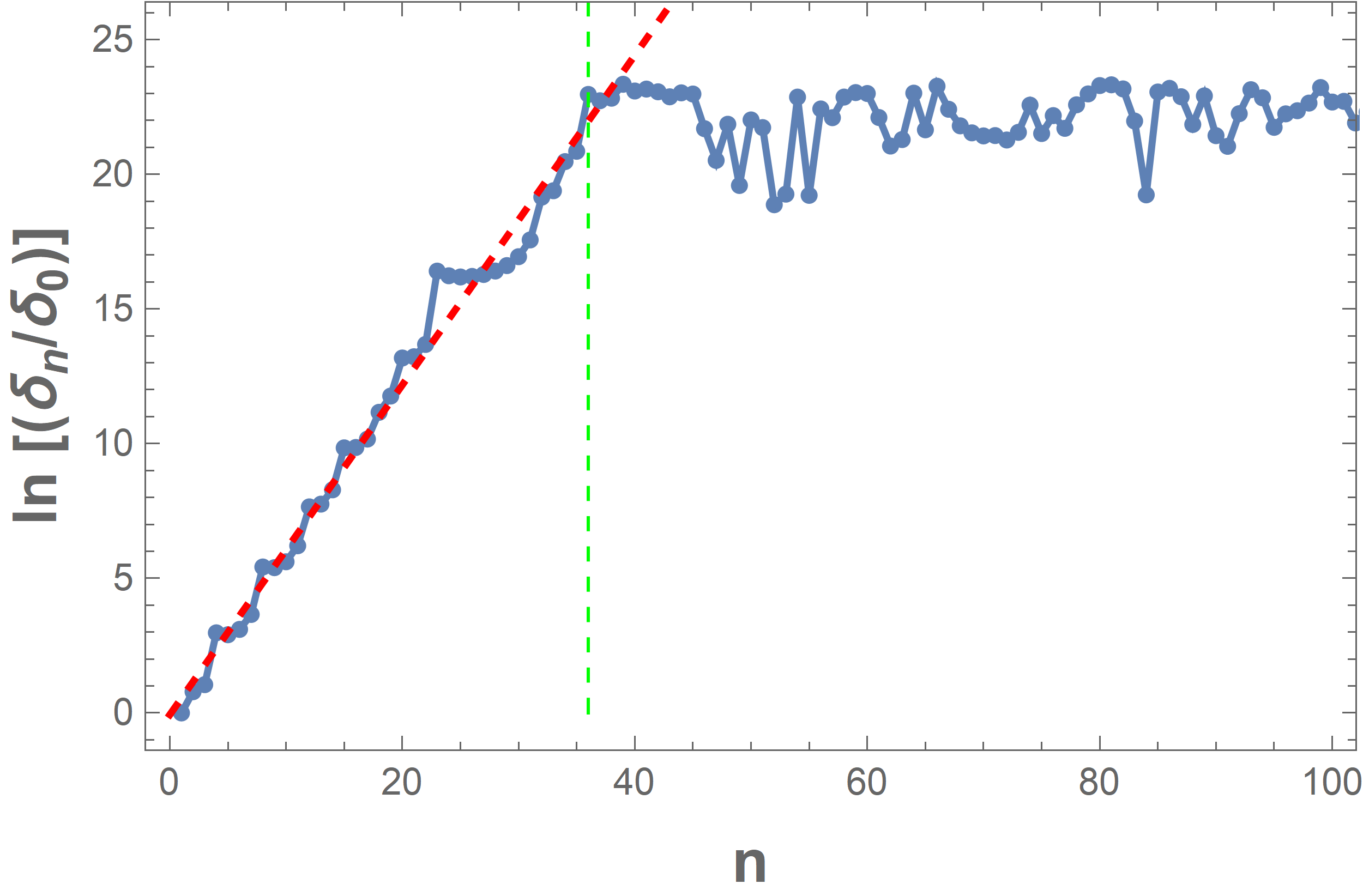}
        \caption{}
        \label{cardioid diamonf trajectory1}
    \end{subfigure}
    \hfil
    \begin{subfigure}[b]{0.45 \textwidth}
        \includegraphics[width=\textwidth]{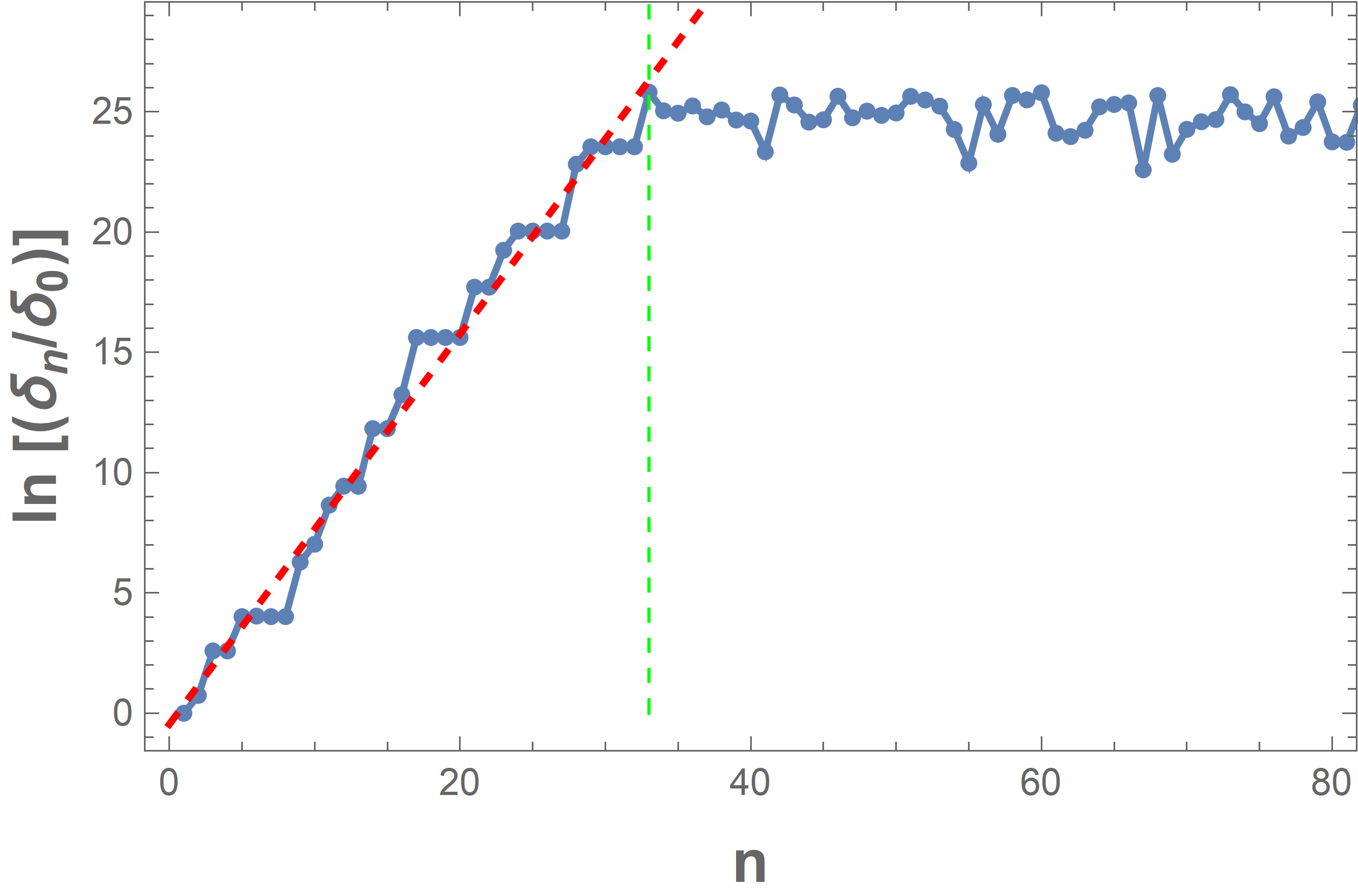}
        \caption{ }
                \label{cardioid diamonf trajectory2}
    \end{subfigure}
    \caption{Growth rate of separation between two trajectories that start out extremely close to one another for the (a) cardioid and (b) diamond billiard systems.}
    \label{cardioid diamonf trajectory}
\end{figure}

Figs.\ \ref{cardioid diamonf trajectory1} and \ref{cardioid diamonf trajectory2} depict the growth rate of separation between a particular pair of trajectories that start out very close to one another for the cardioid and diamond billiards, respectively. Here, the saturation point for the cardioid billiard is at $n=36$, while for the diamond billiard, it is at $n=33$.
\par

For each of our billiards, we generated $2000$ random initial conditions and from them, generated $2000$ pairs of trajectories, each pair consisting of two particles that start out very close to one another. We then calculated $2000$ Lyapunov exponents from these pairs of trajectories using the method described above. Finally, we took the average of all the Lyapunov exponents obtained for each particular billiard to arrive at our final value of the \emph{classical} Lyapunov exponent, $\lc$, for each billiard. Note that in this computation of the Lyapunov exponent we do not use the distance between collisions. The justification for not using the distance between collisions is given by Appendix \ref{appendix_2}.

However, later in the paper when we compute the quantum Lyapunov exponent we utilize the average distance between consecutive collisions of the classical trajectories as a unit in order to facilitate comparison of the quantum Lyapunov exponent with the average classical Lyapunov exponent. We calculated the average distance between two consecutive collision points for the unsaturated part, denoted by $d_{\text{avg}}$ for each typical trajectory and then averaged it across the $2000$ trajectories. In our calculations, the area of the billiard was consistently kept at $1$. Therefore, the average distance between two collisions is calculated for billiards with an area of $1$. In Table \ref{table:classical} we list the average distance between consecutive collisions and the classical Lyapunov exponents of our selected billiards.
\par
\begin{table}[htbp]
\centering
\begin{tabular}{||c | c  | c ||} 
 \hline
 Billiard &   $\lc$ &  $d_{\text{avg}}$ for $A=1$ \\ [0.5ex] 
 \hline\hline
 Sinai &  $0.8048$  & $0.4817$\\ 
 \hline
 Cardioid & $0.6649$ &  $0.8787$ \\
 \hline
Diamond &  $0.6860$  & $0.7805$  \\[1ex] 
 \hline
\end{tabular}
\caption{Average classical Lyapunov exponents $\lc$  and average distance between consecutive collisions $d_{\text{avg}}$ for the Sinai, cardioid, and diamond billiards. }
\label{table:classical}
\end{table}

\section{Numerical calculation of OTOCs}\label{sec:numerics}

\subsection{OTOCs and Wavefunctions}
\begin{figure}[ht!]
    \centering
    \includegraphics[scale=1]{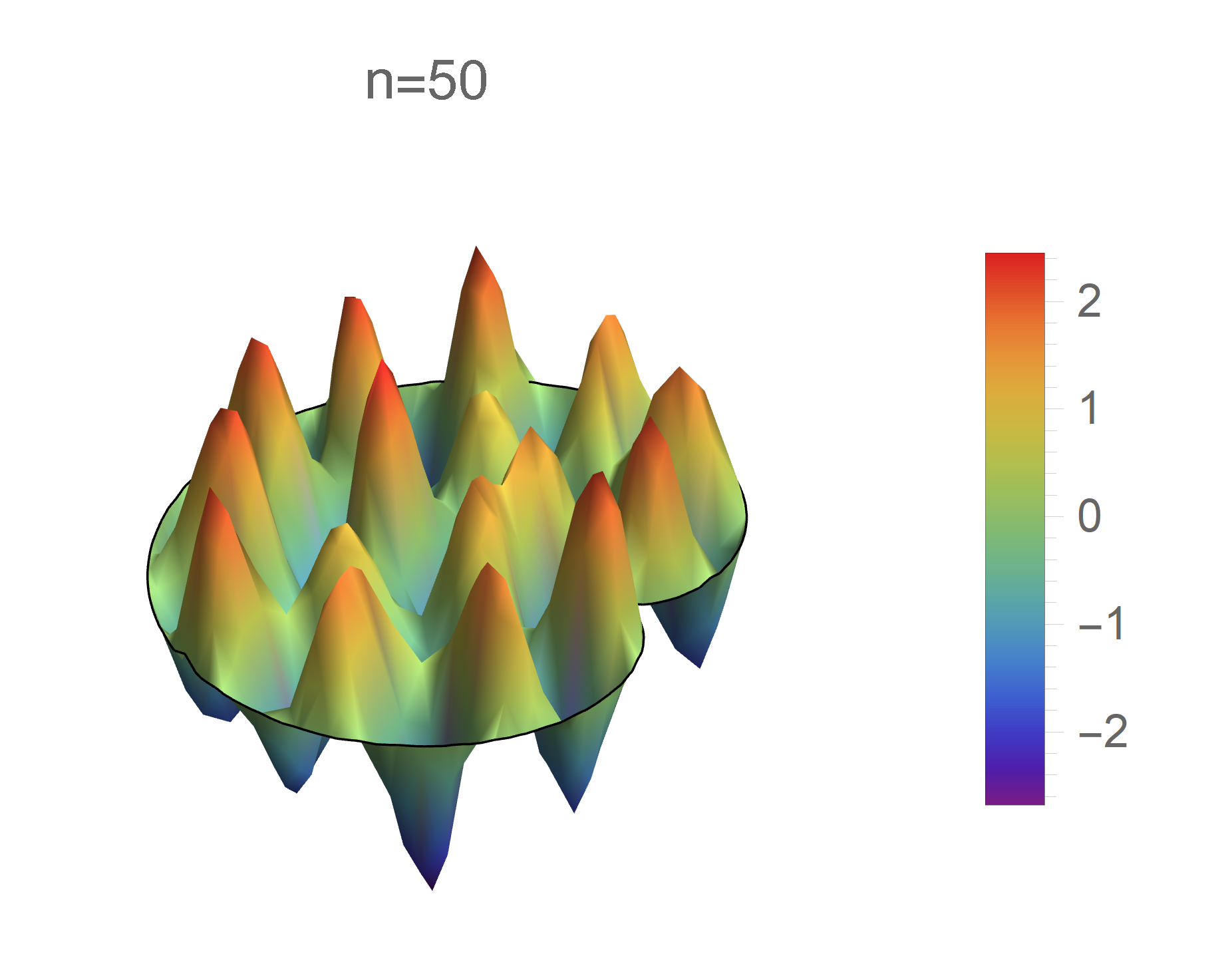}
    \caption{A three-dimensional plot of the cardioid eigenfunction for \(n=50\). The elevation corresponds to the value of the wave function at each point.}
    \label{cardioid_3d}
\end{figure}

To compute the OTOCs of billiards we first need to find the eigenvalues of the Hamiltonian operators of the billiards as was described in Subsection \ref{sec:otoc}. The Hamiltonian of a billiard is given by
\begin{equation}
    H =-\frac{\partial^2}{\partial x^2}-\frac{\partial^2}{\partial y^2}+V_{\text{bill}}(x,y),\quad
  V_{\text{bill}}(x,y)=
  \begin{cases}
  0, & (x,y)\in \Omega\\
  \infty, & \text{elsewhere}
  \end{cases}
\end{equation}
where \(\Omega\) is the region inside the billiard. In Fig.\;\ref{cardioid_3d}, we show a typical wavefunction for the cardioid billiard, $\psi_{50}$, numerically generated by \texttt{Mathematica}. 

\begin{figure}[htbp]
    \centering
    \begin{subfigure}[b]{0.58 \textwidth}
        \caption{Sinai billiard}
        \includegraphics[width=\textwidth]{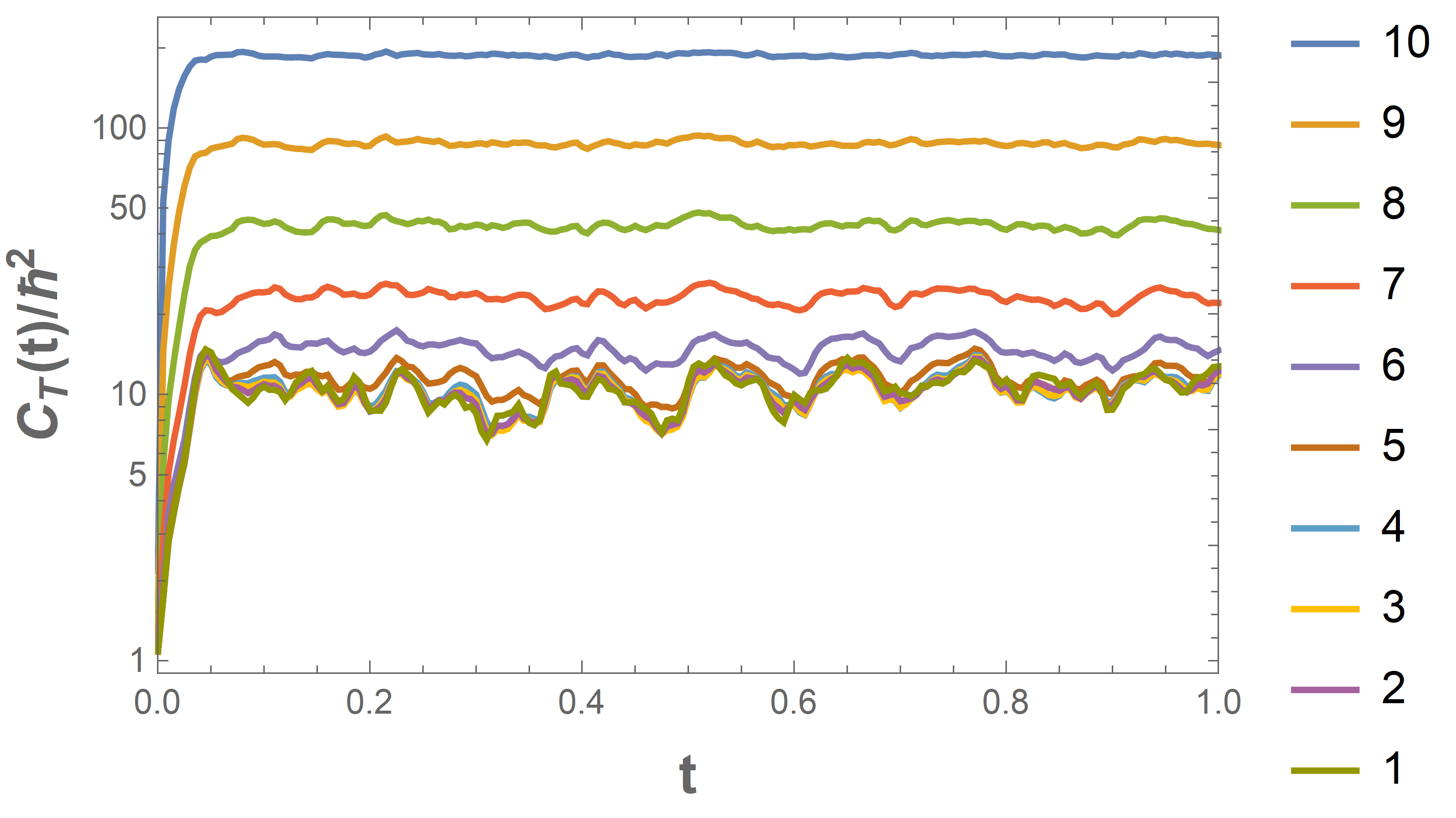}
    \end{subfigure}
    \hfil
    \begin{subfigure}[b]{0.58 \textwidth}
        \caption{Cardioid billiard }
        \includegraphics[width=\textwidth]{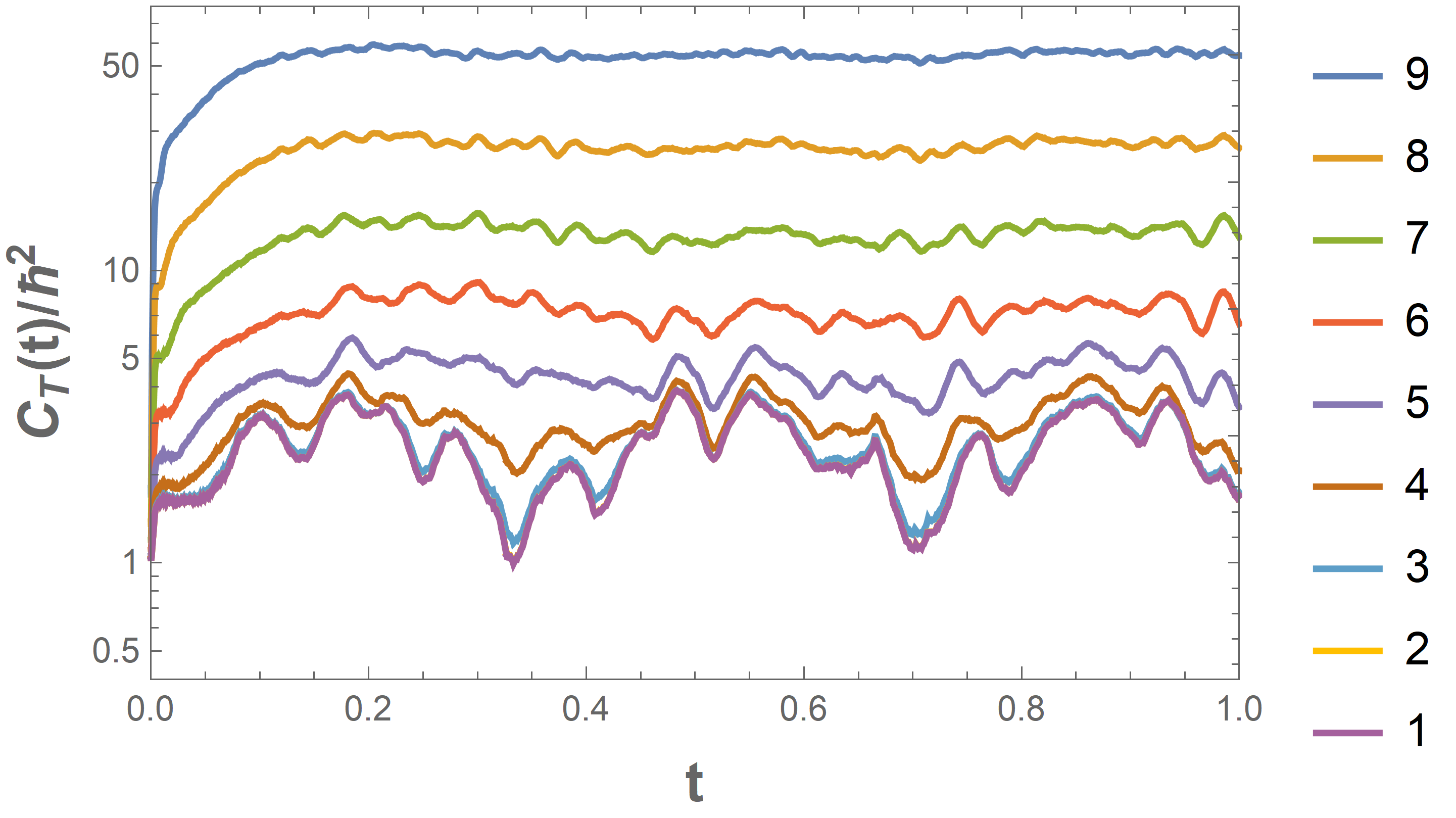}
    \end{subfigure}
    \begin{subfigure}[b]{0.58 \textwidth}
        \caption{Diamond billiard}
        \includegraphics[width=\textwidth]{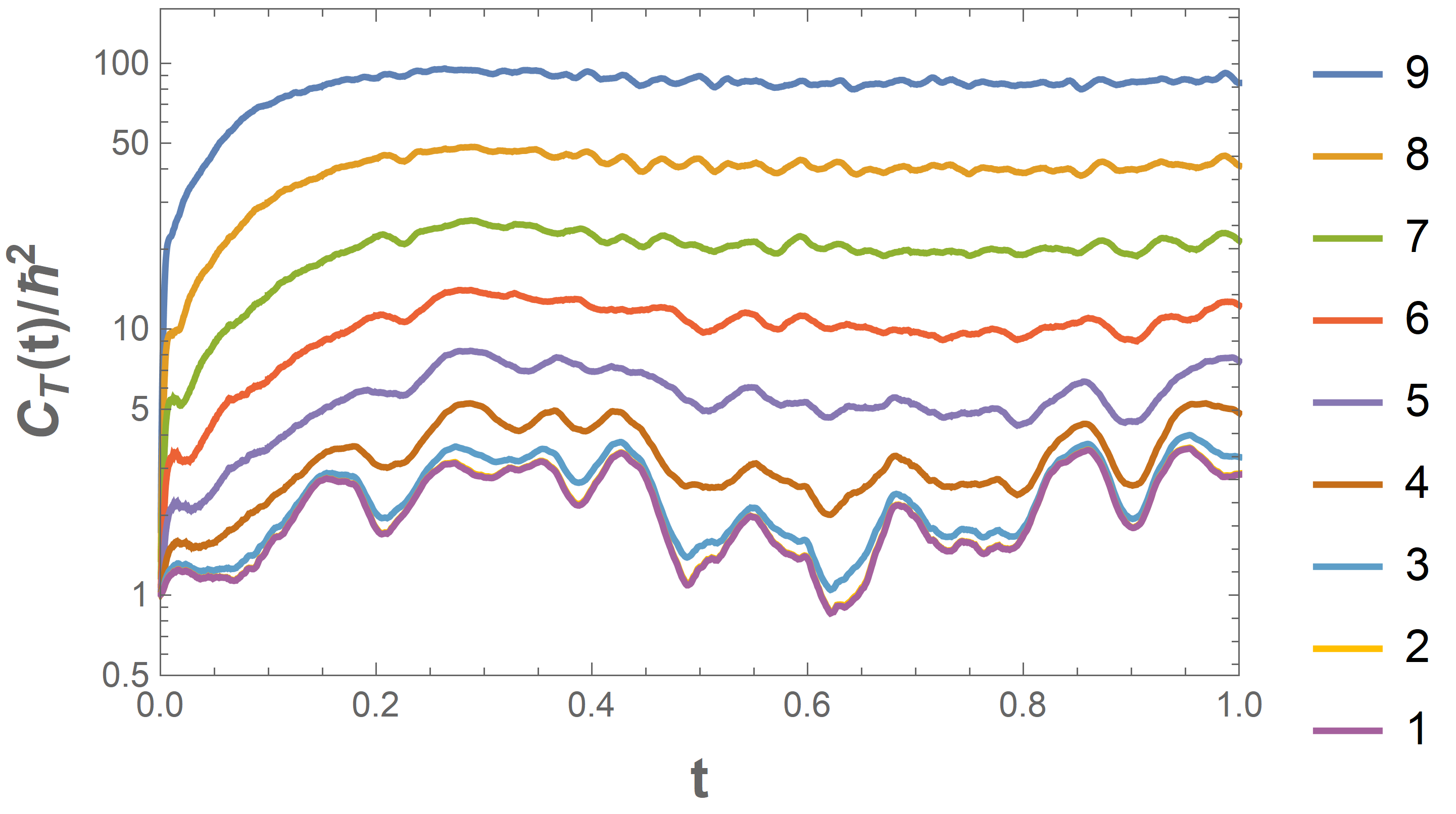}
    \end{subfigure}
    \caption{Log plots of thermal OTOCs for (a) Sinai billiard, (b) cardioid billiard, (c) diamond billiard. The corresponding temperatures are given to the right of the graphs, expressed on a log base \(2\) scale.}
    \label{fig: thermal otocs}
\end{figure}

We follow the same procedure as with the stadium billiard \cite{Hashimoto_2017} to calculate the \(x\)-matrix elements as
\begin{equation}
    x_{nm}=\int_{\Omega} \mathrm{d}x \: \mathrm{d}y \: \psi_n  x \psi_m.
\end{equation}
We then calculate the microcanonical OTOCs by substituting the position matrix elements $x_{nm}$ and energy eigenvalues $E_n$ into Eq.\,(\ref{bnm}) to obtain $b_{nm}(t)$, which we then use in Eq.\,(\ref{cn}) to calculate the microcanonical OTOCs $c_n(t)$ for each energy level $n$. Taking the thermal average of $c_n(t)$ using Eq.\,(\ref{cT}), we obtain the thermal OTOCs. Since numerical calculations require truncation of the summations in Eqs.\,(\ref{bnm}), (\ref{cn}), and (\ref{cT}), we must choose a sufficiently large truncation cut-off. We chose to truncate the sums to \(\It=800\) and we show in Appendix \ref{appendix_1} that this truncation value is large enough so that our results remain sufficiently accurate. We show the log plots of the thermal OTOCs of our selected billiards in Fig.\ \ref{fig: thermal otocs}.

\par There is a noticeable initial growth in the thermal OTOCs of our selected billiards at short times. These graphs are similar to Fig.\;\ref{time-window}, where we presented a schematic diagram of the growth of the thermal OTOCs. At later times, the OTOCs stabilize and oscillate around a constant value.\par
However, at very low temperatures, there is no significant growth in the thermal OTOCs for the diamond and cardioid billiards, and for the cardioid billiard, we observe large oscillations. This behaviour is also observed in the case of low mode microcanonical OTOCs. The reason for this lies in Eq.\,(\ref{cT}), where the Boltzmann factor $ e^{-\beta E_n}$ suppresses the contribution from high modes at low temperatures. Conversely, at higher temperatures, the Boltzmann weight does not suppress the high modes, leading to an increased number of modes contributing to the thermal OTOCs \cite{Hashimoto2}.\par
\begin{figure}[htbp]
    \centering
    \begin{subfigure}[b]{0.43 \textwidth}
        \caption{Sinai billiard}
        \includegraphics[width=\textwidth]{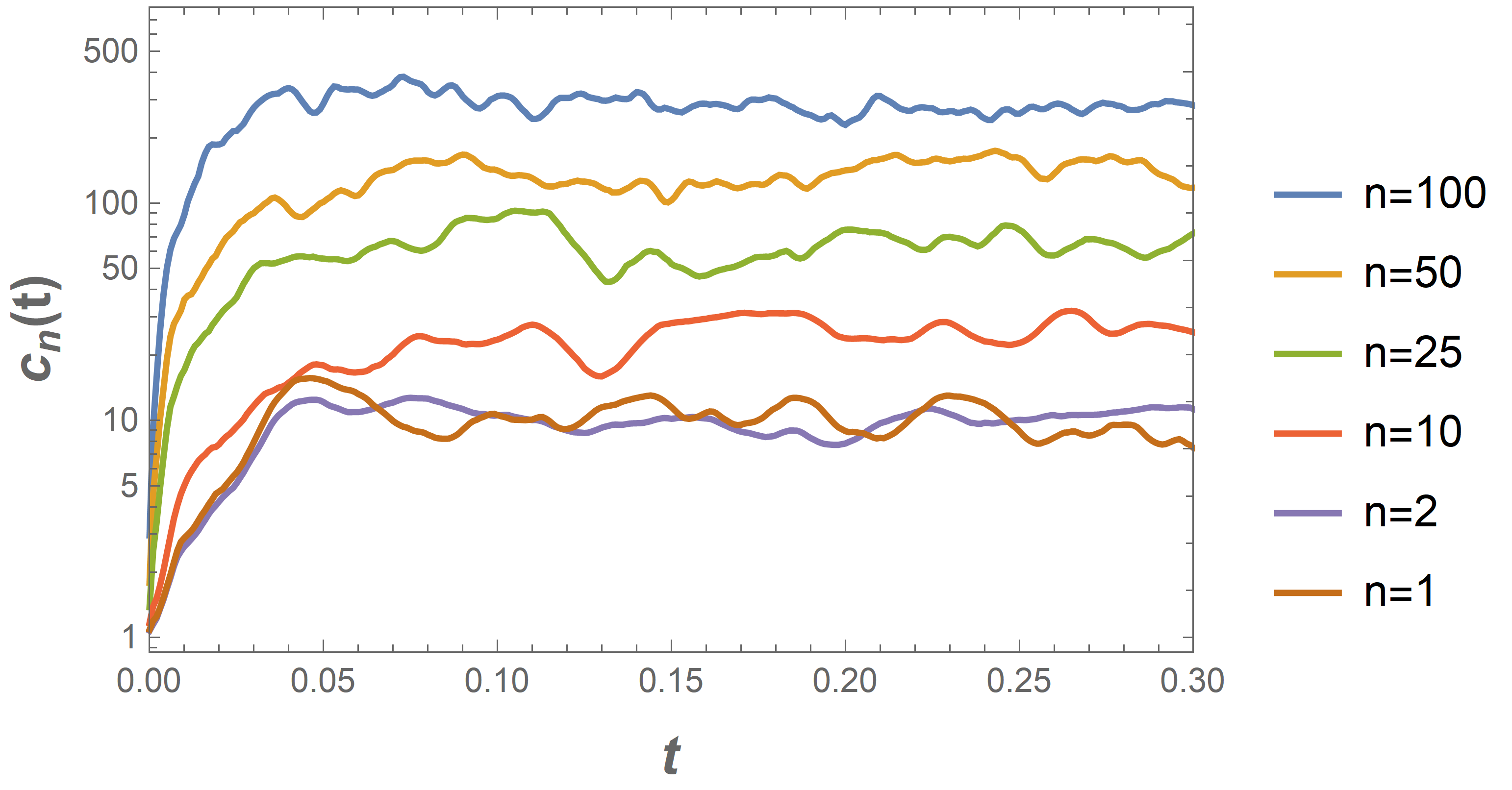}
        
    \end{subfigure}
    \hfil
    \begin{subfigure}[b]{0.43 \textwidth}
    \caption{Cardioid billiard}
        \includegraphics[width=\textwidth]{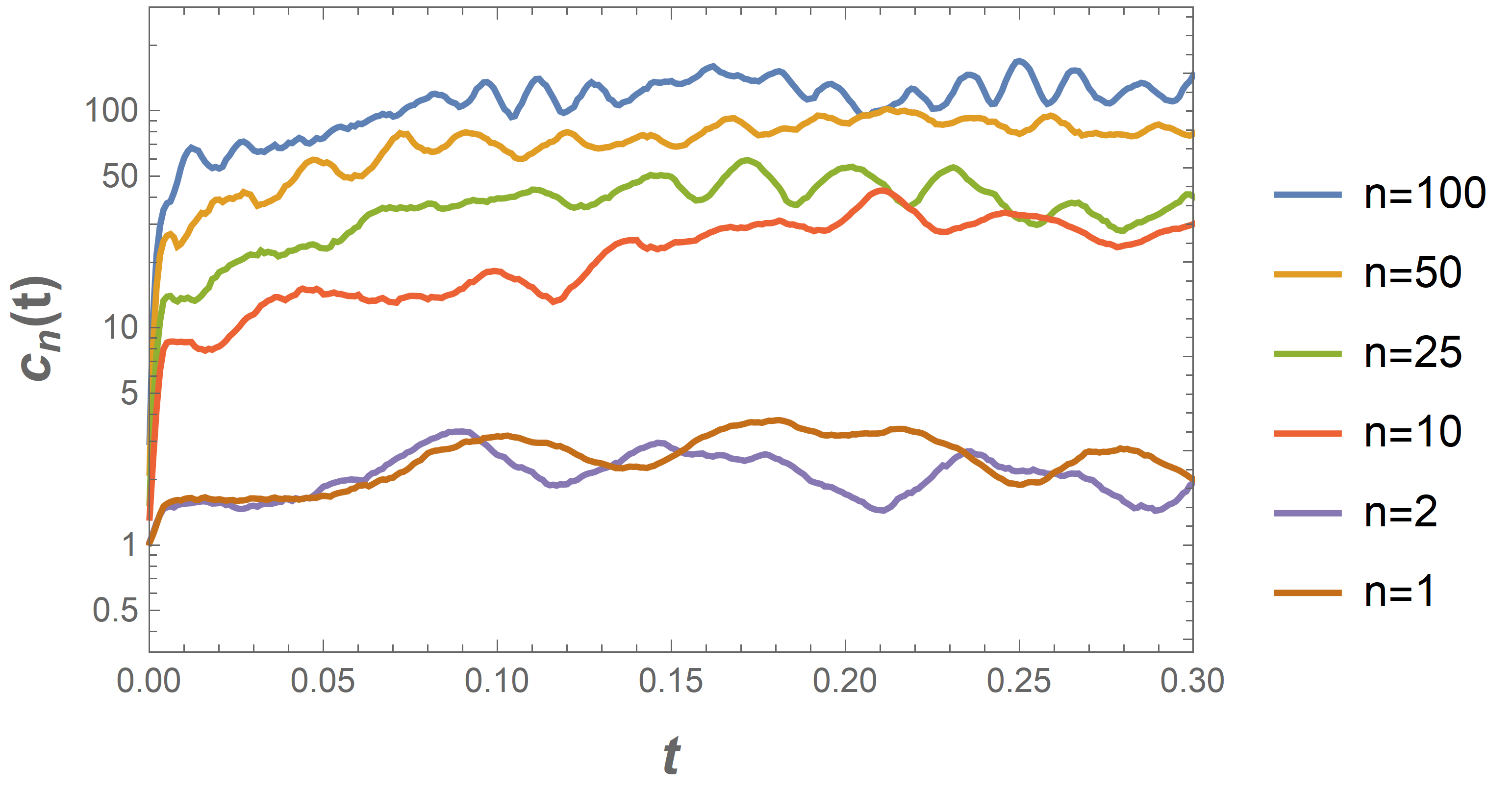}
        
    \end{subfigure}
    \hfil
    \begin{subfigure}[b]{0.43 \textwidth}
        \caption{ Diamond billiard}
        \includegraphics[width=\textwidth]{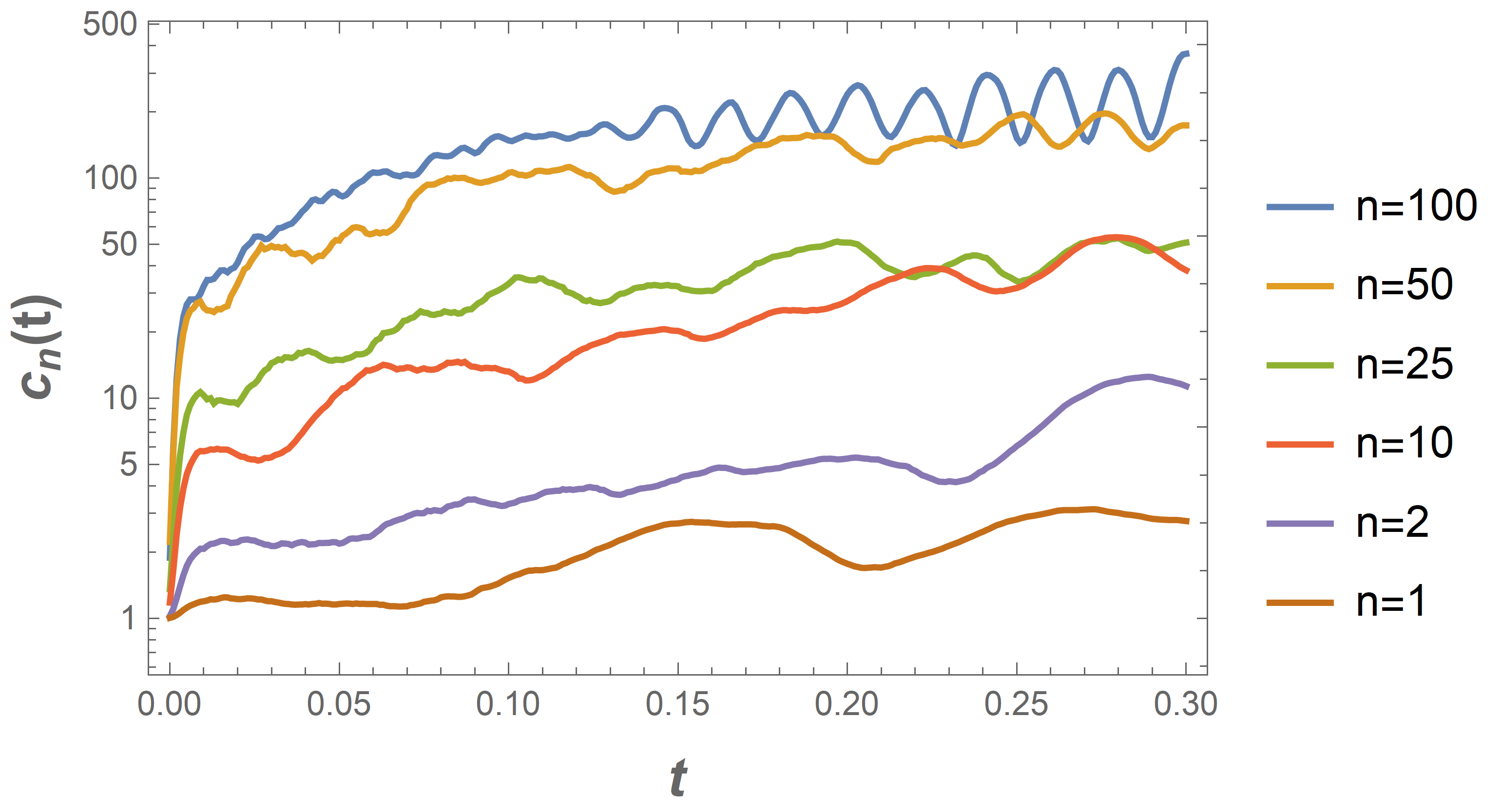}
        
    \end{subfigure}
    \caption{Log plots of microcanonical OTOCs for (a) Sinai billiard, (b) cardioid billiard \& (c) diamond billiard.}
    \label{micro sinai diamond}
\end{figure}

Nevertheless, we observe a strict initial growth in the thermal OTOCs of the Sinai billiard even at $T=2$, despite the suppression of high modes by the Boltzmann factor. This initial growth should also be evident in the microcanonical OTOCs of the same billiard. To verify this, we present the microcanonical OTOCs for the Sinai, as well as the cardioid and diamond billiards in Figure \ref{micro sinai diamond}.\par
\begin{figure}[htbp]
    \centering
    \begin{subfigure}[b]{0.3 \textwidth}
        \includegraphics[width=\textwidth]{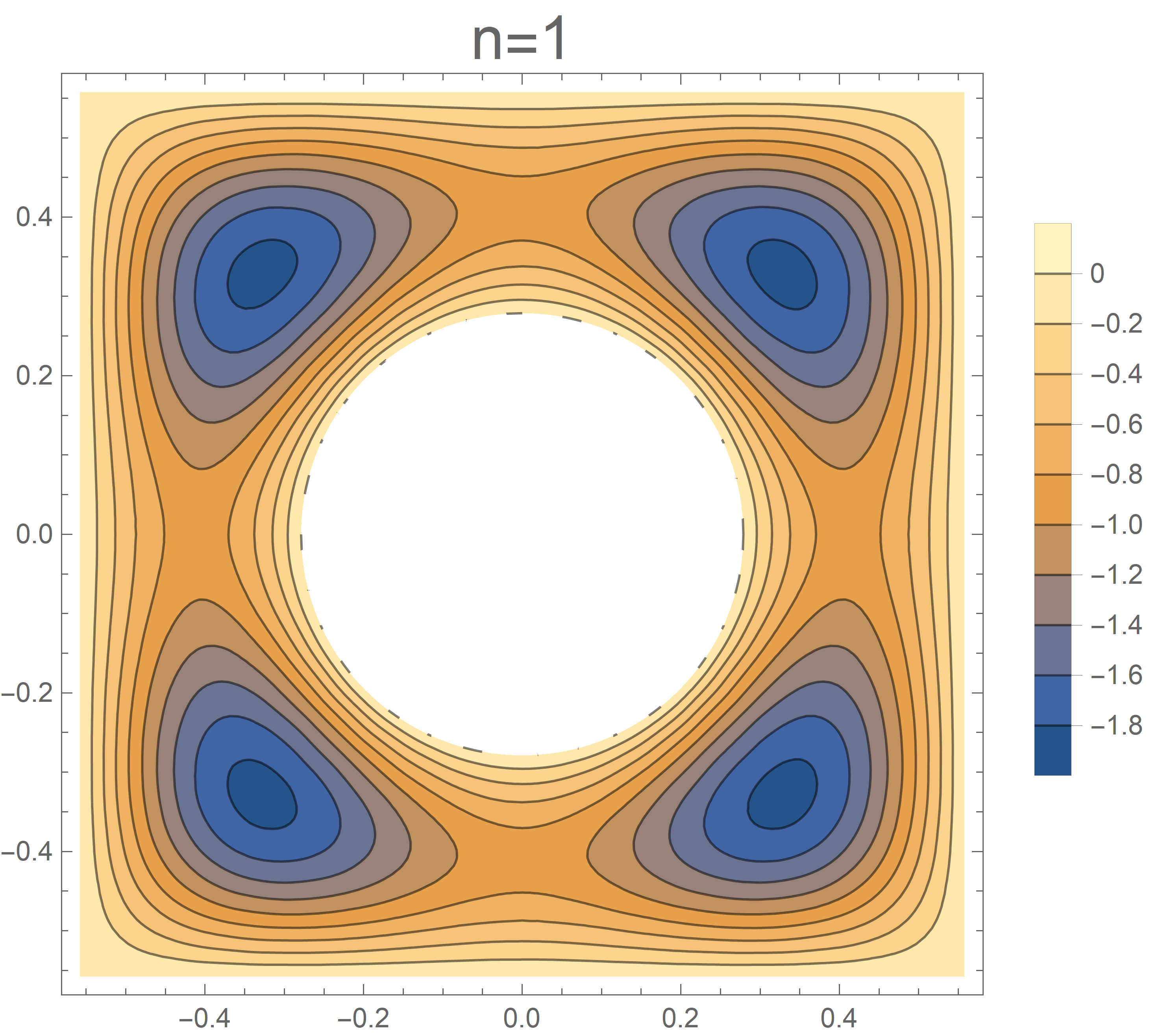}
        \caption*{}
    \end{subfigure}
    \hfil
    \begin{subfigure}[b]{0.3 \textwidth}
        \includegraphics[width=\textwidth]{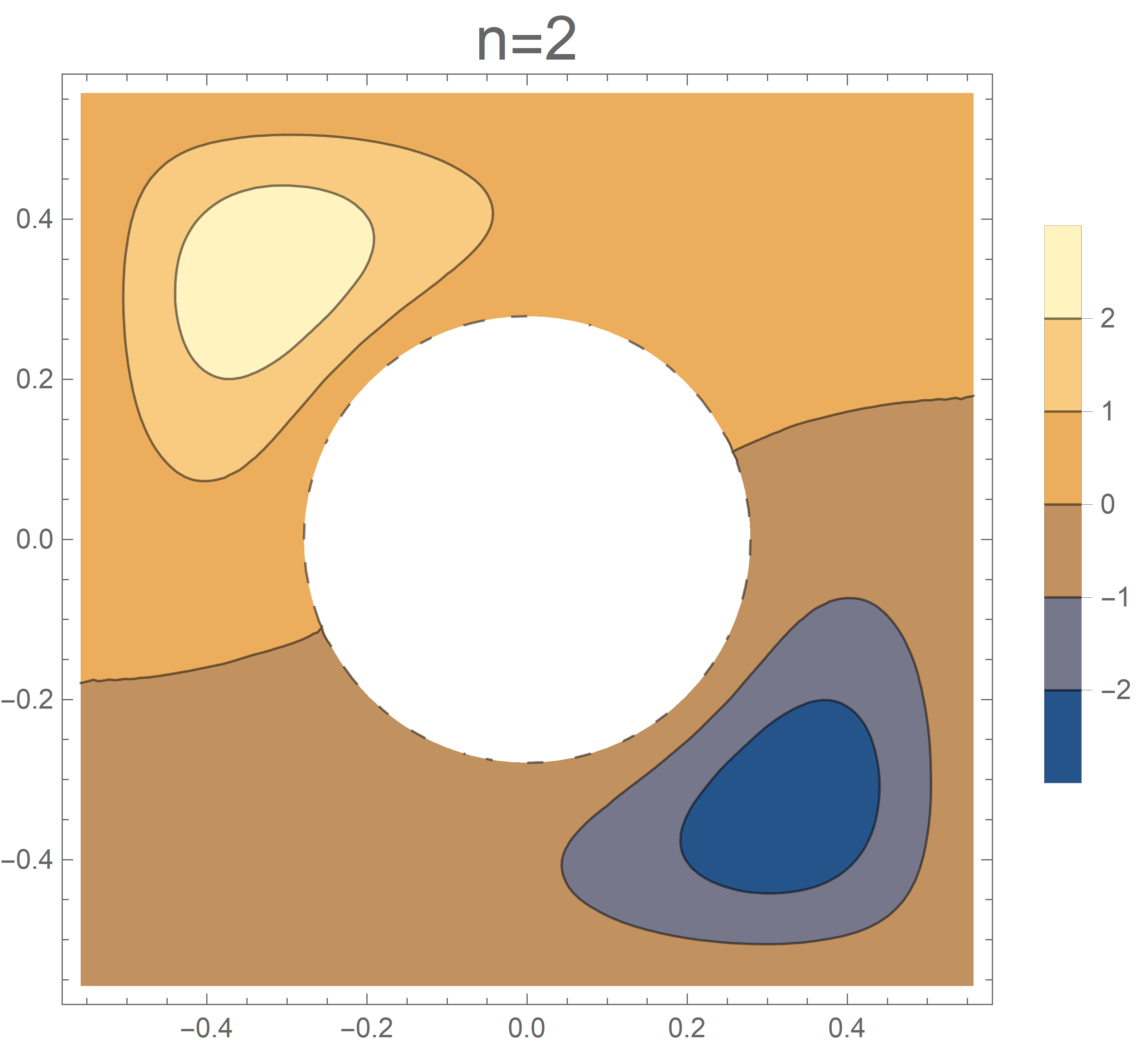}
        \caption*{Sinai wavefunctions}
    \end{subfigure}
    \hfil
    \begin{subfigure}[b]{0.3 \textwidth}
        \includegraphics[width=\textwidth]{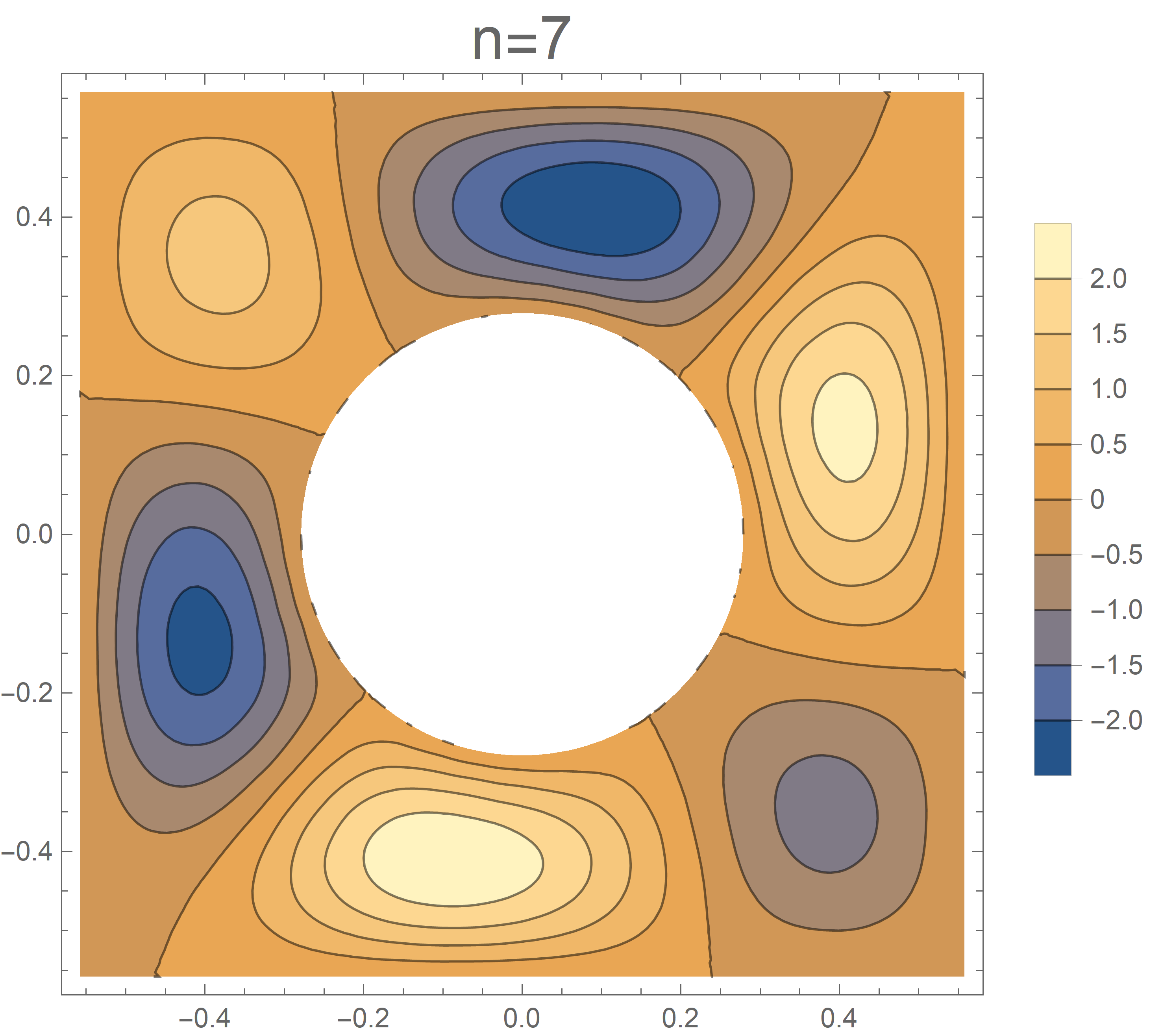}
        \caption*{}
    \end{subfigure}
    \hfil
    \vspace{1cm}
    \begin{subfigure}[b]{0.3 \textwidth}
        \includegraphics[width=\textwidth]{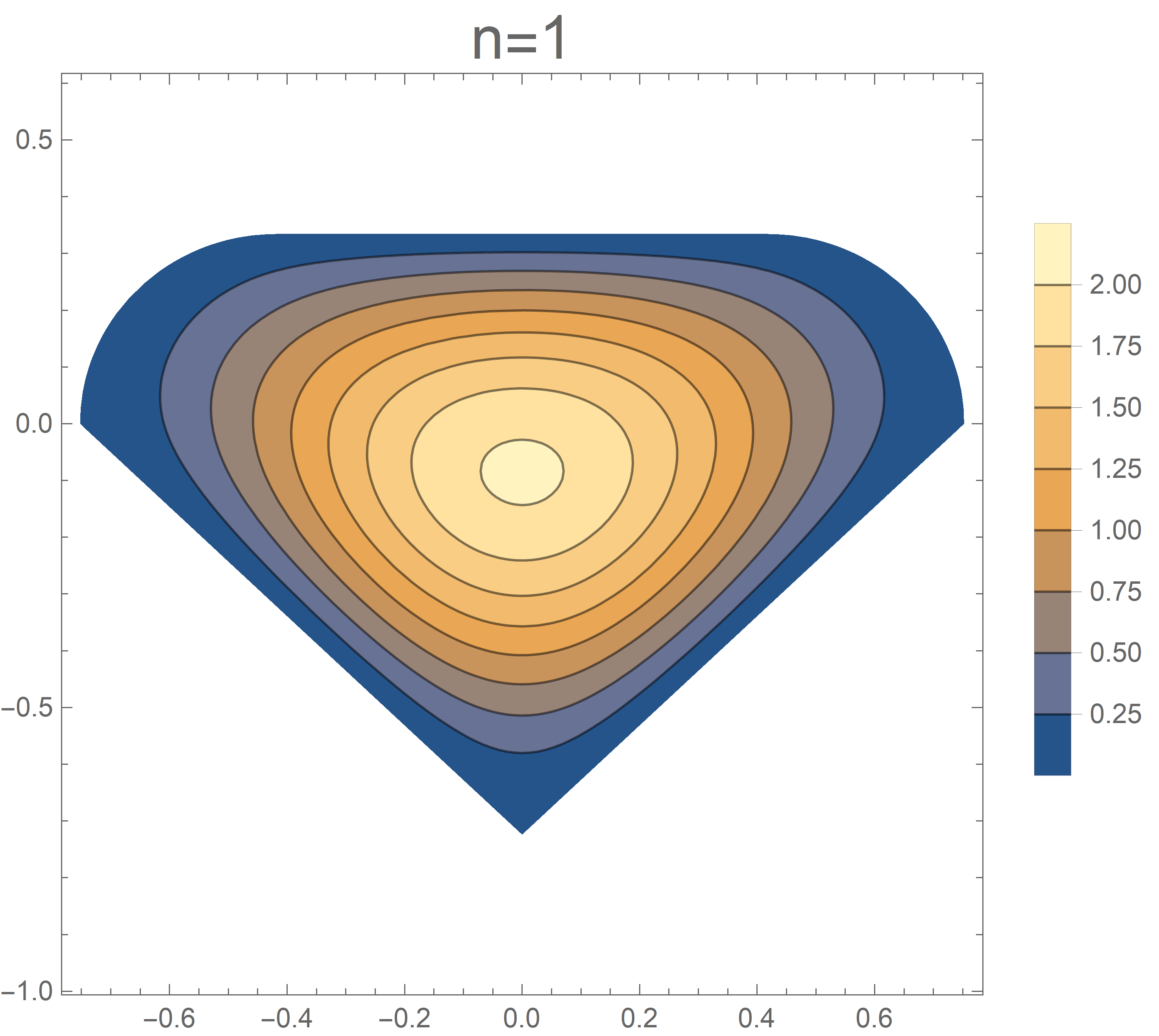}
        \caption*{}
    \end{subfigure}
    \hfil
    \begin{subfigure}[b]{0.3 \textwidth}
        \includegraphics[width=\textwidth]{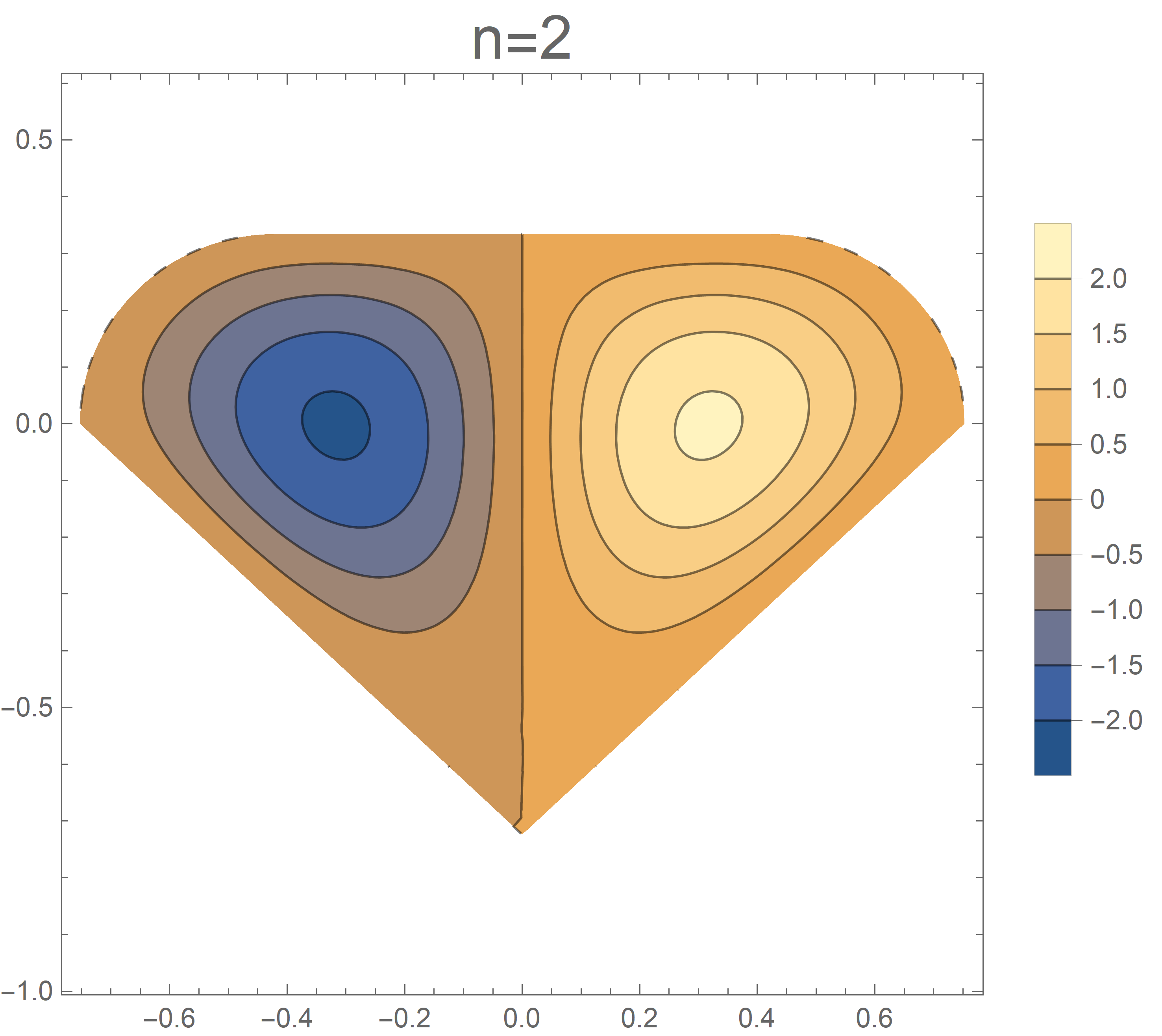}
        \caption*{Superman wavefunctions}
    \end{subfigure}
    \hfil
    \begin{subfigure}[b]{0.3 \textwidth}
        \includegraphics[width=\textwidth]{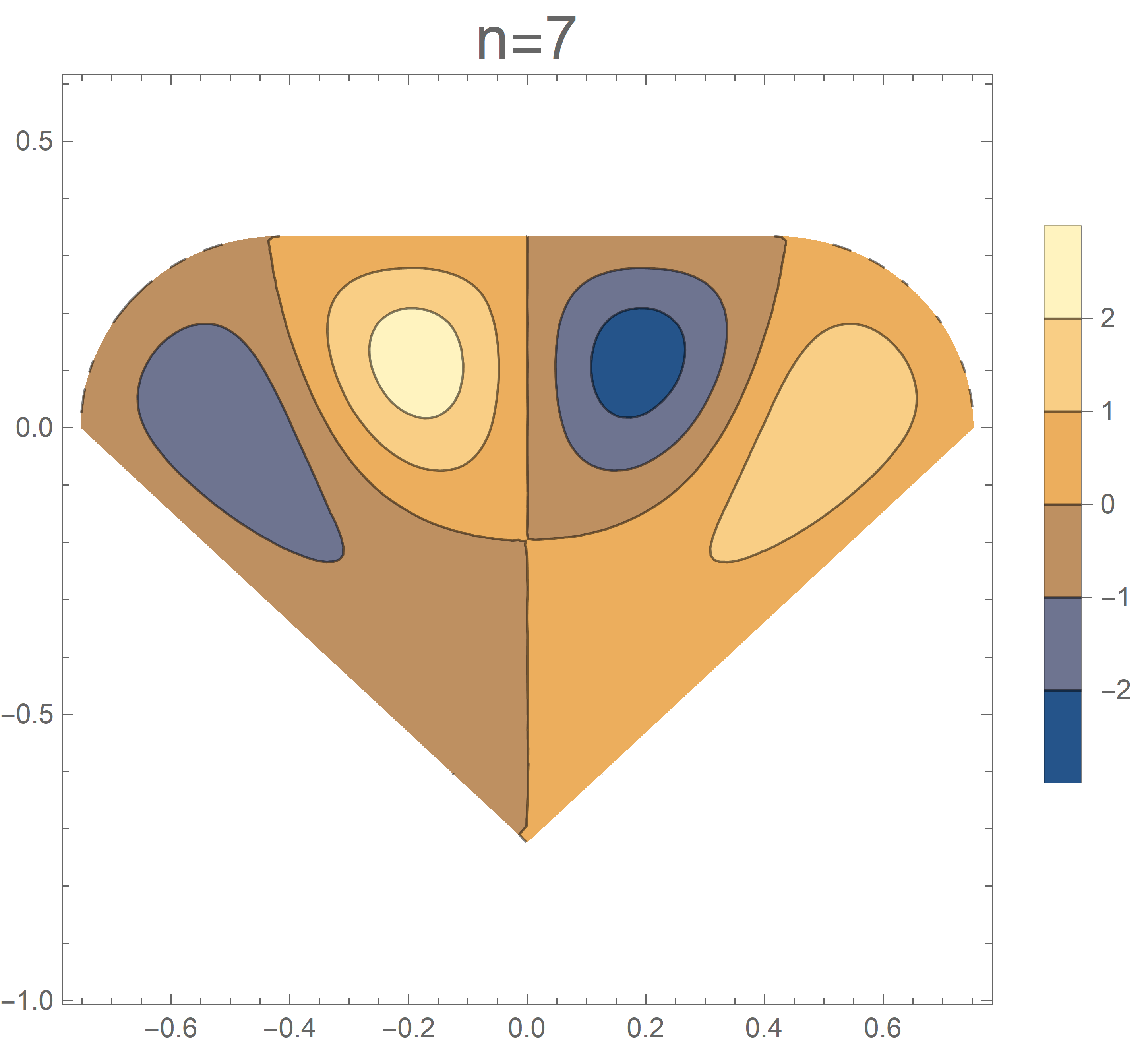}
        \caption*{ }
    \end{subfigure}
    \hfil
    \vspace{1cm}
     \begin{subfigure}[b]{0.3 \textwidth}
        \includegraphics[width=\textwidth]{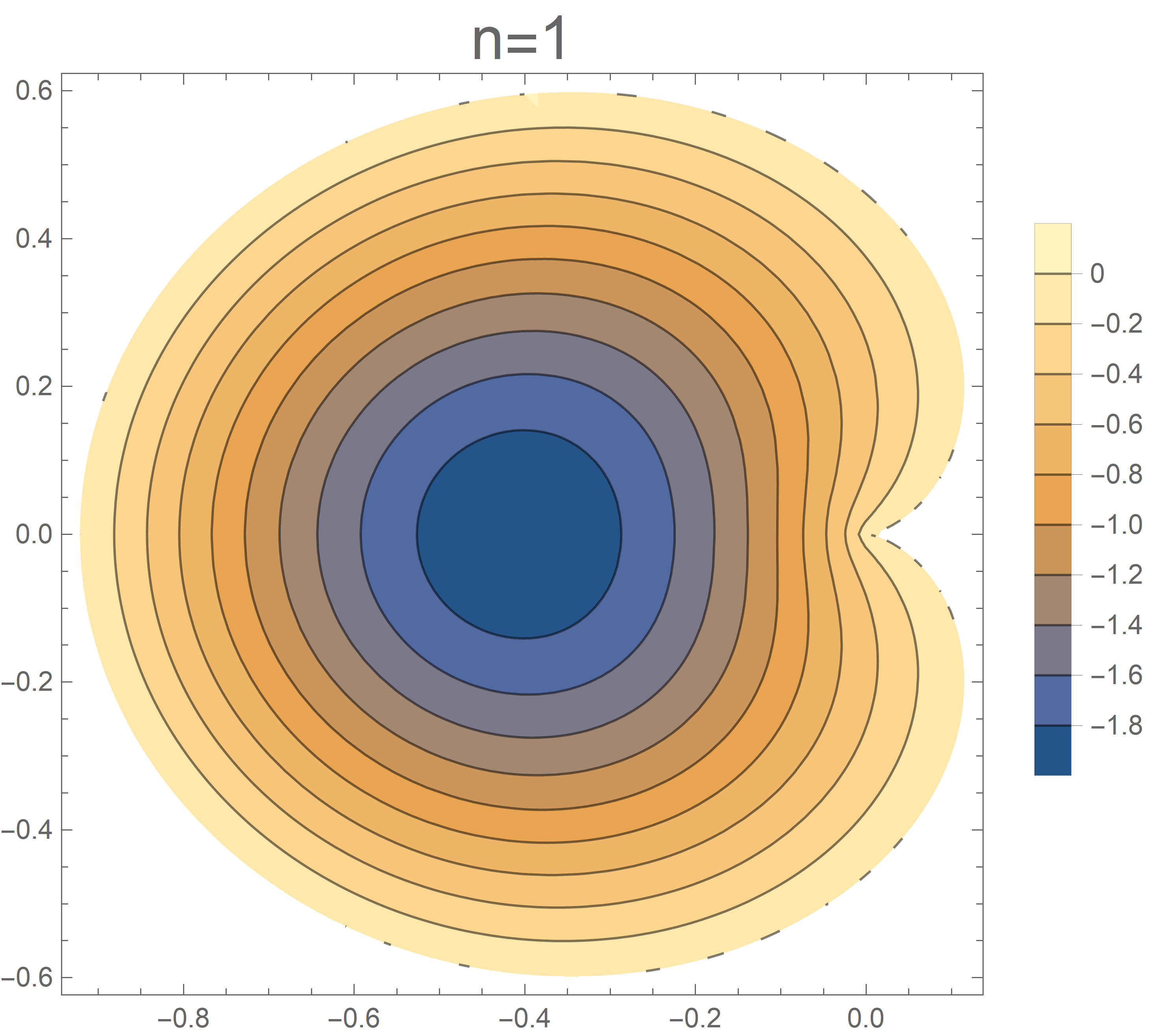}
        \caption*{}
    \end{subfigure}
    \hfil
    \begin{subfigure}[b]{0.3 \textwidth}
        \includegraphics[width=\textwidth]{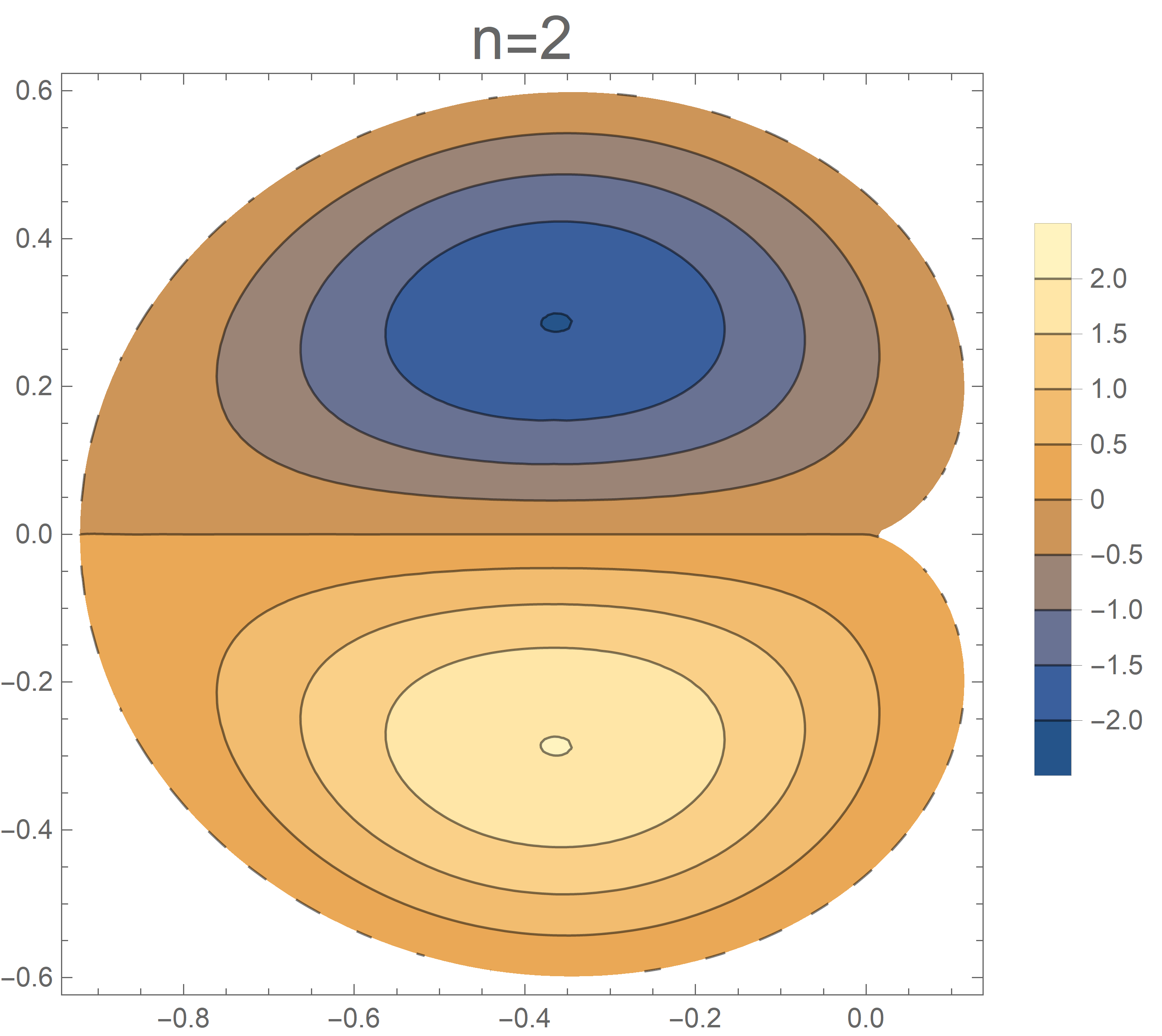}
        \caption*{Cardioid wavefunctions}
    \end{subfigure}
    \hfil
    \begin{subfigure}[b]{0.3 \textwidth}
        \includegraphics[width=\textwidth]{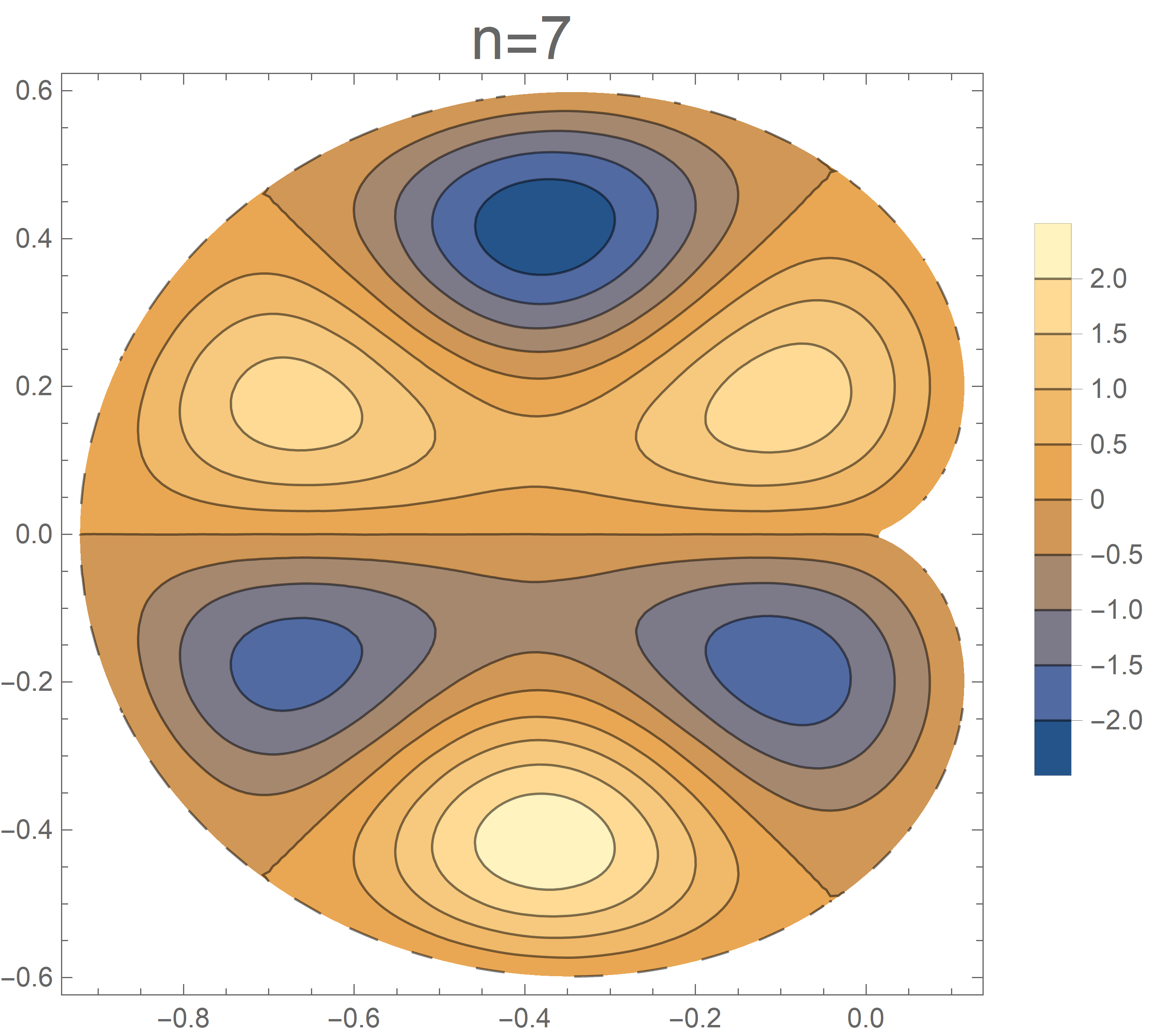}
        \caption*{ }
    \end{subfigure}
    \caption{Contour plots of the wavefunctions for the Sinai, diamond and cardioid billiards for \(n=1,2,7\).}
    \label{wave function sinai diamond}
\end{figure}

Indeed, we see that there is initial growth in microcanonical OTOCs of the Sinai billiard for low modes as well. And so, the question arises: why does this initial growth only occur in the case of the Sinai billiard? The answer lies in the shapes of the wavefunctions of these billiard systems with low \(n\)-values. In Fig.\ \ref{wave function sinai diamond} we can see contour plots of the wavefunctions of the Sinai, cardioid, and diamond billiards for \(n=1,2,7\). It is clear from the contour plots that only the Sinai \(n=1\) wave function consists of multiple peaks and troughs, whereas the other \(n=1\) wavefunctions consist of only one peak (or trough). This means that the typical scale (characteristic length scale over which the probability density of the wave function changes significantly) is smaller (as a fraction of the total size of the system) in the case of the Sinai billiard, in comparison to the typical scales of the wavefunctions of the other billiard systems for \(n=1\).\par
This disparity holds for wavefunctions of these systems corresponding to low \(n\). For the diamond, cardioid and stadium billiards, the typical scales of their low \(n\) wavefunctions are about the same size as the size of the system. Thus, these wavefunctions do not ``feel" the curvature of the walls of their enclosures \cite{Hashimoto_2017}. However, due to the typical scales of the low \(n\) Sinai wavefunctions being significantly smaller than the size of the system, the wavefunctions do experience the curvature of the walls. As a result, the wavefunctions of the Sinai billiard are the only ones out of the four billiard systems to be appreciably affected by the curvature of the walls at low \(n\). This manifests itself as an initial growth in microcanonical OTOCs of the Sinai billiard corresponding to low \(n\) values. It follows that the other three billiard systems do not experience this initial growth in their low-mode microcanonical OTOCs.\par

\subsection{Asymptotic Nature of OTOC at Large Times}
\begin{figure}[htbp]
    \centering
    \begin{subfigure}[b]{0.58 \textwidth}
        \caption{Sinai billiard}
        \includegraphics[width=\textwidth]{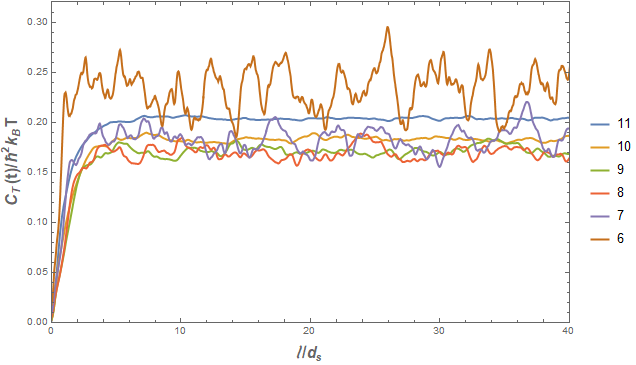}
        
    \end{subfigure}
    \hfil
    \begin{subfigure}[b]{0.58\textwidth}
    \caption{Cardioid billiard}
        \includegraphics[width=\textwidth]{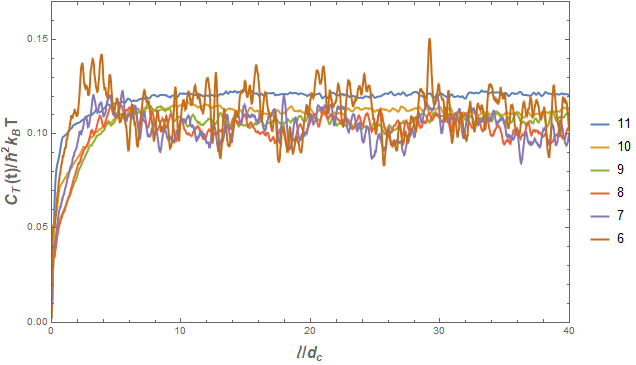}
        
    \end{subfigure}
    \hfil
    \begin{subfigure}[b]{0.58 \textwidth}
        \caption{Diamond billiard}
        \includegraphics[width=\textwidth]{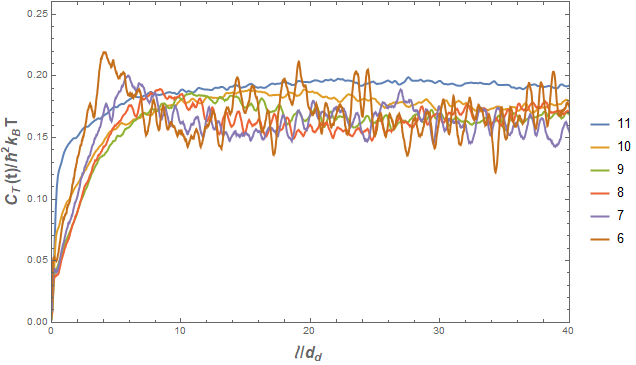}
        
    \end{subfigure}
    \caption{Asymptotic values of OTOCs divided by temperature for the (a) Sinai billiard, (b) cardioid billiard \& (c) diamond billiard. Temperatures are expressed on a log base \(2\) scale. In principle, these graphs have the same information as Fig.\,\ref{fig: thermal otocs}, but here one can compare fluctuations at different temperatures better. We observe that the oscillations of the asymptotic OTOC die down significantly with increasing temperature. The highest temperatures in these graphs are taken to be much higher to show the persistence of the linear growth of the asymptotic values of OTOCs with temperature.}
    \label{asymptotic OTOCs}
\end{figure}

The thermal OTOCs for each billiard reach their \emph{saturation} value after a specific time, which is proportional to temperature and consistent with 
\begin{equation}\label{eq late time}
    C_s \propto m a^2 k_B T,
\end{equation}
as was observed in \cite{Hashimoto_2017,Jalabert_2018}. However, estimating the proportionality constant in the semiclassical framework is challenging due to the various possible pairings \cite{late_time_1} and the influence of trajectory loops \cite{late_time_2,late_time_3}. During the intermediate time window, the OTOC exhibits oscillations with respect to the length parameter, indicating the dynamics of the billiard and periodic-orbit corrections \cite{Jalabert_2018}.

Although (\ref{eq late time}) was rigorously derived in \cite{Jalabert_2018} in the semiclassical limit, \cite{Hashimoto_2017} was able to derive the same result using dimensional analysis. The argument given in \cite{Hashimoto_2017} is that the semiclassical regime ends when the wavefunction spreads over the entire system and since $C_T(t)$ has the dimension of $\hbar^2$, the only reasonable value of the OTOC is given by (\ref{eq late time}). This shows the power of thinking in terms of specific wave functions and dimensional analysis. Although we know that the exact state of the system is given by the Gibbs state which doesn't evolve over time, useful information is still obtained by thinking in terms of specific wavefunctions which do evolve with time. This is because the OTOC is a time-dependent object and there's much value in thinking of the time dependence in terms of states.

The plots of the OTOCs for large times are given in Fig.\,\ref{asymptotic OTOCs}. It is seen that the oscillations of the OTOCs die down significantly with increasing temperature. The linear growth given by Eq.\,(\ref{eq late time}) is also manifest in these graphs. The label for the $x$-axes of these graphs is $\ell/d$, which is a proxy for time. This will be explained in more detail in the following section.

\section{Quantum Lyapunov Exponents and Thermal OTOCs} \label{sec:lyapunov}
When calculating the classical Lyapunov exponent in Section \ref{sec:classical}, we used collision numbers as the parameter instead of time. By doing so, we essentially replaced the time axis with length in units of average distance between two consecutive collisions. Similarly, we can express the scaled time axis of the thermal OTOCs in units of the average collision distance between two successive collisions. We calculated the average distance between collisions ($d_{\text{avg}}$) for our billiards and presented the values in table \ref{table:classical}. This facilitates the comparison between the classical Lyapunov exponent and the quantum Lyapunov exponent. In this section, we derive the quantum Lyapunov exponents for the three billiards.
\par
\begin{figure}[htbp]
    \centering
    \includegraphics[scale=0.8]{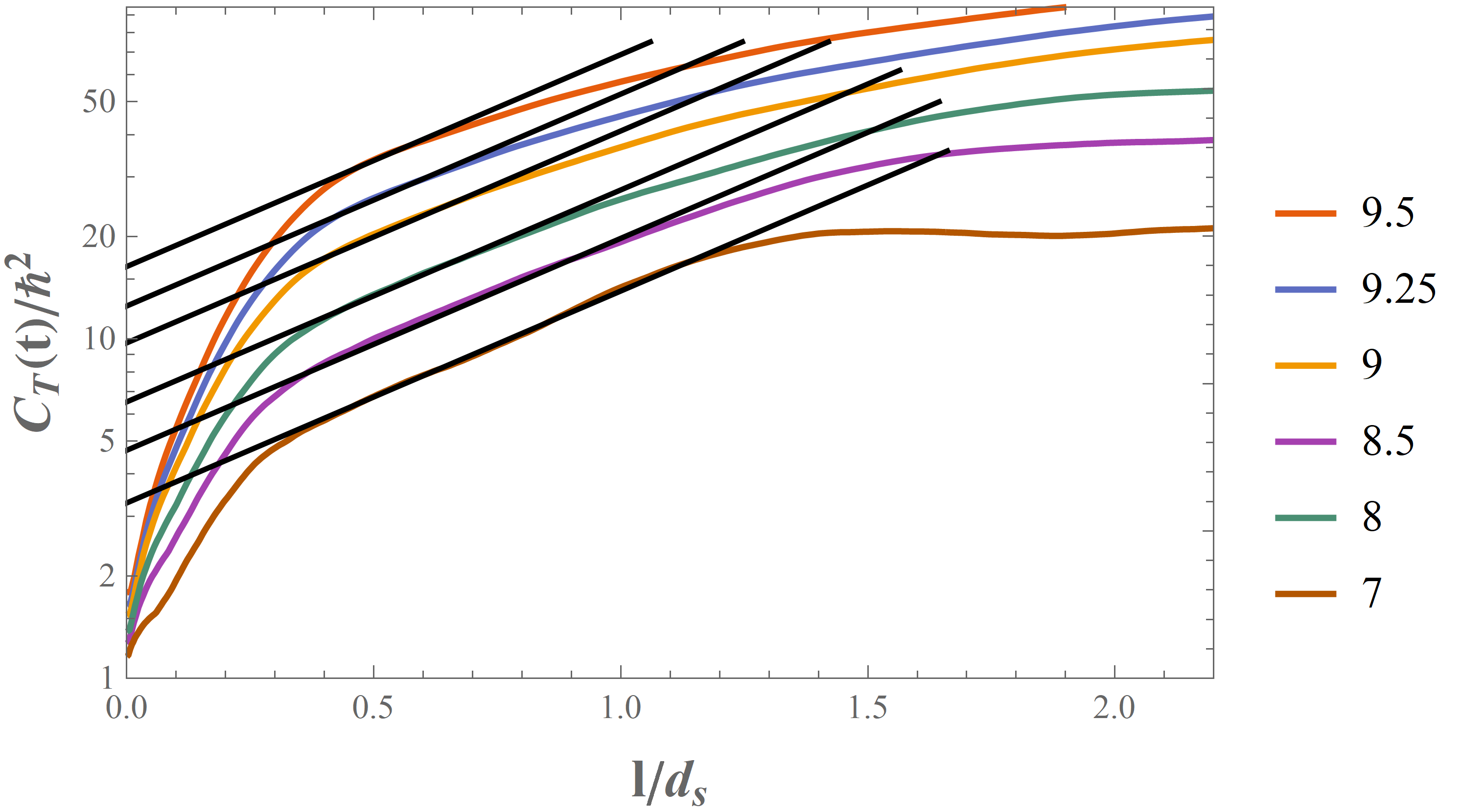}
    \caption{Numerically calculated thermal OTOCs on a logarithmic scale, with respect to the length (scaled time) $\ell=\Tilde{v}t$ (in units of $d_{\textsc{s}}$), where $\Tilde{v} = (\beta m)^{-1/2}$ is the mean-squared $x$-velocity component, and $d_{\textsc{s}}$ is the average collision distance calculated in Sec.\,\ref{sec:classical} for the \textbf{Sinai billiard}. The temperatures are on a logarithmic scale of base $2$. The black straight lines correspond to the exponential growth $a(T) e^{\sqrt{3}\lg  \Tilde{v} t}$, which accurately fits the data within an intermediate time-window $0.4 \leq \ell/d_{\textsc{s}} \leq 1.3$.}
    \label{fig:sinai_early_time2}
\end{figure}
We exhibit the thermal OTOCs for the Sinai billiard as a function of scaled time in Fig.\,\ref{fig:sinai_early_time2}. On the $x$-axis, we have \(\ell/d_{\mathrm{s}}\). Here, $\ell = \Tilde{v} t$ has the dimension of length but it is essentially time scaled by $\Tilde{v}=(k_BT/m)^{1/2}$ which is the mean squared $X$-component of the velocity. The colour code reflects the temperature, which is expressed in a logarithmic scale of base $2$. To achieve a comparable time scale as our classical calculation, the $x$-axis is expressed in units of the average collision distance, which is $d_{\text{avg}}=d_{\textsc{s}}=0.4817$ for the area of the billiard $A=1$. 

\par
In the short time regime, the behaviour of the OTOCs can be divided into two distinct parts. Initially, there is a quadratic increase in the OTOC with respect to time or length, which is characteristic of quantum perturbation theory. This is followed by a rapid growth, leading to a window of exponential increase in the OTOC. The semiclassical approach is valid for the second interval but unsuitable for the initial perturbative or rapid growth periods. This is because they correspond to times much earlier than the time of the first collision with the boundaries when the exponential divergence of classical trajectories has not begun to take place.
\par
To determine the quantum Lyapunov exponent, one first needs to express the OTOC in the semiclassical regime. It turns out the derivation of the semiclassical expression for the OTOC for systems with low degrees of freedom is a non-trivial task. Based on the path-integral treatment of \cite{gutzwiller}, the authors of  \cite{Jalabert_2018,garciamata2022outoftimeorder} derived the following semiclassical expression for the OTOC
\begin{align}
    C_{\mathrm{sc}} (t)= \frac{\beta^2\hbar^2}{64\pi m^2}\int d\mathbf{p} \exp{-\beta\frac{\mathbf{p}^2}{2m}} (e^{2\lambda t}p^2_x).
\end{align}
From this, one gets for low-temperature billiards the following expression
\begin{equation}\label{otoc eq}
    \dfrac{C_{LT}(t)}{\hbar^2} \propto \exp[\sqrt{3}\lg \Tilde{v}t].
\end{equation}
where $\lg$ is known as the ``geometric" Lyapunov exponent. For short time window and low temperatures, the raw exponent $\Lambda$ that one obtains from the graphs is
\begin{equation}\label{Lambda}
    \Lambda =\sqrt{3}\lg \Tilde{v}.
\end{equation}
from which one extracts the geometric Lyapunov exponent.
\par
For the Sinai billiard we have $\lg = 0.83 \hspace{0.1cm} d_{\mathrm{s}}^{-1}$. In Fig.\,\ref{fig:sinai_early_time2}, the solid black lines represent the exponential functions $\mu(T) e^{\sqrt{3}\lg \Tilde{v} t}$, which well approximate the OTO\-Cs within the temperature range $2^7\leq T \leq 2^{9.5}$. As the temperature increases beyond upper limit, the time window of exponential growth starts to shrink until it disappears (more about this below in Subsection \ref{ehrenfest}). Remarkably, although both the value of the L.H.S.\,of Eq. (\ref{otoc eq}) as well as the range of the exponential region vary with temperature, the slope of the curves remains the same. This was also observed in the case of the quarter-stadium billiard in \cite{Jalabert_2018}.
\par
In classical calculations of the Lyapunov exponent, neither the area of the billiard nor the particle's velocity plays a role. In Table \ref{table:classical}, we denoted the average classical Lyapunov exponent extracted from classical calculations as $\lc$. In the quantum calculation, however, the geometric Lyapunov exponent \emph{does} depend on the area of the billiard. Nonetheless, the factor $0.83$ in the geometric Lyapunov exponent for the Sinai billiard remains independent of both the area of the billiard and the velocity of the particle. Given that the distance between two collisions is set to one (see, Appendix \ref{appendix_2}) in our classical calculation, we must multiply the geometric Lyapunov exponent obtained from the quantum calculation by the average collision distance:
\begin{equation}\label{lambdag}
  d_\text{avg}  \lg =\lq,
\end{equation}
where we have introduced $\lq$ as the quantum analog of $\lc$. Note that this way of expressing the Lyapunov exponent makes it a dimensionless quantity. Should one need the traditional dimensions, one can use Eq.\,(\ref{Lambda}).

\begin{figure}[htbp]
    \centering
    \begin{subfigure}[b]{0.45 \textwidth}
        \caption{Cardioid billiard}
        \includegraphics[width=\textwidth]{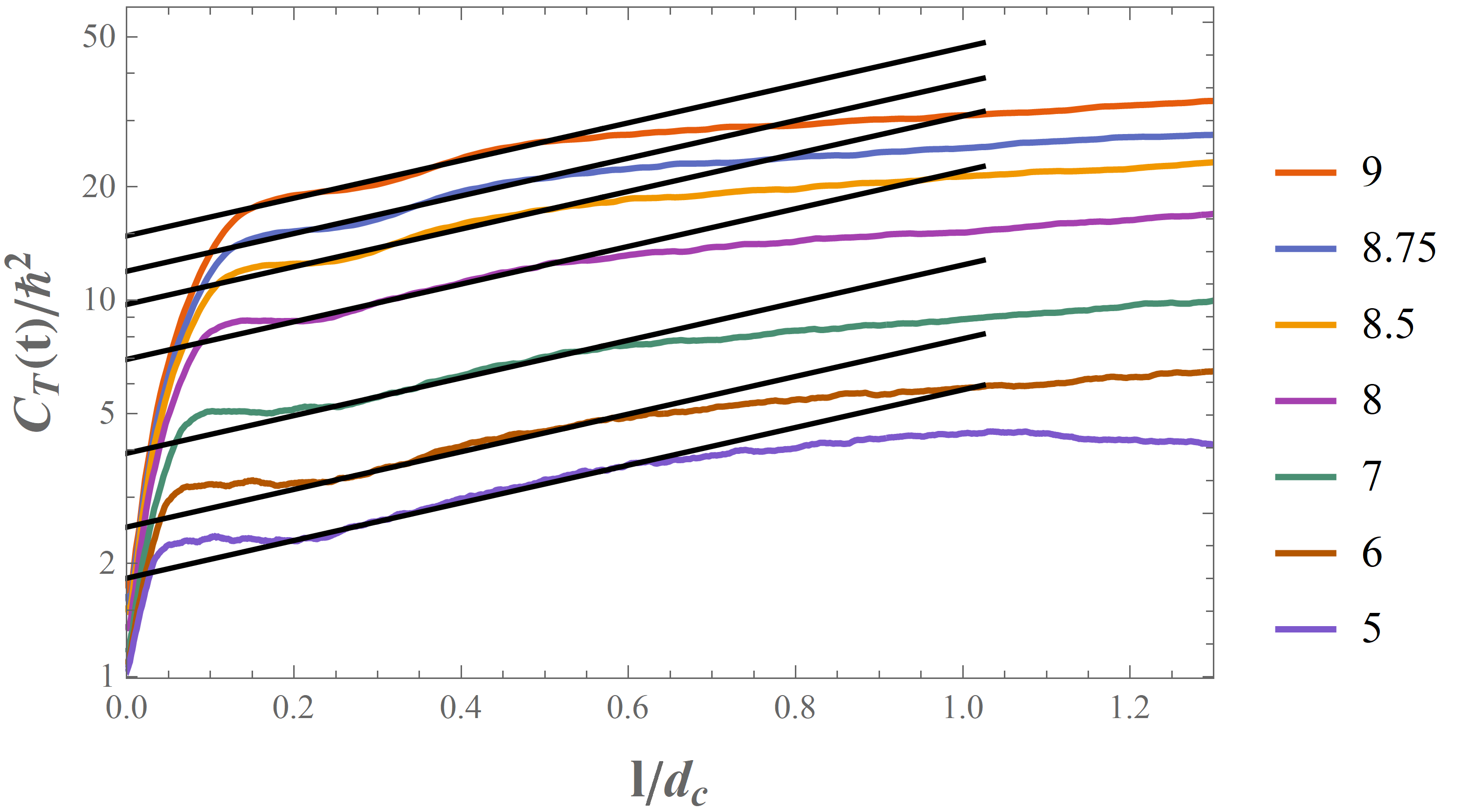}
    \end{subfigure}
    \hfil
    \begin{subfigure}[b]{0.45 \textwidth}
        \caption{Diamond billiard }
        \includegraphics[width=\textwidth]{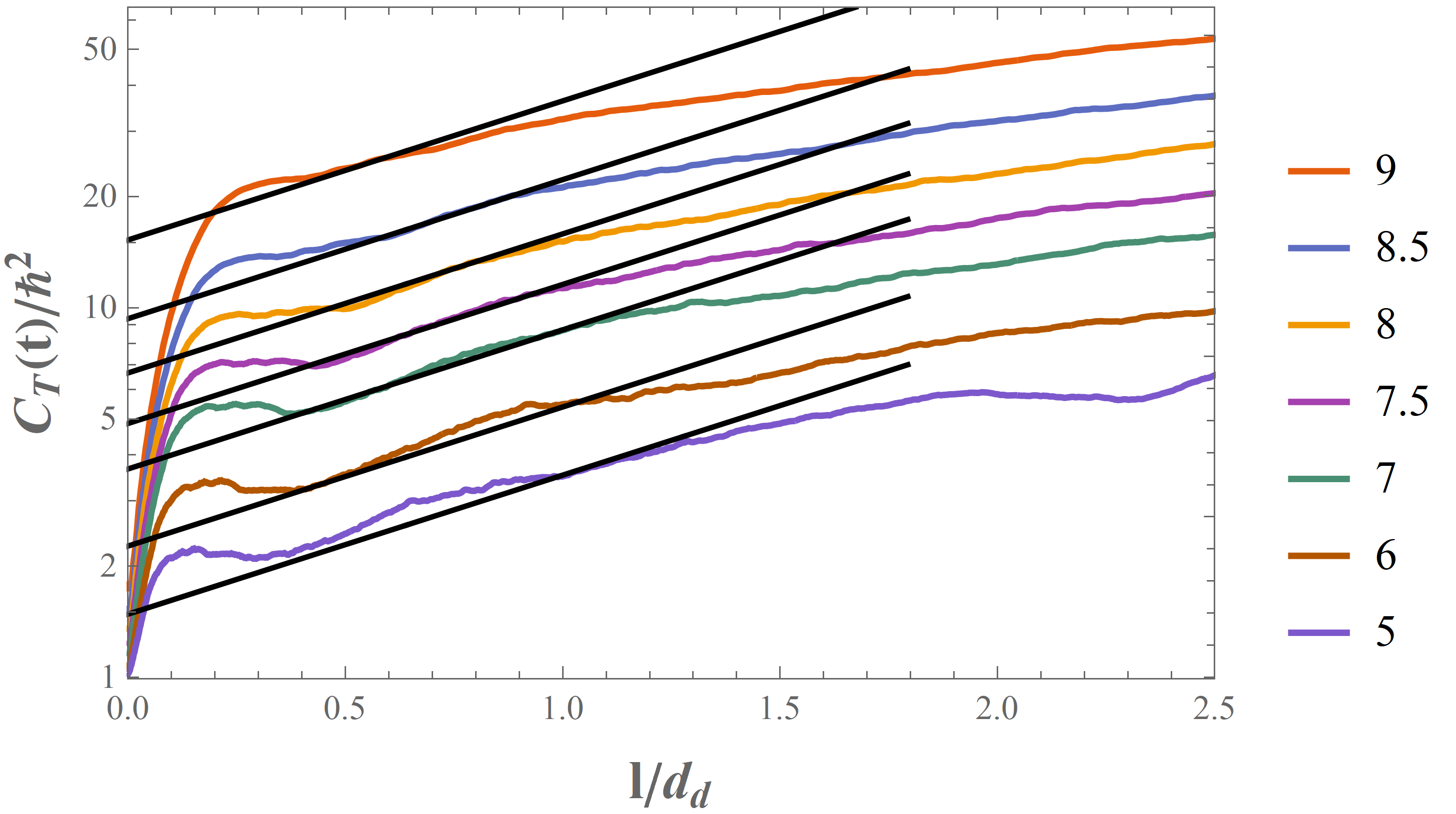}
    \end{subfigure}
    \caption{Numerically calculated thermal OTOCs on a logarithmic scale, with respect to the length (scaled time) $\ell=\Tilde{v}t$ (in units of average collision distance) for (a) cardioid billiard and (b) diamond billiard. The temperatures are on a logarithmic scale with a base of 2. The black straight lines correspond to the exponential growth \(a(T) e^{\sqrt{3}\lg  \tilde{v} t}\), accurately fitting the data within an intermediate time-window of \(0.1 \leq \ell/d_c \leq 0.75\) for the cardioid billiard and \(0.1 \leq \ell/d_d \leq 1.5\) for the diamond billiard.}
    \label{fig: early time cardioid diamond2}
\end{figure}

Figure \ref{fig: early time cardioid diamond2} displays the thermal OTOCs for the cardioid and diamond billiards as a function of scaled time.
The scaled time (length), denoted as $\ell=\Tilde{v}t$, is given in units of average collision distances for $A=1$ and is indicated as $d_{\mathrm{c}}$ for the cardioid billiard and $d_{\mathrm{d}}$ for the diamond billiard. The values of $d_c=0.8787$ and $d_d=0.7805$ are taken from Table \ref{table:classical}. The range of temperatures where the exponential functions provide the best fit is $2^5\leq T \leq 2^9$ for both the cardioid and diamond billiard.\par
The quantum Lyapunov exponent $\lq$ is extracted in a similar way as for the Sinai billiard by fitting the exponential function to the data. Table \ref{table:quantum} compares the classical Lyapunov exponent ($\lc$) and the quantum Lyapunov exponent ($\lq$) for our selected billiards.\par
\begin{table}[htbp]
\centering
\begin{tabular}{||c | c | c ||} 
 \hline
 Billiards &  $\lc$ & $\lq$ \\ [0.5ex] 
 \hline \hline
 Sinai &  $0.805$ &  $0.83$  \\ 
 \hline
 Cardioid & $0.665$ &   $0.665$   \\
 \hline
 Diamond & $0.686$ &   $0.5$  \\ 
 [1ex] 
 \hline
 \end{tabular}
 \caption{ Numerically calculated average geometric classical Lyapunov exponents $\lc$  and average quantum Lyapunov exponents $\lq$ extracted from thermal OTOCs for the Sinai, cardioid, and diamond billiards. }
\label{table:quantum}
\end{table}
Remarkably, we found that the quantum Lyapunov exponent is in excellent agreement to the classical Lyapunov exponent computed in Sec.\,\ref{sec:classical} for the cardioid billiard and the values for the Sinai billiard are also very close. This matching of the classical and quantum exponents are a consequence of Ehrenfest's theorem and is expected to hold until Ehrenfest time. There is, however, a larger difference between the values of the calculated quantum and classical Lyapunov exponents in the case of the diamond billiard. 
\par
\begin{figure}[htbp]
    \centering
    \begin{subfigure}[b]{0.45 \textwidth}
        \caption{}
        \includegraphics[width=\textwidth]{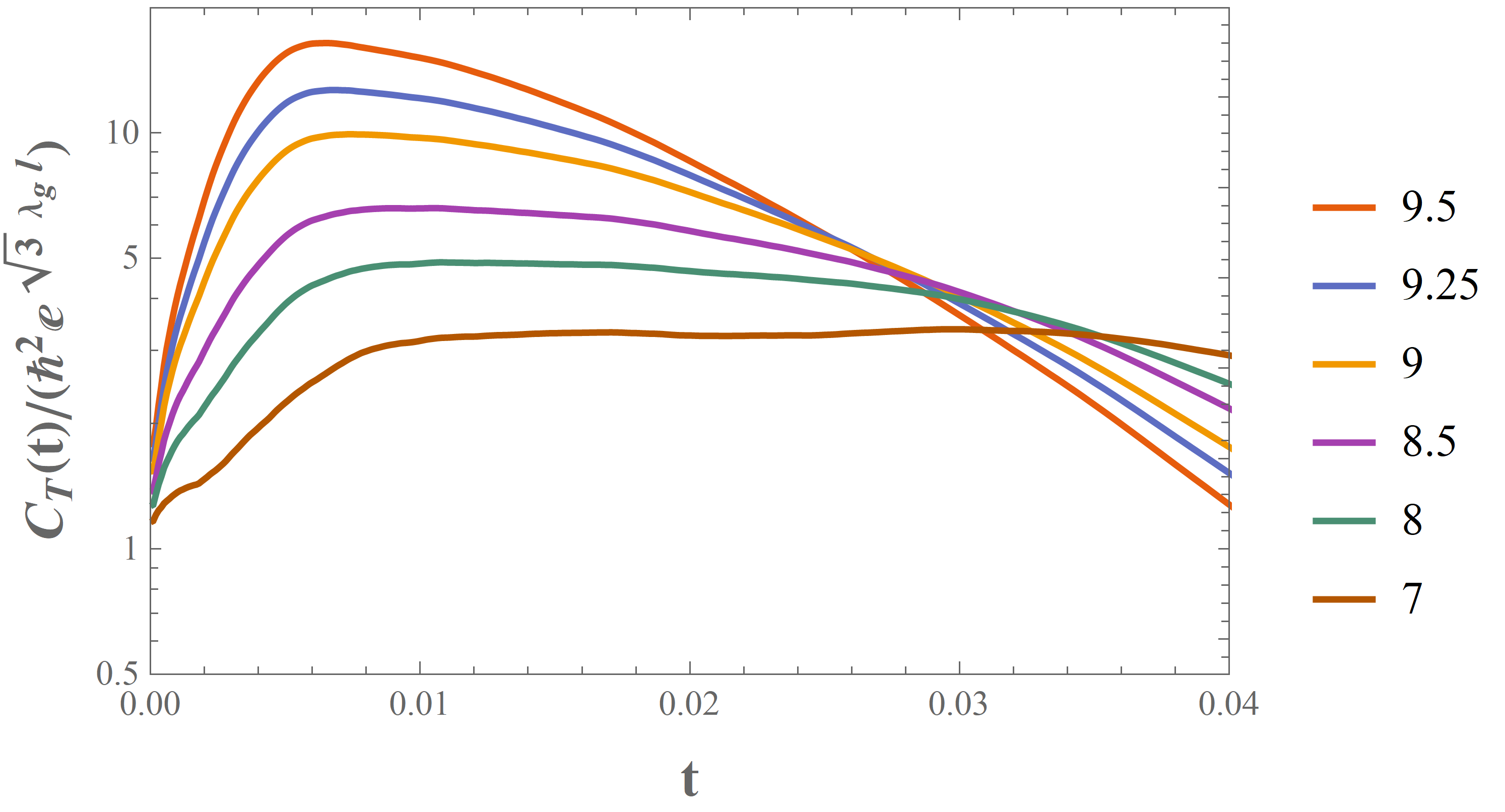}
        \label{fig:sinai temperature a}
    \end{subfigure}
    \hfil
    \begin{subfigure}[b]{0.45 \textwidth}
        \caption{}
        \includegraphics[width=\textwidth]{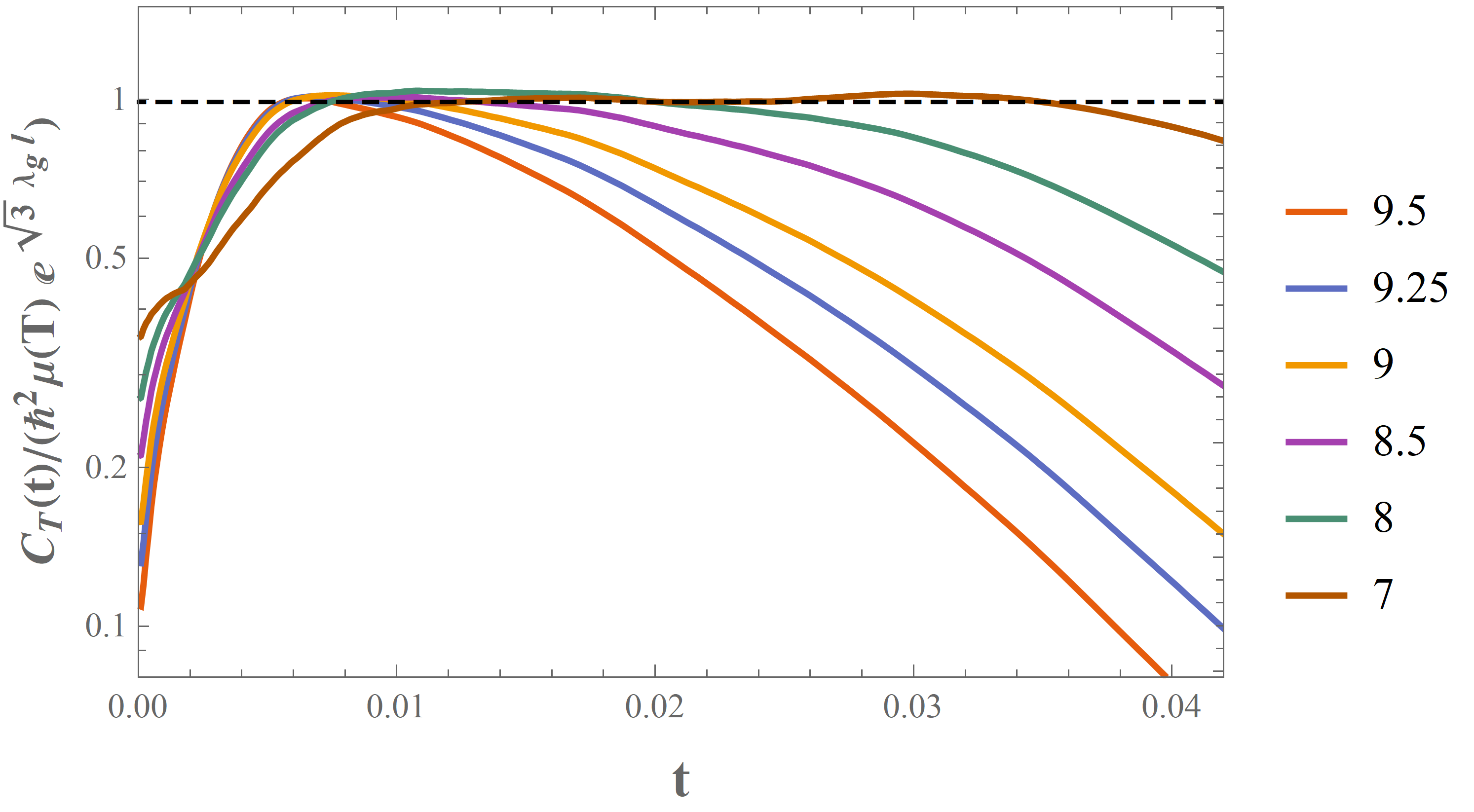}
        \label{fig:sinai temperature b}
    \end{subfigure}
    \caption{Numerically calculated thermal OTOCs of the \textbf{Sinai billiard} on a logarithmic scale: (a) scaled by \(\exp(\sqrt{3} \lg  \ell)\), and (b) scaled by \(\mu(T) \exp(\sqrt{3} \lg  \ell)\)  as a function of time $t$. The temperatures are represented on a logarithmic scale with a base of 2.}
\end{figure}

For the thermal OTOCs the window where we expect exponential growth is temperature-dependent. To observe this window of exponential growth more clearly we plot the OTOCs but divided, first by $\exp(\sqrt{3} \lg \ell)$ and then by \(\mu(T)\times\\\exp(\sqrt{3}\lg \ell)\) where $\mu(T)$ is the temperature-dependent prefactor. These plots are shown in Figs.\,\ref{fig:sinai temperature a} and \ref{fig:sinai temperature b}. Both figures contain the same information, but the first figure, shows the width of the exponential region of each individual OTOC more clearly, whereas the second plot is better for comparison between different curves. In these plots we see that as the temperature increases, the width of the exponential regime decreases until it finally disappears.

\begin{figure}[htbp]
    \centering
    \begin{subfigure}[b]{0.45 \textwidth}
        \caption{ Cardioid billiard}
        \includegraphics[width=\textwidth]{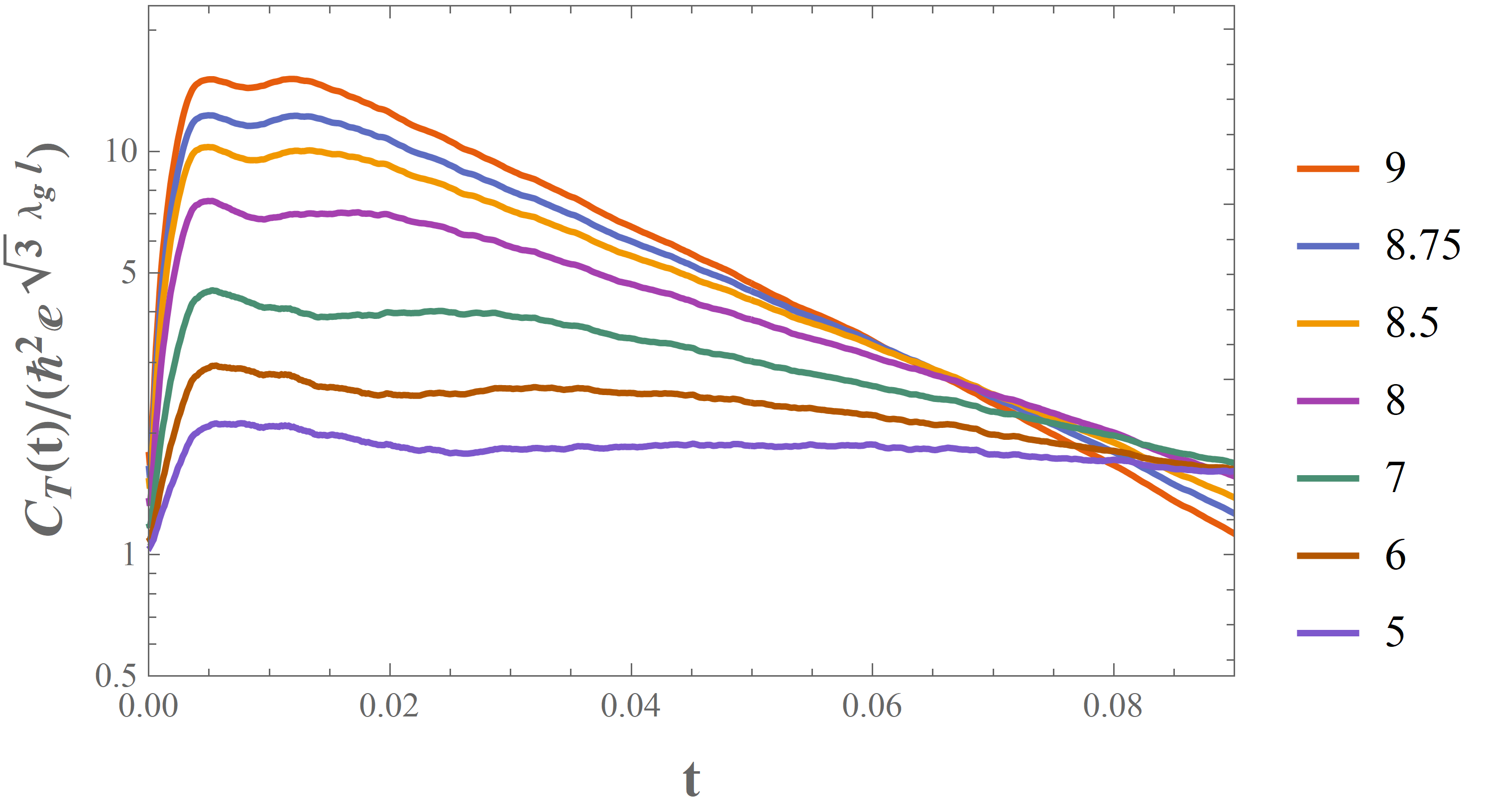}
        \label{fig:card temperature a}
    \end{subfigure}
    \hfil
    \begin{subfigure}[b]{0.45 \textwidth}
        \caption{Cardioid billiard}
        \includegraphics[width=\textwidth]{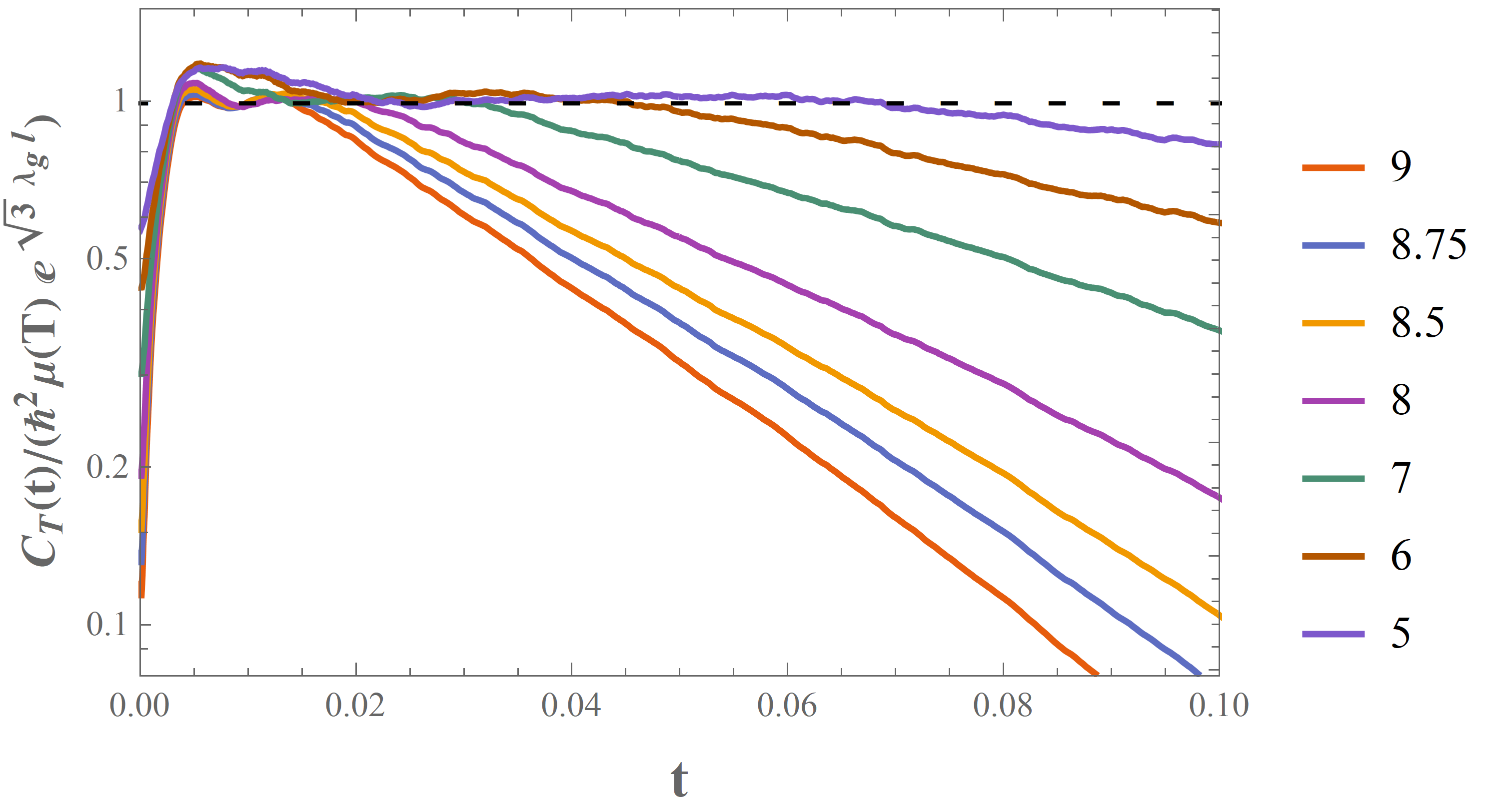}
        \label{fig:card temperature b}
    \end{subfigure}
    \hfil
    \begin{subfigure}[b]{0.45 \textwidth}
        \caption{Diamond billiard}
        \includegraphics[width=\textwidth]{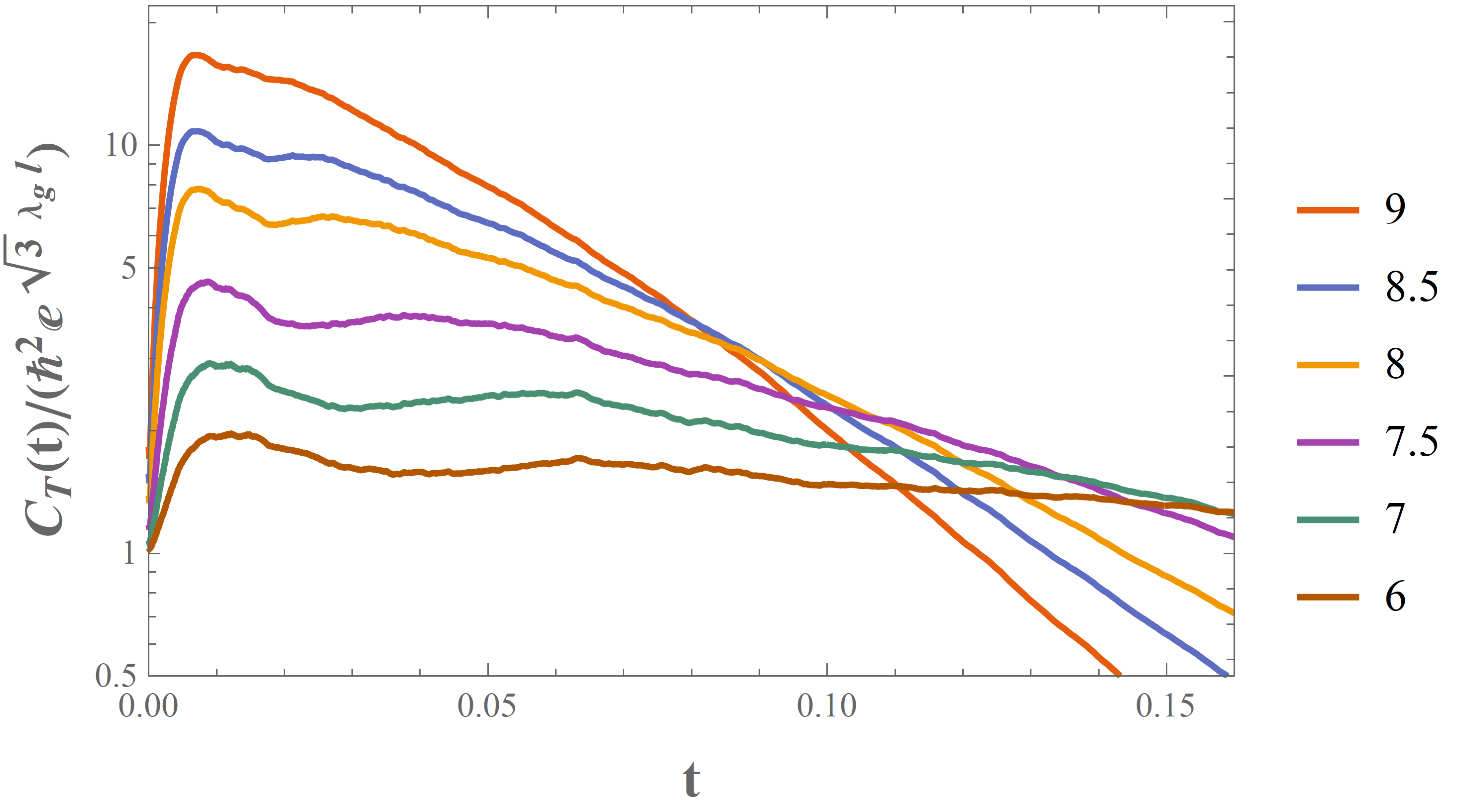}
        \label{fig:diamond temperature a}
    \end{subfigure}
    \hfil
    \begin{subfigure}[b]{0.45 \textwidth}
        \caption{Diamond billiard}
        \includegraphics[width=\textwidth]{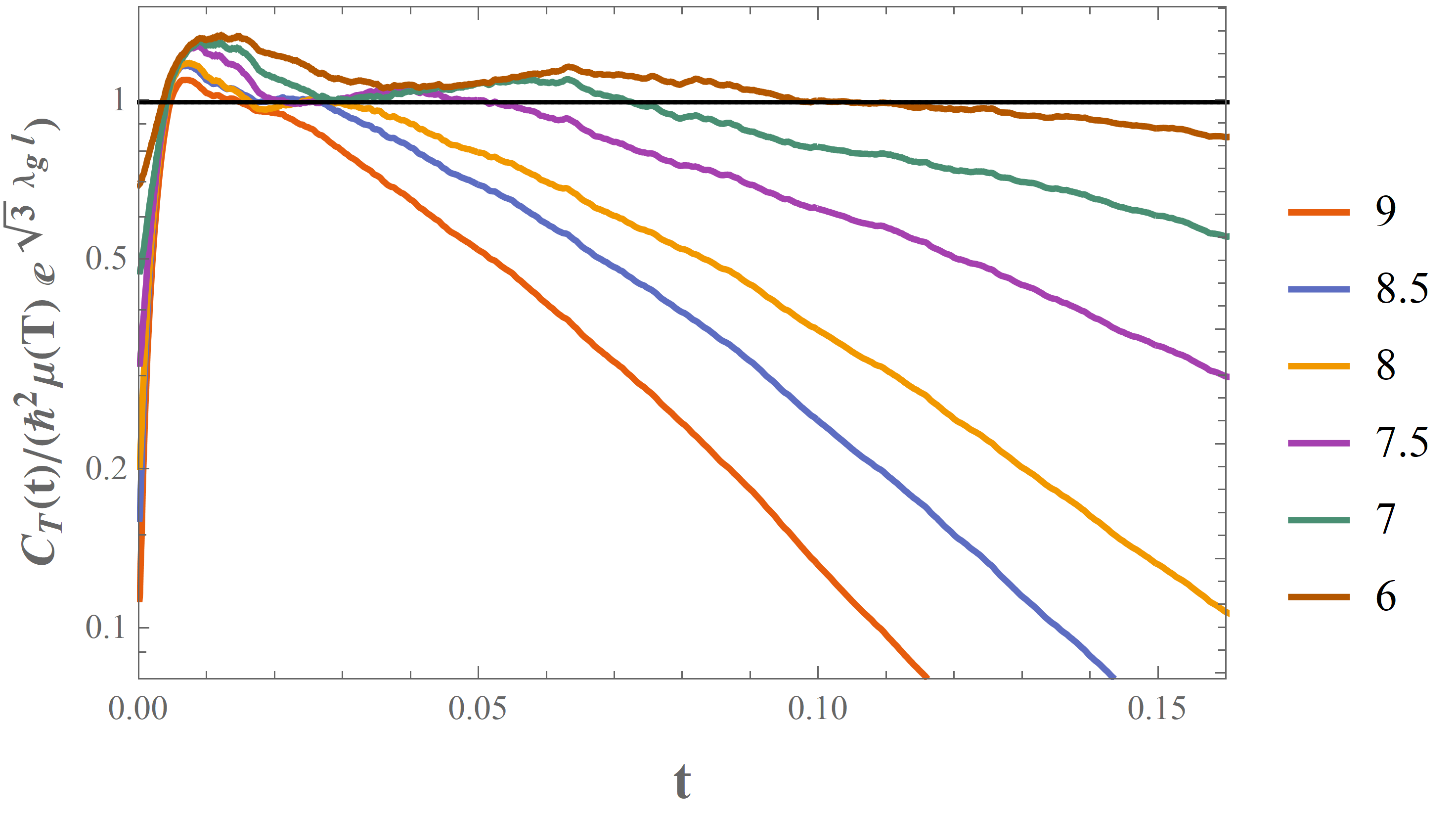}
        \label{fig:diamond temperature b}
    \end{subfigure}
    \hfil
    \caption{Numerically calculated thermal OTOCs of cardioid on a logarithmic scale: (a) scaled by \(\exp(\sqrt{3} \lg  \ell)\), and (b) scaled by \(\mu(T) \exp(\sqrt{3} \lg  \ell)\)  as a function of time $t$. Numerically calculated thermal OTOCs of the diamond billiard on a logarithmic scale: (c) scaled by \(\exp(\sqrt{3} \lg  \ell)\), and (d) scaled by \(\mu(T) \exp(\sqrt{3} \lg  \ell)\)  as a function of time $t$. As in the case of the Sinai billiard, the temperatures are expressed on a \(\log_2\) scale.}
    \label{fig: early time cardioid diamond3}
\end{figure}

The analogous plots of the thermal OTOCs for cardioid and diamond billiards are given in Fig.\ref{fig: early time cardioid diamond3}. The results are consistent with our earlier findings: as the temperature increases, the fit to the exponential function gradually worsens.

\section{Temperature Dependence of Ehrenfest Times}\label{ehrenfest}
\subsection*{Dimensional Analytic Estimates}

In the semiclassical regime, the growth of the OTOC is controlled by the exponent $\Lambda=\sqrt{3} \lg \Tilde{v}$ (Eq.\,\ref{Lambda}). Because of the presence of the rms velocity $\Tilde{v}$, this exponent is proportional to $\sqrt{T}$. As observed in \cite{Jalabert_2018} and also by our data, in the region where one finds exponential growth, the value of $\Lambda$ is consistent with the proposed bound on the growth rate of OTOCs given by $\Lambda \leq 4 \pi \k T/ \hbar$ \cite{Maldacena_2016}. 

However, the region of exponential growth is related to the temperature. Since the end of the exponential growth with respect to time marks the transition from the semiclassical regime to the fully quantum regime, we can mark the transition point as the Ehrenfest time $\te$. We want to understand how the Ehrenfest time depends on the temperature $T$. 
\par
The naive intuition is that the faster a wavepacket moves the quicker it will disperse and spread over the whole billiard. Since we are dealing with a system at a specific temperature, it picks out the rms velocity $\Tilde{v}\sim \sqrt{T}$. Thus our naive, zeroth-order expectation is that the Ehrenfest time should have the form
\begin{align}
    \te \sim \frac{1}{\sqrt{T}}.
\end{align}
This is borne out if we consider the evolution of a Gaussian wavepacket as was done in \cite{Hashimoto_2017}. The reflection of the wave function at the billiard walls will deform its shape, but to the lowest order of approximation we can still consider it to be localized (this corresponds to the validity of the particle approximation). As is well known from elementary quantum mechanics \cite{Powell}, the uncertainty of a Gaussian wavepacket evolves with time according to 
\begin{align}\label{evolution}
\Delta x (t) = \Delta x_0\sqrt{1+\left( \frac{t}{\tau}\right)^2}
\end{align}
where $\Delta x_0 \equiv \Delta x(0)$ and $\tau \equiv 2m\Delta x^2_0/\hbar$ is the time scale after which the width of the Gaussian changes significantly. From the thermal de Broglie wavelength for a quantum particle we set 
\begin{align}\label{deBroglie}
\Delta x_0 =\sqrt{\frac{2\pi \hbar^2}{m \k T}}\,.
\end{align}
The Gaussian starts to change shape significantly when $t\simeq \tau$ and so we can expect $\te$ to be given by (switching to our units)
\begin{align}\label{ehrenfest-temp}
\te \sim \frac{L}{\sqrt{T}}
\end{align}
where $L = \Delta x(\te)$ is the typical system size. 
\par
One might object to this dimensional analysis argument saying that the uncertainty of the thermal state is time-independent and thus its uncertainty has no ``evolution." But, observe that one can decompose a thermal state in a basis of (over-complete) squeezed states and track their evolution in time. While it is true that the Gibbs state is invariant under time evolution, it is the nature of OTOC to introduce dynamics in this scenario. For the evolution of the OTOC, one expects that it would be dominated by the Gaussian with the length scale set by the temperature of the system. One can miss the semiclassical growth of the OTOC (see \cite{Rozenbaum_2019}, for example) that was found in \cite{Jalabert_2018} if one ignores the ``dynamics" in the thermal state.  

\begin{figure}[ht!]
    \centering
    \begin{subfigure}[b]{0.45 \textwidth}
        \caption{Sinai billiard}
        \includegraphics[width=\textwidth]{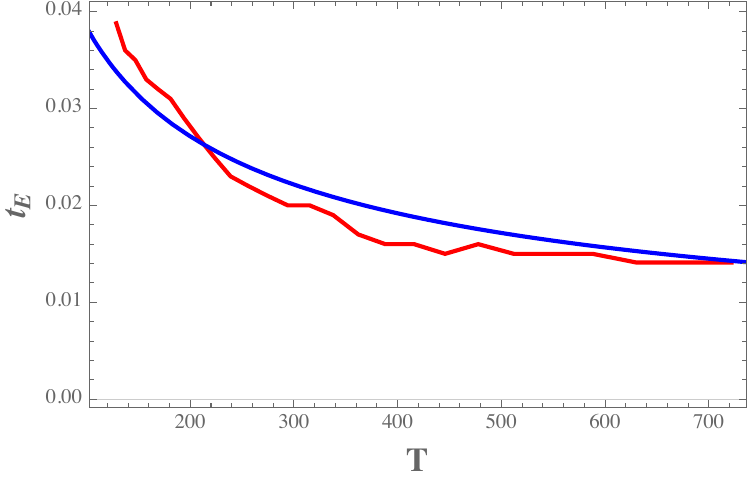}
        \label{fig:sinai ehrenfest}
    \end{subfigure}
    \hfil
    \begin{subfigure}[b]{0.45 \textwidth}
        \caption{Cardioid billiard}
        \includegraphics[width=\textwidth]{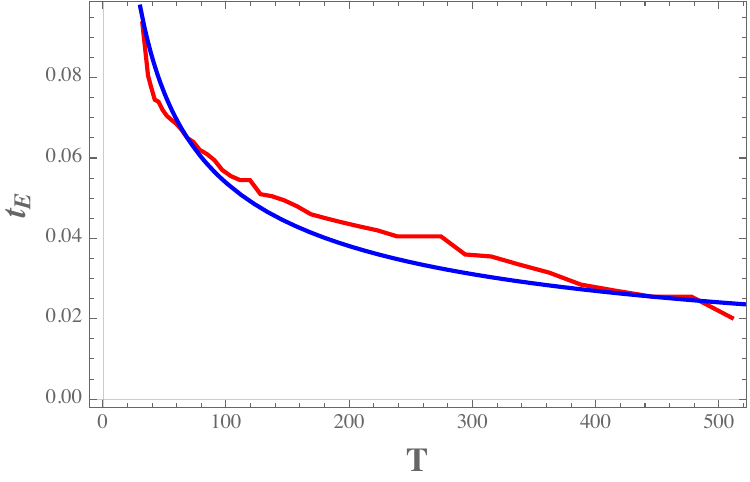}
        \label{fig:card ehrenfest}
    \end{subfigure}
    \hfil
    \begin{subfigure}[b]{0.45 \textwidth}
        \caption{Diamond/Superman billiard}
        \includegraphics[width=\textwidth]{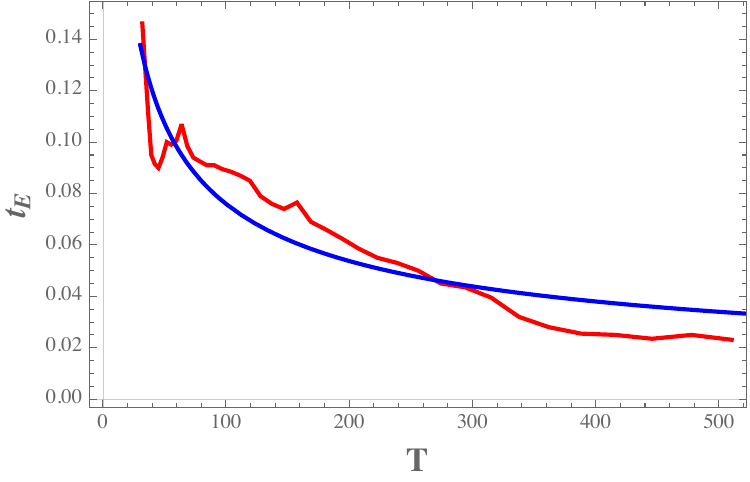}
        \label{fig:diamond ehrenfest}
    \end{subfigure}
    \hfil
    \caption{The dependence of the \textbf{Ehrenfest time} $t_E$ on the \textbf{temperature} $T$ for each billiard. The red curves represent the data obtained by numerics while the blue curve is the fitted curve. The fitting was done with the function $\te = c/\sqrt{T}$.}
    \label{ehrenfest-times-vs-temperature}
\end{figure}

There is another, \emph{independent} argument as to why the Ehrenfest time for quantum billiards should have $1/\sqrt{T}$ dependence. We can interpret the commutator $[x(t), p]$ as representing the failure in the simultaneous precise measurement of the position of the particle at time $t$ after an initial measurement of its momentum $p$ at $t = 0$ and vice versa\footnote{Note that unitarity means that we can reverse evolve our state from time $t$ to 0. What we are proposing here is reminiscent of the Loschsmidt echo.}. Thus one can make the following crude approximation
\begin{align}
\langle [x(t), p]^2\rangle \simeq \left(\Delta x(t) \Delta p_0\right)^2
\end{align}
One can use (\ref{evolution}), (\ref{deBroglie}), and $\Delta x_0 \Delta p_0 \simeq \hbar$ on the right-hand side and the observation that $C(t\rightarrow \infty) \sim T$ from (\ref{eq late time}) on the left-hand side, to obtain the \emph{same} temperature dependence $\te\sim 1/\sqrt{T}$.\par
Note that, in this second argument, the asymptotic dependence of the OTOC on temperature, as given by (\ref{eq late time}) (and verified by our data), is a \emph{new} input which makes it independent from the previous argument.\par
In Fig.\,\ref{ehrenfest-times-vs-temperature}, we see a general agreement between the data and our approximation given by $\te \simeq c/\sqrt{T}$, with $c(\text{Sinai}) = 0.383614$, $c(\text{cardioid}) = 0.538536$, and $c(\text{Superman}) = 0.759939$. We see that these numbers are of the same order of magnitude as the length scale of the billiards.

\section{Discussion}\label{sec:discussion}
The OTOC represents a novel and interesting approach to studying quantum chaos. Although mainly used to study many-body systems, following \cite{Hashimoto_2017, Jalabert_2018, Rozenbaum_2017, Rozenbaum_2019} we found that it is also a good tool for studying the quantum chaos of systems with a small number of degrees of freedom, namely chaotic billiards. In this paper, we were mainly concerned with the quantum Lyapunov exponents and the semiclassical regime of three quantum billiards which are known to be chaotic classically. We exclude a detailed examination of the stadium billiard as it has been examined extensively in recent papers using OTOCs as the primary tool \cite{Hashimoto_2017,Jalabert_2018,Rozenbaum_2019}.

We  found that the agreement between the classical and quantum exponents is excellent for the Sinai and cardioid billiards but for the diamond billiard it is not as good (see Table \ref{table:quantum}). It is also worth noting that, try as we might, our computation of the classical Lyapunov exponent for the diamond/Superman could not reproduce the results of \cite{salazar2012classical}. Interestingly, it is their \emph{classical} value that is close to the \emph{quantum} Lyapunov exponent that we computed. Since we used the same code and technique to compute the Lyapunovs of all the billiards in this paper, we cannot dismiss the result of the diamond/Superman billiard as a fluke. This is a discrepancy that needs to be looked into more carefully.

Even in such simple systems, we found interesting phenomena in terms of the growth of OTOC of the Sinai billiard at low temperature. We traced this anomalous behaviour to the symmetry of the ground state of the wave function of the Sinai billiard which reflects the geometry of the billiard in contrast to the low-energy states of the other billiards. This observation provides a counterexample to the point of view that the microcanonical OTOCs for low $n$ may not see the curvature of the billiard because of the typical scales of the wave function for small energy being of the same size as the system \cite{Hashimoto_2017}. We also see a lot more structure in the long-time OTOC of the Sinai billiard at low temperatures compared to the other billiards at low temperatures (Fig.\,\ref{asymptotic OTOCs}).

As argued in \cite{swingle2018unscrambling}, the OTOC measures the ballistic growth of Heisenberg operators. Thus for systems of finite sizes, there comes a time when this growth is saturated. The time at which this happens is the scrambling time. Here we prefer to call it Ehrenfest time $\te$, as it is the time at which the wave function of a quantum particle has spread out over the whole system so that one can no longer expect the semiclassical regime (\emph{i.e.,} Ehrenfest's theorem) to hold. Using, admittedly, very rough dimensional analysis we presented \emph{two} logically independent arguments as to why the Ehrenfest time $\te$ for the three billiards presented in this paper should depend on temperature as $1/\sqrt{T}$. We have also presented numerical data to back up this claim. This is the zeroth order approximation and the data of the Ehrenfest time's dependence on temperature has a lot of structure in it. It would be an interesting exercise to look at this relationship in greater detail in a future publication.
\acknowledgments
We are very grateful to the authors of \cite{Jalabert_2018} for insightful and helpful correspondence.

\appendix

\section{Assessment of the error resulting from level truncation}\label{appendix_1}
At various points during our calculations of OTOCs, namely Eqs. (\ref{bnm}), (\ref{cn}), and (\ref{cT}), we encounter infinite sums. As we evaluate these sums via numerical calculations, the infinite sums in these equations must be truncated to a certain cut-off value, \(\It\). In this segment, we determine the effect of taking different values of \(\It\) on the OTOCs with the aim of ascertaining a suitable cut-off value. We shall focus on the Sinai billiard and compute the microcanonical OTOC for \(n=100\) for various truncation values \(\It\). The microcanonical OTOCs for \(\It = 100, 150, 200, 400, 600, 800\) are shown below.
\begin{figure}[htbp]
    \centering
    \includegraphics[scale=0.6]{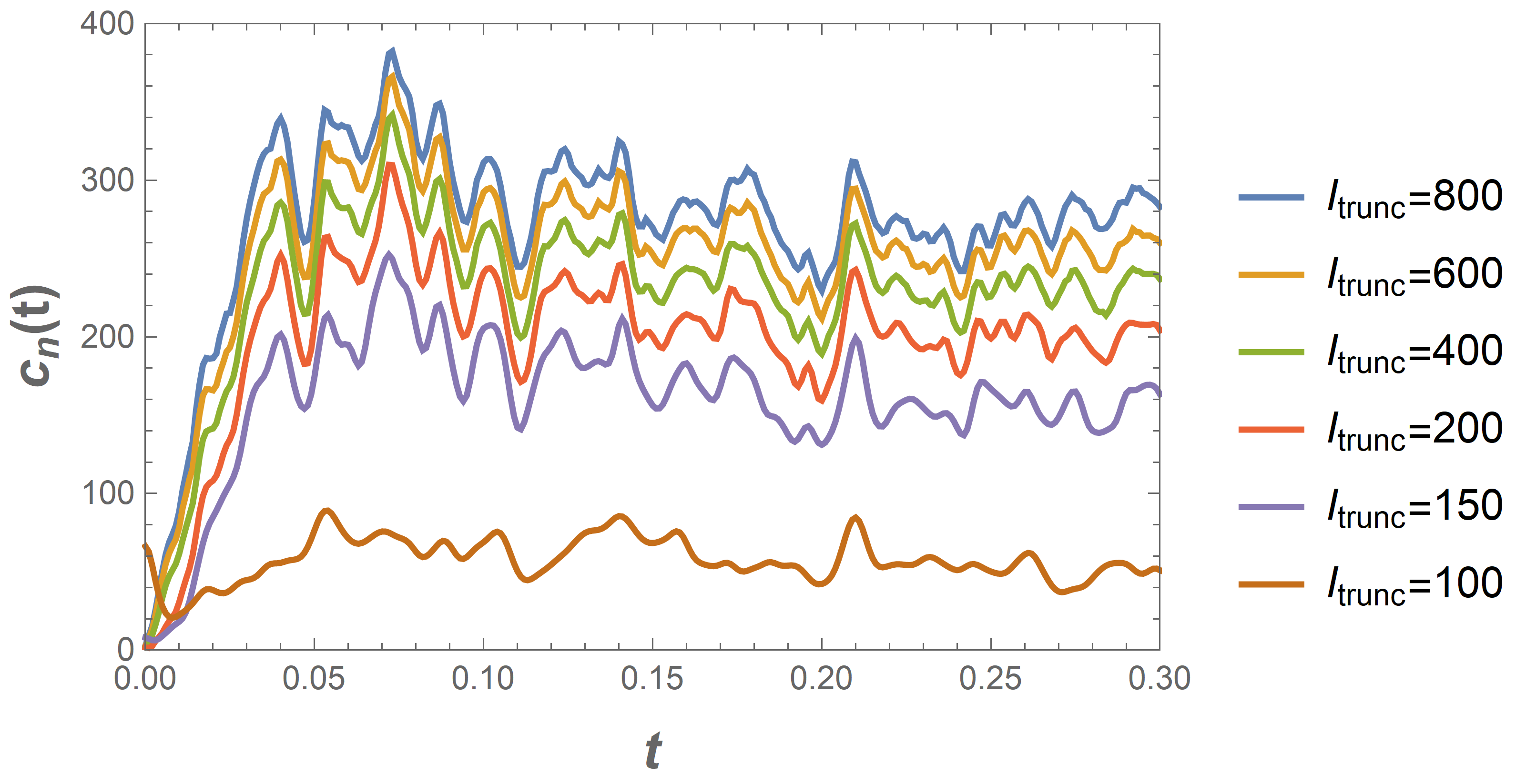}
    \caption{Microcanonical OTOCs of the Sinai billiard corresponding to \(n=100\) for \(\It =100, 150, 200, 400, 600, 800\).}
    \label{fig:truncation}
\end{figure}

As we can see, the microcanonical OTOCs for \(n=100\) converge fairly well as the value of \(\It \) is increased from \(100\) to \(800\). We found similar convergences for microcanonical OTOCs with \(n<100\). As \(n\) increases above 100, the corresponding microcanonical OTOCs do not contribute to the thermal OTOCs to a significant extent. This is because the \(\exp(\frac{-E_n}{T})\) term in Eqn. (\ref{cT}) suppresses the contributions of microcanonical OTOCs corresponding to large \(n\). As a result, the microcanonical OTOCs for \(n \le 100\) converging well at our truncation value \(\It \) is sufficient for our purposes. Therefore, we picked \(\It =800\) for our calculations of the Sinai OTOCs. Furthermore, we found, through similar analyses, that \(\It =800\) was a suitable truncation value for the cardioid and diamond billiard systems as well. Thus, we truncated the infinite sums to \(\It =800\) during the calculations of the OTOCs of those systems as well.   

\section{Averaging Collision Distance to Scale Length Parameters}\label{appendix_2}
\begin{figure}[htbp!]
    \centering
    \begin{subfigure}[b]{0.45 \textwidth}
        \caption{}
        \includegraphics[width=\textwidth]{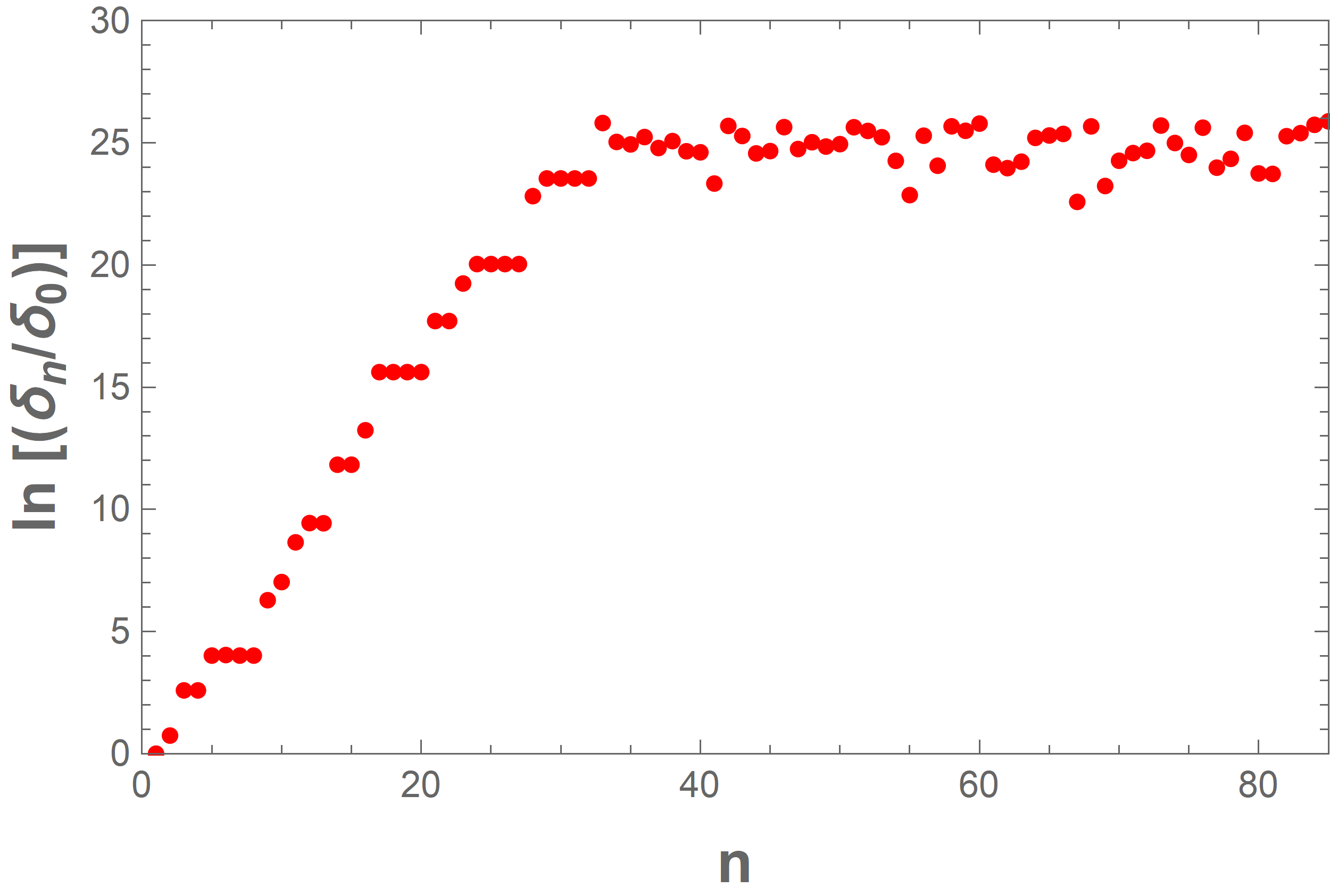}
        \label{fig:dots a}
    \end{subfigure}
    \hfil
    \begin{subfigure}[b]{0.45 \textwidth}
        \caption{}
        \includegraphics[width=\textwidth]{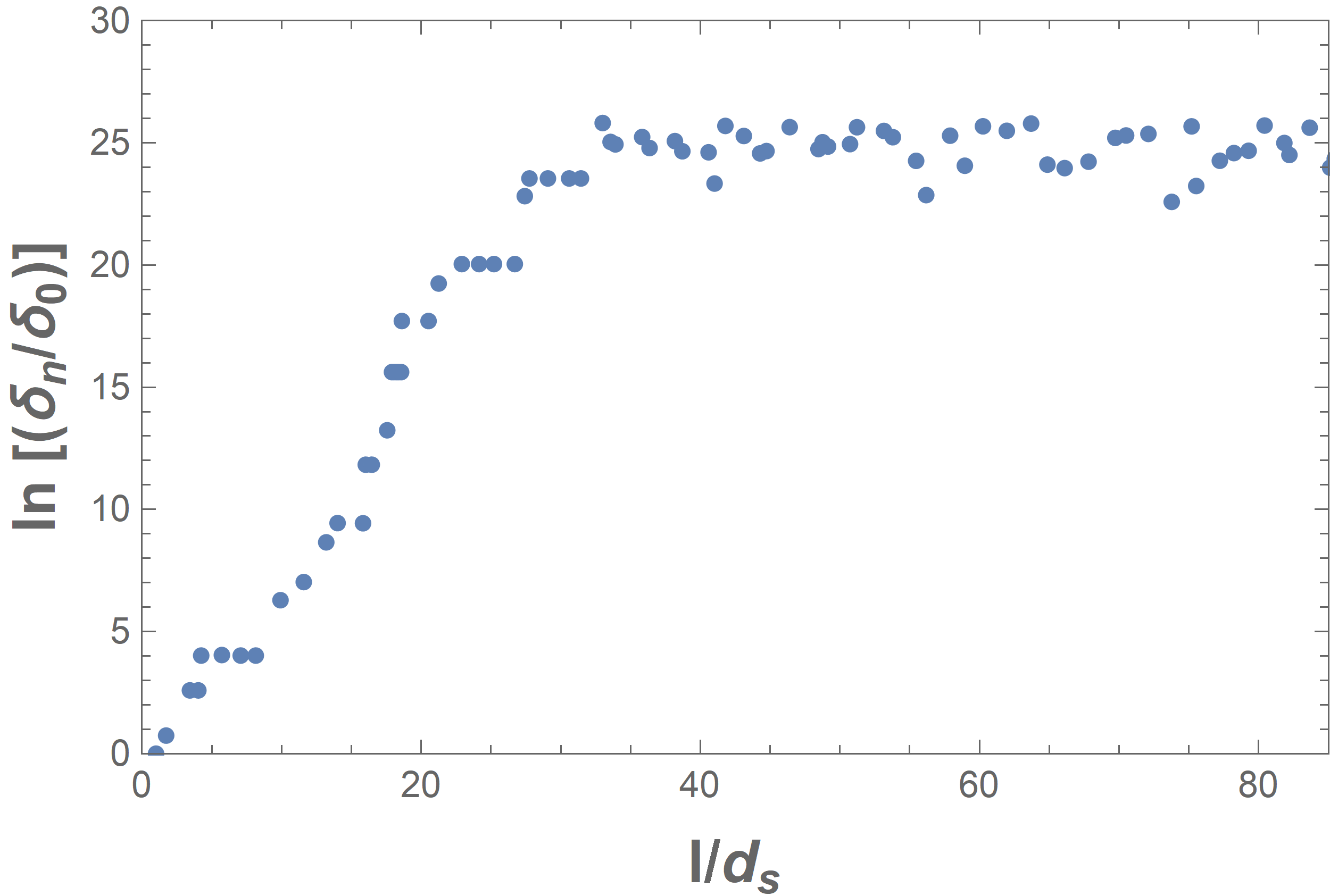}
        \label{fig:dots b}
    \end{subfigure}
    \caption{The growth rate of separation between two trajectories inside the Sinai billiard that start out extremely close to one another is depicted as follows: (a) with respect to collision number $n$, and (b) with respect to length scaled with the average distance between two consecutive collisions.}
    \label{fig: Classical Sinai }
\end{figure}

We plot a graph of a random trajectory in a classical billiard in Fig.\,\ref{fig:dots a}, where the slope of the unsaturated part represents the Lyapunov exponent. The $x$-axis denotes the collision number $n$. Although the \emph{actual} distances between \(n_1\) and \(n_2\), and between \(n_2\) and \(n_3\), differ, we consider them equal and set the separation to $1$ when plotting the points.\par
In Fig.\,\ref{fig:dots b}, we plot a graph where we recorded the exact distance between each pair of collisions, compiled a list, and plotted the points accordingly. This accurately represents the position of each point on the y-axis based on the distance traveled from the initial point. However, we scaled the $x$-axis by dividing it by the average collision distance up to the unsaturated part for this specific trajectory. This adjustment is necessary because the previous graph has a unit distance between \(n_1\) and \(n_2\), \(n_2\) and \(n_3\), and so forth. To achieve comparable scaling on the $x$-axis in the second graph, we needed to divide it by the average collision distance.\par
\begin{figure}[htbp!]
    \centering
    \includegraphics[scale=0.6]{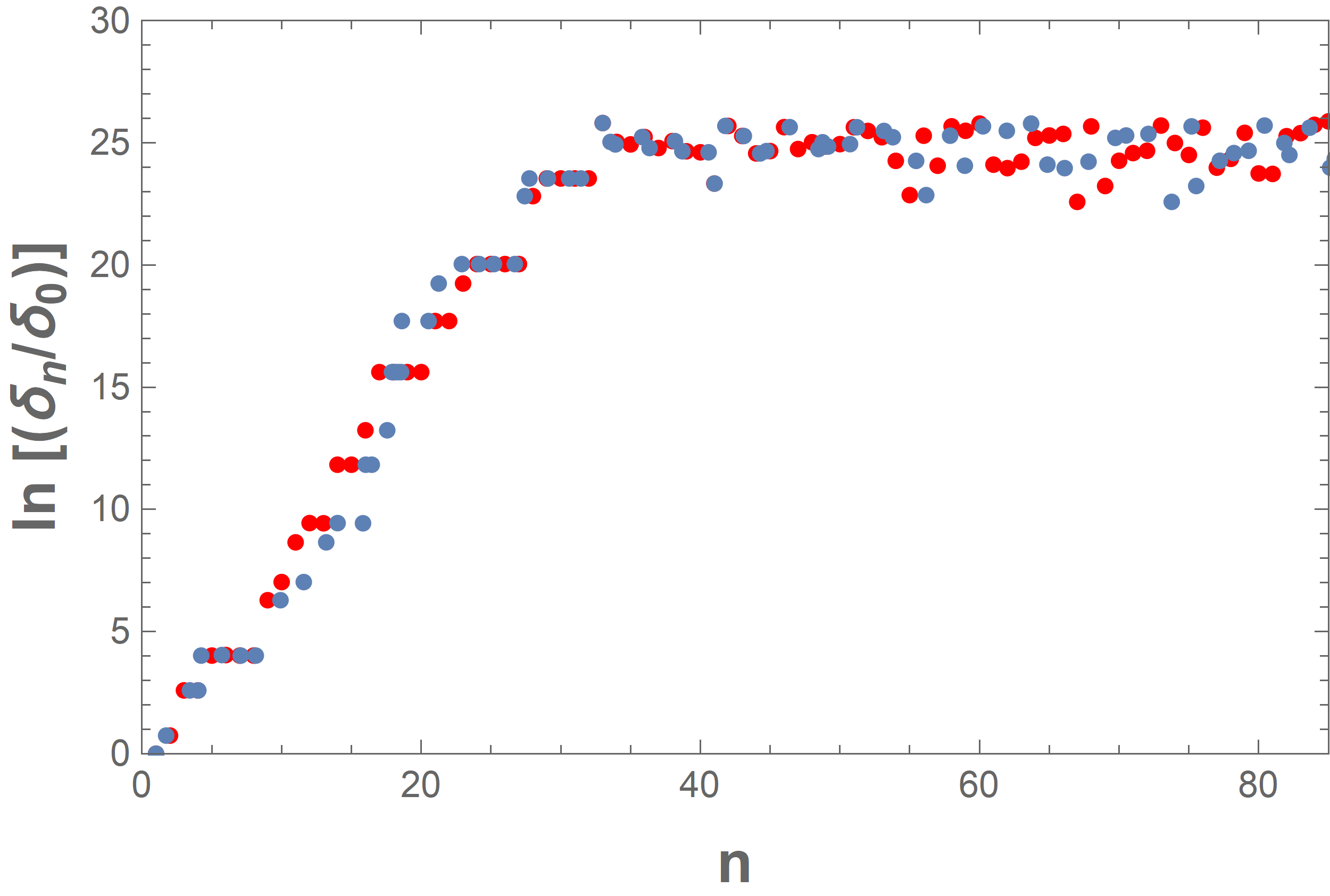}
    \caption{Fig.\,\ref{fig:dots a} and Fig.\,\ref{fig:dots b} plotted together.}
    \label{fig:truncation2}
\end{figure}

Now, in Fig.\,\ref{fig:truncation2}, we plot the graphs together, observing that both follow a narrow lane and exhibit very similar growth rates. This observation has motivated the use of the average collision distance as a unit for OTOCs when comparing the growth rates of the OTOCs to the classical growth rates of separation. As the number of random initial points increases, the average Lyapunov exponent appears to converge to a more accurate value.\par

\bibliographystyle{JHEP}
\bibliography{billiard.bib}

\providecommand{\href}[2]{#2}\begingroup\raggedright\begin{thebibliography}{10}

\bibitem{larkin1969quasiclassical}
A.I.~Larkin and Y.N.~Ovchinnikov, \emph{Quasiclassical method in the theory of
  superconductivity}, {\emph{Soviet Physics JETP} {\bfseries 28} (1969) 1200}.

\bibitem{Shenker_2014}
S.H.~Shenker and D.~Stanford, \emph{Black holes and the butterfly effect},
  \href{https://doi.org/10.1007/jhep03(2014)067}{\emph{Journal of High Energy
  Physics} {\bfseries 2014} (2014) }.

\bibitem{Shenker2_2014}
S.H.~Shenker and D.~Stanford, \emph{Multiple shocks},
  \href{https://doi.org/10.1007/jhep12(2014)046}{\emph{Journal of High Energy
  Physics} {\bfseries 2014} (2014) }.

\bibitem{kitaev}
A.~Kitaev, \emph{Hidden correlations in the hawking radiation and thermal
  noise}, {\emph{talk given at Fundamental Physics Prize Symposium} (2014) }.

\bibitem{shenker2015stringy}
S.H.~Shenker and D.~Stanford, \emph{Stringy effects in scrambling},  2015.

\bibitem{Maldacena_2016}
J.~Maldacena, S.H.~Shenker and D.~Stanford, \emph{A bound on chaos},
  \href{https://doi.org/10.1007/jhep08(2016)106}{\emph{Journal of High Energy
  Physics} {\bfseries 2016} (2016) }.

\bibitem{Rozenbaum_2017}
E.B.~Rozenbaum, S.~Ganeshan and V.~Galitski, \emph{Lyapunov exponent and
  out-of-time-ordered correlator’s growth rate in a chaotic system},
  \href{https://doi.org/10.1103/physrevlett.118.086801}{\emph{Physical Review
  Letters} {\bfseries 118} (2017) }.

\bibitem{Gharibyan_2019}
H.~Gharibyan, M.~Hanada, B.~Swingle and M.~Tezuka, \emph{Quantum lyapunov
  spectrum}, \href{https://doi.org/10.1007/jhep04(2019)082}{\emph{Journal of
  High Energy Physics} {\bfseries 2019} (2019) }.

\bibitem{Hashimoto_2017}
K.~Hashimoto, K.~Murata and R.~Yoshii, \emph{Out-of-time-order correlators in
  quantum mechanics},
  \href{https://doi.org/10.1007/jhep10(2017)138}{\emph{\href{https://doi.org/10.1007/JHEP10(2017)138}{Journal
  of High Energy Physics}} {\bfseries 2017} (2017) }.

\bibitem{Jalabert_2018}
R.A.~Jalabert, I.~García-Mata and D.A.~Wisniacki, \emph{Semiclassical theory
  of out-of-time-order correlators for low-dimensional classically chaotic
  systems}, \href{https://doi.org/10.1103/physreve.98.062218}{\emph{Physical
  Review E} {\bfseries 98} (2018) }.

\bibitem{swingle2018unscrambling}
B.~Swingle, \emph{Unscrambling the physics of out-of-time-order correlators},
  \href{https://doi.org/10.1038/s41567-018-0295-5}{\emph{\href{https://doi.org/10.1038/s41567-018-0295-5}{Nature
  Physics}} {\bfseries 14} (2018) 988}.

\bibitem{garciamata2022outoftimeorder}
I.~Garc\'\i{}a-Mata, R.A.~Jalabert and D.A.~Wisniacki, \emph{{Out-of-time-order
  correlators and quantum chaos}},
  \href{https://doi.org/10.4249/scholarpedia.55237}{\emph{Scholarpedia}
  {\bfseries 18} (2023) 55237}
  [\href{https://arxiv.org/abs/2209.07965}{{\ttfamily 2209.07965}}].

\bibitem{Rozenbaum_2019}
E.B.~Rozenbaum, S.~Ganeshan and V.~Galitski, \emph{Universal level statistics
  of the out-of-time-ordered operator},
  \href{https://doi.org/10.1103/physrevb.100.035112}{\emph{Physical Review B}
  {\bfseries 100} (2019) }.

\bibitem{time-window}
I.~Garc\'{\i}a-Mata, M.~Saraceno, R.A.~Jalabert, A.J.~Roncaglia and
  D.A.~Wisniacki, \emph{Chaos signatures in the short and long time behavior of
  the out-of-time ordered correlator},
  \href{https://doi.org/10.1103/PhysRevLett.121.210601}{\emph{\href{https://link.aps.org/doi/10.1103/PhysRevLett.121.210601}{Phys.
  Rev. Lett.}} {\bfseries 121} (2018) 210601}.

\bibitem{gutzwiller}
M.~Gutzwiller, \emph{The semi-classical quantization of chaotic hamiltonian
  systems},  in \emph{Chaos and Quantum Physics}, Z.-J.~Giannoni, Voros, ed.,
  pp.~201--250, North Holland, Amsterdam (1991).

\bibitem{Markovi__2022}
D.~Markovi{\'{c}} and M.~{\v{C}}ubrovi{\'{c}}, \emph{Detecting few-body quantum
  chaos: out-of-time ordered correlators at saturation},
  \href{https://doi.org/10.1007/jhep05(2022)023}{\emph{Journal of High Energy
  Physics} {\bfseries 2022} (2022) }.

\bibitem{PhysRevResearch.3.023214}
N.~Anand, G.~Styliaris, M.~Kumari and P.~Zanardi, \emph{Quantum coherence as a
  signature of chaos},
  \href{https://doi.org/10.1103/PhysRevResearch.3.023214}{\emph{Phys. Rev.
  Res.} {\bfseries 3} (2021) 023214}.

\bibitem{Sinai_1970}
Y.G.~Sinai, \emph{Dynamical systems with elastic reflections},
  \href{https://doi.org/10.1070/RM1970v025n02ABEH003794}{\emph{Russian
  Mathematical Surveys} {\bfseries 25} (1970) 137}.

\bibitem{infinite_horizon_billiards}
L.~Zarfaty, A.~Peletskyi, E.~Barkai and S.~Denisov, \emph{Infinite horizon
  billiards: Transport at the border between gauss and l\'evy universality
  classes}, \href{https://doi.org/10.1103/PhysRevE.100.042140}{\emph{Phys. Rev.
  E} {\bfseries 100} (2019) 042140}.

\bibitem{Sinai1970}
Y.G.~Sinai, \emph{Dynamical systems with elastic reflections},
  \href{https://doi.org/10.1070/RM1970v025n02ABEH003794}{\emph{Russian
  Mathematical Surveys} {\bfseries 25} (1970) 137}.

\bibitem{altmann2007intermittent}
E.G.~Altmann, \emph{Intermittent chaos in Hamiltonian dynamical systems}, Ph.D.
  thesis, Verlag nicht ermittelbar, 2007.

\bibitem{cardioid1}
M.P.~Wojtkowski, \emph{Principles for the design of billiards with nonvanishing
  lyapunov exponents}, {\emph{Communications in Mathematical Physics}
  {\bfseries 105} (1986) 391}.

\bibitem{cardioid2}
D.~{Sz{\'a}sz}, \emph{{On the K-property of some planar hyperbolic billiards}},
  \href{https://doi.org/10.1007/BF02099399}{\emph{Communications in
  Mathematical Physics} {\bfseries 145} (1992) 595}.

\bibitem{cardioid3}
R.~Markarian, \emph{New ergodic billiards: exact results},
  \href{https://doi.org/10.1088/0951-7715/6/5/009}{\emph{Nonlinearity}
  {\bfseries 6} (1993) 819}.

\bibitem{salazar2012classical}
R.~Salazar, G.~Téllez, D.~Jaramillo and D.L.~González, \emph{Classical and
  quantum chaos in the diamond shaped billiard},  2012.

\bibitem{salazar2}
D.J.~R.~Salazar, G.~Tellez and D.~Gonzalez, \emph{Chaos in the diamond-shaped
  billiard with rounded crown},
  \href{https://doi.org/https://doi.org/10.18257/raccefyn.99}{\emph{Rev. Acad.
  Colomb. Cienc. Ex. Fis. Nat} {\bfseries 39} (2015) }.

\bibitem{Hashimoto2}
T.~Akutagawa, K.~Hashimoto, T.~Sasaki and R.~Watanabe, \emph{Out-of-time-order
  correlator in coupled harmonic oscillators},
  \href{https://doi.org/10.1007/jhep08(2020)013}{\emph{Journal of High Energy
  Physics} {\bfseries 2020} (2020) }.

\bibitem{late_time_1}
M.~Guti\'errez and A.~Goussev, \emph{Long-time saturation of the loschmidt echo
  in quantum chaotic billiards},
  \href{https://doi.org/10.1103/PhysRevE.79.046211}{\emph{Phys. Rev. E}
  {\bfseries 79} (2009) 046211}.

\bibitem{late_time_2}
M.~Sieber and K.~Richter, \emph{Correlations between periodic orbits and their
  role in spectral statistics},
  \href{https://doi.org/10.1238/Physica.Topical.090a00128}{\emph{Physica
  Scripta} {\bfseries 2001} (2001) }.

\bibitem{late_time_3}
B.~Gutkin, D.~Waltner, M.~Guti\'errez, J.~Kuipers and K.~Richter, \emph{Quantum
  corrections to fidelity decay in chaotic systems},
  \href{https://doi.org/10.1103/PhysRevE.81.036222}{\emph{Phys. Rev. E}
  {\bfseries 81} (2010) 036222}.

\bibitem{Powell}
J.L.~Powell and B.~Crasemann, \emph{Quantum Mechanics}, Addison-Wesley (1961).

\end{thebibliography}\endgroup

\end{document}